\documentclass[12pt,a4paper,preprint,superscriptaddress,aip,pop,showkeys]{revtex4-1}

\usepackage{float}
\usepackage{amsmath,amssymb}
\usepackage[utf8]{inputenc}
\usepackage{graphicx}
\usepackage[english]{babel}
\makeatletter\AtBeginDocument{\let\@elt\relax}\makeatother
\usepackage{tabularx}
\newcolumntype{L}[1]{>{\hsize=#1\hsize\raggedright\arraybackslash}X}
\newcolumntype{R}[1]{>{\hsize=#1\hsize\raggedleft\arraybackslash}X}
\newcolumntype{C}[2]{>{\hsize=#1\hsize\columncolor{#2}\centering\arraybackslash}X}

\usepackage{hyperref}

\usepackage{xcolor}

\makeatletter
\let\@fnsymbol\@fnsymbol@latex
\@booleanfalse\altaffilletter@sw
\makeatother

\begin{document}

\date{\today}

\title{Kinetic simulation of electron cyclotron resonance assisted gas breakdown in split-biased waveguides for ITER collective Thomson scattering diagnostic}

\author{Jan Trieschmann}
\email[]{jan.trieschmann@b-tu.de}
\affiliation{Brandenburg University of Technology Cottbus-Senftenberg, Chair for Electrodynamics and Physical Electronics, 03046 Cottbus, Germany}
\author{Axel Wright Larsen}
\affiliation{Technical University of Denmark, Department of Physics, 2800 Kgs. Lyngby, Denmark}
\author{Thomas Mussenbrock}
\affiliation{Ruhr University Bochum, Applied Electrodynamics and Plasma Technology, 44801 Bochum, Germany}
\author{Søren Bang Korsholm}
\affiliation{Technical University of Denmark, Department of Physics, 2800 Kgs. Lyngby, Denmark}

%\keywords{non-linear phenomena, computer simulation: Monte Carlo, measurement methods: RF and microwave, Other}

\begin{abstract}
For the measurement of the dynamics of fusion-born alpha particles $E_\alpha \leq 3.5$\,MeV in ITER using collective Thomson scattering (CTS), safe transmission of a gyrotron beam at mm-wavelength (1\,MW, 60\,GHz) passing the electron cyclotron resonance (ECR) in the in-vessel tokamak `port plug' vacuum is a prerequisite. Depending on neutral gas pressure and composition, ECR-assisted gas breakdown may occur at the location of the resonance, which must be mitigated for diagnostic performance and safety reasons. The concept of a split electrically biased waveguide (SBWG) has been previously demonstrated in [C.P.\ Moeller, U.S. Patent 4,687,616 (1987)]. The waveguide is longitudinally split and a kV bias voltage applied between the two halves. Electrons are rapidly removed from the central region of high radio frequency electric field strength, mitigating breakdown. As a full scale experimental investigation of gas and electromagnetic field conditions inside the ITER equatorial port plugs is currently unattainable, a corresponding Monte Carlo simulation study is presented. Validity of the Monte Carlo electron model is demonstrated with a prediction of ECR breakdown and the mitigation pressure limits for the above quoted reference case with $^1$H$_2$ (and pollutant high $Z$ elements). For the proposed ITER CTS design with a 88.9\,mm inner diameter SBWG, ECR breakdown is predicted to occur down to a pure $^1$H$_2$ pressure of 0.3\,Pa, while mitigation is shown to be effective at least up to 10\,Pa using a bias voltage of 1\,kV. The analysis is complemented by results for relevant electric/magnetic field arrangements and limitations of the SBWG mitigation concept are addressed.

Copyright (2021) Author(s). This article is distributed under a Creative Commons Attribution (CC BY) License.
\end{abstract}

\maketitle

\section{Introduction}
ITER will be the first fusion reactor to achieve a fusion power gain of $Q \geq 1$ (goal $Q=10$). Collective Thomson scattering (CTS) has been proposed as a primary diagnostic for the measurement of fusion-born alpha particles ($E_\alpha < 3.5$\,MeV).\cite{korsholm_design_2019,salewski_alpha-particle_2018} It relies on scattering of an intense gyrotron beam at mm-wavelength (1\,MW, 60\,GHz) by fluctuations in the plasma, which are captured by receiver mirrors. The fast ions' velocity distribution is inferred from the measured spectrum. The gyrotron beam must pass the electron cyclotron resonance (ECR, where $\omega_\text{rf} = \omega_\text{ce}$) within the in-vessel waveguide of the diagnostic, subject to the tokamak 'port plug' vacuum. However, the intense radio frequency (RF) radiation of the beam may cause an ECR-assisted electric gas breakdown within the waveguide, which may lead to failure and damage of the diagnostic. This phenomenon is qualitatively depicted by the Paschen law. \cite{paschen_uber_1889,townsend_theory_1910} Gas breakdown and the development of a discharge occurs when the gas pressure is (i) high enough so that ionizing collisions of electrons (which are accelerated by the driving electric field) with the neutral gas atoms are frequent enough to compensate for diffusion and losses to the walls, but (ii) low enough that the electrons gain sufficient energy for ionization (from the electric field) in between subsequent collisions. A positive net balance between electron count gain and loss results in an exponential growth in the free electron population and a corresponding ionization avalanche,
\begin{align}
\frac{\partial n_\text{e}}{\partial t} &= n_\text{e} (\nu_\text{iz} - \nu_\text{loss}), \label{eq:balance} \\
n_\text{e}(t) &= n_\text{e,0} \exp\{(\nu_\text{iz} - \nu_\text{loss})\,t\}.
\end{align}
The minimum voltage to maintain a discharge is correspondingly specified at break even. On breakdown, the resulting transition from an under- to an over-dense regime, where a RF electric field is substantially affected by the plasma, corresponds to an angular frequency $\omega_\text{rf}$ approximately equal to the electron plasma frequency, $\omega_\text{rf} \approx \omega_\text{pe}(n_\text{e})$. It is marked by the critical electron density $n_\text{e,crit}=\omega_\text{rf}^2 \epsilon_0 m_\text{e}/e^2$. \cite{lieberman_principles_2005} $e$ is the elementary charge, $m_\text{e}$ is the electron mass, and $\epsilon_0$ is the vacuum permittivity. While static and RF breakdown are conceptually related, they differ quantitatively. \cite{paschen_uber_1889,townsend_theory_1910,macdonald_high_1949-1,lieberman_principles_2005}

Two contributions govern breakdown. On the one hand, the electron count gain is determined by the mean ionization frequency
\begin{align}
\nu_\text{iz} = \frac{\langle \sigma_\text{iz}(v_\text{e}) v_\text{e} \rangle \, p}{k_\text{B} T},
\label{eq:meancollisionrate}
\end{align}
which scales proportional to gas pressure $p$, electron velocity $v_\text{e}$, and velocity dependent ionization collision cross section $\sigma_\text{iz}$. Therein, parentheses $\langle . \rangle$ denote the mean value, averaged over the electron energy distribution, $T$ is the neutral gas temperature, and $k_\text{B}$ is Boltzmann constant. On the other hand, the electron count loss is determined by electron diffusion loss. It is governed by the rate $\nu_\text{loss} = \langle v_\text{e} \rangle / L $ with which electrons are lost from the volume, where breakdown is investigated, at a typical distance $L$ (e.g., waveguide diameter $2L = D = 88.9$\,mm). Assuming an energy independent ionziation collision cross section, the balance of the processes  is approximately determined by the growth ratio $\nu_\text{iz} / \nu_\text{loss} \propto p L$. Consequently, with $\nu_\text{iz} / \nu_\text{loss} < 1$, diffusion loss typically dominates at low pressures (below 1\,Pa), resulting in a negative net electron balance and thus no breakdown. Due to the factual dependence of the ionization collision cross section on energy (cf.\ Section~\ref{ssec:collisions}), an optimum electron energy (respectively velocity) may be determined. It maximizes $\langle \sigma_\text{iz}(v_\text{e}) v_\text{e} \rangle$ and therefore, above this energy, electrons are subject to overheating (i.e., are increasingly unlikely to cause ionization).

The presence of a magnetic field (e.g., of a fusion device) significantly alters the gas breakdown dynamics in two respects: (i) diffusive transport is inhibited across magnetic field lines, reducing wall losses; and (ii) the RF beam may need to be transmitted through the ECR region. The first mechanism predominantly influences the gain/loss balance, enabling gas breakdown at much lower pressures. Regarding the second aspect, when electrons are accelerated subject to the ECR (if present) this brings about a substantial increase in electron heating (but little influence on the particle balance for electron energies above the ionization threshold). Resonance occurs when the frequency $\omega_\text{rf}$ equals an integer multiple $k$ of the electron cyclotron frequency $k \omega_\text{ce} = k e B_\perp / m_\text{e}^* = \omega_\text{rf}$ (where $B_\perp$ is the magnetic field component transverse to the RF electric field and $m_\text{e}^* = \gamma m_\text{e}$ is the relativistic electron mass; $\gamma \to 1$ is a safe approximation for ITER CTS, but should be treated with caution in ECR heating schemes). Electrons are accelerated in phase with their gyration about a magnetic field line and rapidly gain energy from the RF electric field component perpendicular to the magnetic field, $\vec{E}_\text{rf} \perp \vec{B}$. The related phenomenon of ECR-assisted gas breakdown has been studied theoretically, based on analytical model formulations, since the 1950s. \cite{lax_effect_1950,lax_cyclotron_1973} The approaches utilize global particle and energy balance relations, paired with considerations on the high frequency electron kinetics. This phenomenon is greatly exploited in heating of fusion plasmas.\cite{bornatici_electron_1983,strauss_nearing_2019}

While the afore-mentioned considerations and models approximately capture the global phenomena, they do not provide any detailed spatio-temporal information. Spatially resolved kinetic models may accurately predict these dynamics, limited by their computational requirements. To the best of our knowledge, no corresponding simulation study has been previously conducted investigating ECR-assisted gas breakdown. Conceptually similar previous studies have been mostely concerned with the decisively different DC gas breakdown and streamer regime.\cite{li_comparison_2012,teunissen_3d_2016} In addition, non-magnetized microwave air breakdown was recently investigated by means of Monte Carlo simulations, taking into account electron-surface interaction.\cite{mao_monte_2020} In particular, in the frame of RF diagnostics for fusion devices or heating schemes, such as ITER CTS or ECR heating, no theoretical study has been concerned with the analysis of mitigation schemes to prevent undesired ECR-assisted breakdown within the diagnostic's waveguide. This aspect has been addressed experimentally in previous works by Moeller et al.,\cite{moeller_avoidance_1987,[{}][{, U.S. Patent 4,687,616, August 18, 1987.}]moeller_method_1987,dellis_experience_1987,moeller_trip_1987,moeller_high_1987} who proposed a longitudinally-split electrically-biased waveguide (SBWG) design to avoid in-waveguide ECR breakdown by promoting the removal of electrons from the central region of high RF electric field strength and amplifying wall loss (detailed later).

To address the aspects of ECR breakdown and SBWG mitigation in the context of the ITER CTS SBWG design,\cite{larsen_mitigation_2019} the onset of a gyrotron pulse and the corresponding ECR breakdown dynamics are simulated and analyzed in this work, using a spatially resolved Monte Carlo electron model. Initially, a reference setup described by Moeller et al.\ \cite{moeller_avoidance_1987,[{}][{, U.S. Patent 4,687,616, August 18, 1987.}]moeller_method_1987,dellis_experience_1987,moeller_trip_1987,moeller_high_1987} and the ITER CTS configuration \cite{larsen_mitigation_2019} are introduced in Section~\ref{sec:setup}. Thereafter, the simulation approach and the fundamental prerequisites and model inputs are presented in Section~\ref{sec:model}. This is followed by simulation results for the `Moeller' reference case, which are established for verification and validation in Section~\ref{ssec:Moeller}. The ITER CTS setup is addressed subsequently in Section~\ref{ssec:ITER}, and the effectiveness of SBWG breakdown mitigation is demonstrated for this setup. After a discussion of the results, the work is concluded.

\begin{table*}[b!]
\centering
\begin{tabularx}{\textwidth}{ X X X }
& Moeller \cite{moeller_avoidance_1987,[{}][{, U.S. Patent 4,687,616, August 18, 1987.}]moeller_method_1987,moeller_high_1987,moeller_trip_1987,dellis_experience_1987} & ITER CTS\cite{larsen_mitigation_2019} \\
\hline
\hline
Gas & Hydrogen (+ outgassing) & Hydrogen\\
Pressure $p$ & $10^{-2} \dots 0.133$\,Pa & $<2$\,Pa\\
Temperature $T$ & $400$\,K & $400$\,K\\
Wall material & Stainless steel & Cu (coating or CuCrZr)\\
Magnetic field $\vec{B}$ & Solenoid $B_\mathsf{z} \approx 2.14$\,T & Toroidal $B \approx 2.14$\,T\\
Gyrotron power $P_\text{gyr}$ & 200\,kW & 1\,MW\\
SBWG inner diameter $D$ & 19.1\,mm & 88.9\,mm\\
Mode & TE$_{11}$ & LP$_{01}$/HE$_{11}$\\
1st zero of mode $x$ & $1.8411$ & $2.4045$\\
Pulse duration $\tau$ & 5\,ms & 1 \dots 1000\,ms \\
Pulse rise time $\tau_\text{rise}$ & & $\gtrsim 100\,\mu$s\\
Maximum RF field $E_0$ \cite{gould_handbook_1956} & $\approx 1500$\,kV/m & $\approx 670$\,kV/m\\
Bias potential $V_\text{bias}$ & 1 \dots 2.3\,kV & 1 \dots 2\,kV\\
\hline
\hline
\end{tabularx}
\caption{Summary of parameters specified for the respective configuration and approximated in the simulation for the two scenarios of concern.}
\label{tab:configurations}
\end{table*}

\section{Setup}
\label{sec:setup}
In the following, the two relevant configurations are presented. The main parameters and features are collected in Table~\ref{tab:configurations}. Schematics of the configurations are depicted in Figure~\ref{fig:configurations}. When not specified otherwise, a homogeneous magnetic field magnitude $B=2.14$\,T is imposed in both cases (note the different magnetic field directions for the `Moeller' and ITER CTS cases).

\subsection{`Moeller' configuration}
\begin{figure*}[t!]
\centering
\includegraphics[width=0.32\textwidth]{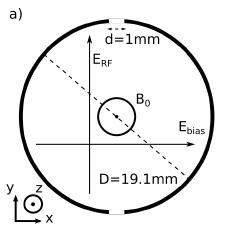}
\includegraphics[width=0.32\textwidth]{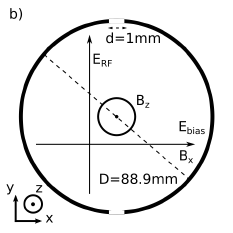}
\hspace*{1ex}
\includegraphics[width=0.26\textwidth]{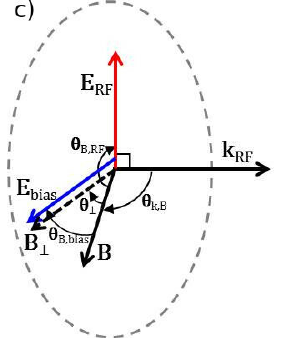}
\caption{Schematics of the configurations: a) 2D `Moeller' scenario, b) 2D ITER CTS scenario ($E_\text{rf} \perp B$), and c) 3D ITER CTS field arrangement. The dominant directions of the respective fields are indicated by arrows. $D$ is the waveguide inner diameter, $d$ the gap between the SBWG halves.}
\label{fig:configurations}
\end{figure*}

The results due to Moeller et al. \cite{moeller_avoidance_1987,[{}][{, U.S. Patent 4,687,616, August 18, 1987.}]moeller_method_1987,dellis_experience_1987,moeller_trip_1987,moeller_high_1987} were obtained considering a circular smooth waveguide with $D=19.1$\,mm inner diameter and TE$_{11}$ microwave mode propagation. \cite{balanis_advanced_1989} The waveguide surface material was stainless steel. \cite{moeller_high_1987} The magnetic field was created by a solenoid magnet, therefore, pointing predominantly in the axial direction (cf.\ Figure~\ref{fig:configurations}a). A two dimensional model well represents the system inside the solenoid. It takes the invariant axial direction as $z$ and assumes a homogeneous axial magnetic field $B_\text{z}=2.14$\,T. The TE$_{11}$ polarization is taken in the $y$ direction. The bias electric field $\vec{E}_\text{bias}$ (consistently calculated numerically) points predominantly in the $x$ direction. Therewith, effective ECR heating, as well as breakdown mitigation mainly based on an $\vec{E}_\text{bias}\times\vec{B}$ drift are realized. For simplification, the configuration is simulated in a two dimensional transversal cross section of the waveguide. Moeller reports on gas breakdown and mitigation in hydrogen $^1$H$_2$. However, as noted outgassing from the walls may have had an important contribution. \cite{moeller_high_1987}

\subsection{ITER CTS configuration}
\begin{figure}[t!]
\includegraphics[width=8cm]{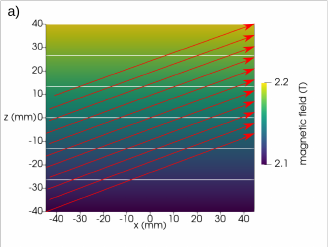}
\includegraphics[width=8cm]{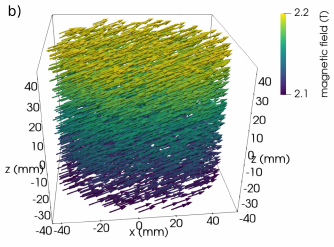}
\caption{a) Magnetic field plotted over a cross sectional cut in the $x-z$ plane. The white lines indicate magnetic field strength isocurves and the red lines the direction of representative field lines. b) Structure of the magnetic field within the waveguide section used in the model.}
\label{fig:ITERmagneticfield}
\end{figure}

The ITER CTS configuration differs from the `Moeller' case in several respects. Firstly, the proposed corrugated waveguide with inner diameter $D=88.9$\,mm supports LP$_{01}$ (HE$_{11}$ respectively) mode propagation. \cite{kowalski_linearly_2010} Moreover, the waveguide surface material is most certainly CuCrZr (or Cu coating), which among other aspects suppresses outgassing from the walls (stainless steel retains much H$_2$O).\cite{moeller_high_1987,moeller_trip_1987} Secondly, the magnetic field structure within the waveguide is neither axial nor transversal, but entails contributions in both directions. The relevant section of the waveguide is subject to a magnetic field strength of $B \approx 2.14$\,T to be affected by ECR heating. This is schematically depicted in Figures~\ref{fig:configurations}b) and c) and Figure~\ref{fig:ITERmagneticfield}, whereas the waveguide axis is aligned with the $z$ axis. The linearly polarized RF electric field $\vec{E}_\text{rf}$ points in the $y$ direction, whereas the magnetic field $\vec{B}$ has components in the $x$ and $z$ direction. The bias electric field $\vec{E}_\text{bias}$ points predominantly in the $x$ direction. Two model representations are investigated both assuming a hydrogen $^1$H$_2$ neutral gas background:

(i) A two dimensional setup which neglects the axial magnetic field component $B_\text{z}$, but maintains the magnetic field magnitude $B=2.14$\,T and includes all governing mechanisms concerning ECR breakdown as well as mitigation (cf.\ Figure~\ref{fig:configurations}b).

(ii) A three dimensional setup depicting a waveguide section ($L=80$\,mm) and which includes a more realistic magnetic field (including relevant $B \approx 2.14$\,T; cf.\ Figures~\ref{fig:configurations}b and \ref{fig:ITERmagneticfield}). While the axial magnetic field is $B_\text{z}=0.735$\,T, the transverse magnetic field $B_\text{x}$ varies with $dB_\text{x}/dz = 1.674$\,T/m. Magnetic field lines are oblique at an angle with the $z$ axis. The 3D magnetic field structure is approximated from the complicated 3D structure of the ITER tokamak baseline scenario.\cite{noauthor_iter_d_q2j6me_nodate} Due to computational limitations, only a limited number of representative cases were simulated for this setup.

\subsection{Electron cyclotron resonance and gyrotron excitation}
The RF excitation frequency is $f=60$\,GHz for both cases discussed. The corresponding RF period is $T=1/f\approx16.67$\,ps. The ECR frequency $\omega_\text{ce}$ defines the intrinsic timescale of the electron dynamics which needs to be resolved. At the gyrotron beam frequency $\omega_\text{rf} \approx \omega_\text{ce}$, the magnetic field magnitude of the fundamental resonant mode is $B \approx 2.14$\,T, whereas the second harmonic resonant mode is excited when $\omega_\text{rf} \approx 2\omega_\text{ce}$ at $B \approx 1.07$\,T.

Moeller reports pulses of $\tau=5$\,ms, while ITER CTS design specifies typical pulse lengths of $\tau=10$\,ms (between $\tau=1\,\text{ms} \text{~to~} 1$\,s depending on operating conditions). The pulse rise time is typically on the order of $\tau_\text{rise} \approx 100\,\mu$s, \cite{kartikeyan_gyrotrons_2004} which is in line with the design specifications for ITER CTS. Moeller does not specify the pulse rise time.

\begin{figure}[b!]
\includegraphics[width=8cm]{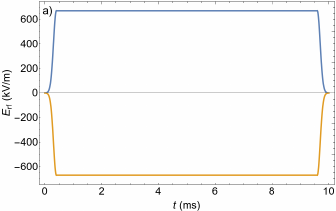}
\includegraphics[width=8cm]{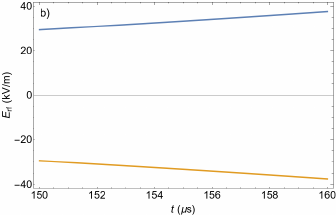}
\caption{a) Schematic of a typical gyrotron electric field pulse envelope with rise time of $\tau_\text{rise}=100\,\mu$s, a total pulse duration of $\tau=10$\,ms, and maximum RF electric field of $E=670$\,kV/m, b) zoom to the relevant $\mu$s timescale $150\,\mu\text{s} \leq t \leq 160\,\mu$s  (blue upper, yellow lower bound). For illustration, the pulse form is evaluated from Eq.~\eqref{eq:pulseform} with $\sigma=\tau_\text{rise}$ and ITER CTS parameters.}
\label{fig:pulseform}
\end{figure}

Both the pulse length and the rise time are orders of magnitude larger than the intrinsic timescale of the ECR heating dynamics (see above paragraph). Moreover, while the electron heating timescale needs to be resolved, the dominant timescale for gas breakdown is governed by electron impact ionization collisions with the gas background. The mean collision time for ionization
\begin{align}
\tau_\text{c} = \frac{1}{\nu_\text{iz}}
\label{eq:meancollisiontime}
\end{align}
provides the relevant timescale. It represents an intermediate timescale between the ECR heating time and the pulse rise time and duration. As tabulated in Table~\ref{tab:collisiontime}, it is below $\tau_\text{c} \lesssim 5\,\mu$s for the relevant pressure range.

The excitation pulse envelope is schematically depicted in Figure~\ref{fig:pulseform} for a simplified gyrotron pulse. It is approximated by a Gaussian ramp up phase defined by the rise time $\tau_\text{rise}=100\,\mu$s, a plateau defined by the pulse duration $\tau = 10$\,ms, and an analogue decay phase $\tau_\text{decay}=\tau_\text{rise}$. Notably, the pulse magnitude is approximately constant on the timescale of gas breakdown ($10\,\mu$s depicted in Figure~\ref{fig:pulseform}b). Hence, the RF modulated pulse waveform can be considered a continuous wave (CW) concerning the intrinsic electron heating dynamics, as well as the collisional processes leading to gas breakdown. The time interval which is most relevant for gas breakdown should be considered. The latter is dictated by the RF electric field magnitude $E_0$ which is required to heat electrons to sufficient energies -- this is detailed in Section~\ref{ssec:modeling}.

\begin{figure}[t!]
\includegraphics[width=5.4cm]{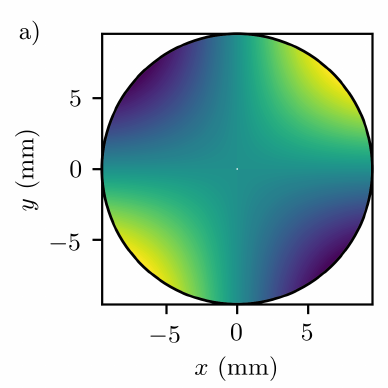}
\includegraphics[width=5.4cm]{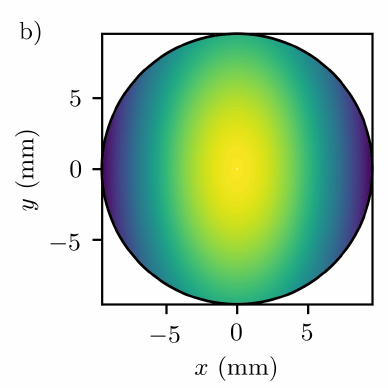}
\includegraphics[width=5.4cm]{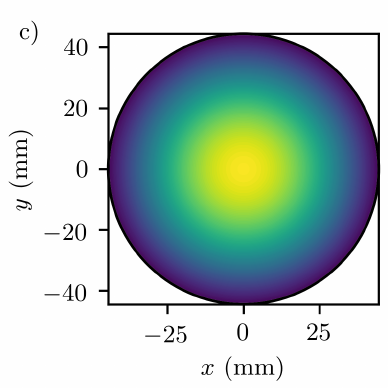}
\caption{Electric field patterns in `Moeller' waveguide for TE$_{11}$ a) $E_\text{x}$, b) $E_\text{y}$ and ITER CTS waveguide for LP$_{01}$ c) $E_\text{y}$, $E_\text{x}$ zero. The former has a high RF electric field in a larger fraction of the waveguide.}
\label{fig:modes}
\end{figure}
 
\begin{table}[b!]
\centering
\begin{tabularx}{0.5\textwidth}{ X X }
$p$ (Pa) & $\tau_\text{c}$ ($\mu$s) \\
\hline
\hline
0.01 & $5.5$ \\
\hline
0.1 & $5.5 \cdot 10^{-1}$ \\
\hline
1 & $5.5 \cdot 10^{-2}$ \\
\hline
10 & $5.5 \cdot 10^{-3}$ \\
\hline
\hline
\end{tabularx}
\caption{Mean time for electron impact ionization collisions in hydrogen obtained from Eq.~\eqref{eq:meancollisiontime}. Values are given for an electron velocity $v_\text{e}=v_\text{max} \approx 10^7$\,m/s ($E_\text{max} \approx 285$\,eV), where $\max(\sigma_\text{iz}(v_\text{e}) v_\text{e}) = \sigma_\text{iz}(v_\text{max}) v_\text{max} \approx 10^{-13}\,\text{m}^3/s$ (cf.\ Section~\ref{ssec:collisions}). Neutral gas temperature is $T=400$\,K.}
\label{tab:collisiontime}
\end{table}

The field pattern within the waveguide is defined by the propagation modes TE$_{11}$ for `Moeller' and LP$_{01}$ for ITER CTS. In the latter case, a corrugated waveguide is used. Corrugations on the waveguide surface are located, where the field magnitude is minimum (i.e., close to zero). Consequently, their effect on gas breakdown is negligible and corrugations have been omitted in the geometry of the simulation. The electric field patterns of the relevant modes are depicted in Figure~\ref{fig:modes} for both configurations. While the TE$_{11}$ mode contains two transverse electric field contributions $E_\text{x}$ and $E_\text{y}$, LP$_{01}$ entails only an $E_\text{y}$ contribution. In both cases, the $y$ component dominates and is peaked in the center of the waveguide. The region of high RF electric field strength is more distributed for TE$_{11}$ compared to LP$_{01}$, resulting in a smaller peak electric field at equal intensity and waveguide inner diameter. See references \cite{balanis_advanced_1989,kowalski_linearly_2010} for more details.

\subsection{Longitudinally-split electrically-biased waveguide}
\label{ssec:SBWG}
The SBWG mitigation scheme has been originally proposed and detailed by Moeller. \cite{moeller_avoidance_1987,[{}][{, U.S. Patent 4,687,616, August 18, 1987.}]moeller_method_1987,dellis_experience_1987,moeller_trip_1987,moeller_high_1987} The fundamental reasoning is to enhance drift-diffusion of electrons from the central resonant region of high RF electric field strength, before these electrons initiate or participate in a gas breakdown avalanche. This is achieved by longitudinally splitting the waveguide in two half-cylinders and applying a DC bias voltage, $V_\text{bias}$, between them. (Note that the symbol $V_\text{bias}$ is used to denote the electric potential difference $V_\text{bias}=\phi_\text{anode} - \phi_\text{cathode}$.) For `Moeller' a bias voltage in the range $V_\text{bias} = 1 \text{ to } 2.3$\,kV has been reported, while ITER CTS is specified with $V_\text{bias} = 1 \text{ to }2$\,kV. Depending on the geometry and the resulting bias electric field $\vec{E}_\text{bias}$, two mechanisms drive electrons out of the central waveguide region and to the walls:

(i) Electron acceleration and transport along magnetic field lines may proceed freely due to $\vec{E}_\text{bias}\parallel\vec{B}$. Assuming a uniform bias electric field $E_\text{bias} = V_\text{bias}/D$ and neglecting collisions, the maximum sweep time can be approximated by
\begin{align}
\tau_\text{sweep} = \sqrt{\frac{2 D^2 m_\text{e}}{e V_\text{bias}}}.\label{eq:sweeptime}
\end{align}
This assumes an electron acceleration and trajectory from one side of the waveguide to the opposite side and provides an upper bound. For ITER CTS with 1\,kV bias voltage this evaluates to $\tau_\text{sweep} \approx 10$\,ns, for 2\,kV it corresponds to $\tau_\text{sweep} \approx 6.7$\,ns.

(ii) Electron transport may follow an $\vec{E}_\text{bias} \times \vec{B}$ drift with velocity $v_\text{sweep} = |\vec{E}_\text{bias} \times \vec{B}|/B^2$ omitting collisions. This results in an approximate maximum time of travel to the wall $\tau_\text{sweep} = D/v_\text{sweep}$. For the `Moeller' scenario this gives $\tau_\text{sweep} \approx 0.8\,\mu$s. \cite{moeller_avoidance_1987,[{}][{, U.S. Patent 4,687,616, August 18, 1987.}]moeller_method_1987}

Electron transport is principally inhibited by collisions with the gas background. By comparison with the approximate mean collision times from Table~\ref{tab:collisiontime}, in the proposed scenario the influence is negligible when $\tau_\text{sweep}/\tau_\text{c} \lesssim 1$. This relates to low pressures of $p \lesssim 0.1$\,Pa for `Moeller' and $p \lesssim 1$\,Pa for ITER CTS. When collisions are frequent, they not only contribute to slowdown but also to multiplication of electrons, due to ionization. Hence, at higher pressures especially mechanism (i) may additionally contribute to the gas breakdown dynamics. The influence of the latter, as well as nonuniform bias electric fields cannot straightforwardly be included in the presented approximations. This stresses the need for accurate numerical simulation predictions.

\section{Model}
\label{sec:model}
\subsection{Kinetic electron model}
\label{ssec:modeling}
The utilized Monte Carlo electron model was developed within the OpenFOAM framework. \cite{weller_openfoam_2020,scanlon_open_2010} The underlying particle in cell particle (PIC) code has been used for pure simulation studies, \cite{bobzin_continuum_2013,trieschmann_transport_2015,trieschmann_neutral_2018} but has also been validated with experiments. \cite{layes_species_2017,trieschmann_combined_2018,kirchheim_improved_2019} In addition, the code has been validated against the benchmarked reference PIC code \textit{yapic}. \cite{turner_simulation_2013,trieschmann_particle_2017} By considering an ensemble of pseudo-electrons  (each representing a given number of physical electrons) a kinetic description is established. The pseudo-electrons are traced in a given 3D geometry based on an unstructured tetrahedral mesh. Their individual motion is subject to Newton's laws, following microscopic Coulomb and Lorentz force terms. Consequently, average force terms such as the ponderomotive force are intrinsically included. \cite{boris_relativistic_1970,zenitani_boris_2018} Two or three spatial coordinates are considered, depending on the geometry and setup, whereas three velocity components are maintained throughout (2D-3V or 3D-3V components per electron).

Electron collisions with the gas background are included in a Monte Carlo collision scheme. The neutral gas is assumed as a stationary Maxwell-Boltzmann distributed background with temperature $T$ and adjustable gas composition. Collisional processes are incorporated using a modified no-time counter. \cite{bird_molecular_1994,trieschmann_transport_2015} The selection of different collision processes further uses a null-collision approach. \cite{skullerud_stochastic_1968} The magnetic field and the RF electric field are imposed within the domain, whereas the static bias electric field is calculated using the finite volume method.\cite{leveque_finite_2002} Feedback of charged species onto the electromagnetic fields is neglected. This approach is valid in the underdense regime with $n_\text{e} \ll n_\text{e,crit}$. Initially, a homogeneous electron density of $n_\text{e}(t=0\,\text{s}) = 10^{10}\,\text{m}^{-3}$ is imposed. The previous assumption is well justified for the subsequent evolution, since simulations starting with this low density are conducted only until gas breakdown is detected from a noticeable rise in $n_\text{e}$. The electric field due to local space charge effects can be estimated from Poisson's equation. In 1D Cartesian coordinates, the maximum electric field of a uniform charge density $\rho \approx e \cdot 10^{11}\,\text{m}^{-3}$ over the length of a waveguide diameter $D=88.9$\,mm is estimated to 160\,V/m. This is several orders of magnitude smaller than the bias electric field and the RF electric field imposed in this work, and consequently negligible. Moreover, due to the low charge carrier density, collective effects such as quasi-neutrality and Debye shielding occur on the length scale of the geometric configuration, and are negligible as well.

The pseudo-particle weight was chosen to maintain sufficient statistics ($\gtrsim$ 80 electrons per mesh cell $\gg 1$). The time step was set to $\Delta t = 200$\,fs, to capture the gyration of electrons around magnetic field lines ($\gtrsim$ 80 time steps per gyration $\gg 1$).

The computational effort of the simulation and the divergent physical timescales render an evaluation of the complete pulse waveform infeasible. As schematically depicted in Figure~\ref{fig:pulseform}, for a time interval less than approximately $10\,\mu$s the pulse magnitude is nearly time invariant. This is inherently the case for the intrinsic electron dynamics. Hence, a modified CW waveform is considered in the simulation without loss of generality. The rise of the pulse is not evaluated to scale, but an initial start-up phase using an RF modulated Gaussian
\begin{align}
E_\text{rf}(t) = E_0 \exp\left(-\frac{(t-4\sigma)^2}{2\sigma^2}\right) \, \cos(\omega_\text{rf} t),\qquad t \leq 4\sigma
\label{eq:pulseform}
\end{align}
with $\sigma = 10$\,ns and a subsequent CW signal specifies the RF electric field waveform. The maximum electric field strength $E_0$ is chosen such that a gas breakdown avalanche is initiated most effectively, taking into account electron impact ionization in the volume as well as electron-induced secondary electron emission and reflection at the surface.

Electrons need to be heated to sufficient energies, but not overheated. The latter may occur once the product of $\langle \sigma_\text{iz}(v_\text{e}) v_\text{e} \rangle$ -- which defines the collision probability -- attains a negative slope and decreases with increasing kinetic energy. For energies in the range $E \approx 70 \text{~to~} 700$\,eV, $\langle \sigma_\text{iz}(v_\text{e}) v_\text{e} \rangle$ varies less than 10\,\% from the maximum at $E \approx 285$\,eV (cf.\ Sections~\ref{ssec:collisions}). Consequently, strongest initiation of an ionization cascade is expected for the mentioned energy range. The exact energy, however, is of subordinate relevance, as long as a minimum energy of $E \gtrsim 70$\,eV is maintained.

In addition to electron impact ionization in the volume, however, an electron count balance more realistic than Equation~\eqref{eq:balance} also depends on surface processes in a realistic scenario. That is, electron-induced secondary electron emission and reflection contribute respectively. Due to markedly different energy dependencies (e.g., maximum emission at 400 to 600\,eV; cf.\ Section\,\ref{ssec:surface}), these alter the optimum RF electric field strength $E_0$, which most effectively causes gas breakdown. For the `Moeller' scenario an RF electric field magnitude, $E_0 = 80$\,kV/m, and for the ITER CTS case, $E_0 = 150$\,kV/m (for 2D) and $E_0 = 35$\,kV/m (for 3D), have been determined to most effectively initiate gas breakdown. These values have been iteratively estimated for the reference cases with $p=0.133$\,Pa for `Moeller' and $p=1$\,Pa for ITER CTS.

Following this reasoning, the proposed simulations operate at a relatively low RF electric field strength, corresponding to the start-up phase of the gyrotron pulse. This circumstance does not impose any limitation on the validity for the case with higher RF electric field strengths encountered later during the `experimental realization'. The RF electric field strength and the corresponding energy window were selected in the simulations to provide a conservative gas breakdown estimate. Higher RF electric field strengths cause electron overheating, suggesting raised pressure limits for gas breakdown. The same reasoning applies to the situation when the pressure increases during operation and becomes critically high only after the RF modulated pulse magnitude has reached its maximum.

Compared to an experimental realization, the simulation procedure differs in two aspects:

(i) The simulated RF modulated pulse increases within $\tau_\text{rise}=\sigma=10$\,ns to the CW electric field magnitude. This rise imposes a rapid excitation of the system of electrons (which are initially Maxwell-Boltzmann distributed). As apparent from the results, this is associated with a noticeable `ringing' in the average electron energy, due to a slower time response of the system. The rise time $\tau_\text{rise}=10$\,ns is chosen as a compromise, minimizing these `ringing' oscillations, but also the computational load (i.e., the time duration to be simulated).

(ii) In the proposed ITER CTS realization, the mitigation bias is designed to be constantly active. In contrast, the simulations are initially evolved until a noticeable gas breakdown occurs (exponential rise in $n_\text{e}$), and thereafter the mitigation bias voltage is switched on. On the one hand, this procedure is used to demonstrate the effectiveness of the SBWG mitigation scheme. On the other hand, it is also used because the onset of the gyrotron pulse cannot feasibly be simulated due to the simulation run-time. This and the inclusion of background ionization processes (instead of a pre-specified initial electron density) would, however, be required to capture the early phase of ECR breakdown and mitigation appropriately.

Both of the above raised aspects signify limitations of the simulation approach. However, as the conditions for ECR breakdown depicted by the procedure are more severe than in an experimental realization, the above points do not seem to entail any implications regarding the validity of the conclusions. Hence, the limits determined by the approach can be regarded as conservative bounds. This reasoning is supported by the agreement between experimental and simulation results, as elaborated for the `Moeller' reference case in Section~\ref{ssec:Moeller}.

\subsection{Surface Coefficients}
\label{ssec:surface}
Particles interacting with bounding walls are subject to several interaction mechanisms. In the present context, in particular electron-induced secondary electron emission (e-SEE, denoted by $\delta$) and electron reflection (denoted by $\eta$) are important. \cite{kollath_sekundarelektronen-emission_1956,ruzic_secondary_1982,janev_atomic_1991} Notably, e-SEE can produce a net gain in electron number count. The ITER CTS design proposes CuCrZr or ITER-grade 316LN stainless steel (SS), as material for the waveguide. Regarding the surface coefficients, CuCrZr is well approximated by Cu, SS will be Cu coated. For the `Moeller' case SS was reported as the material of choice. \cite{moeller_high_1987}

e-SEE from Cu has been investigated in numerous publications. \cite{kollath_sekundarelektronen-emission_1956,ruzic_secondary_1982,janev_atomic_1991,walker_secondary_2008,tolias_secondary_2014} In contrast, the literature basis for SS is comparably weak. \cite{janev_atomic_1991} In this work, most conservative (i.e., highest) emission values are used with the goal to represent the worst-case scenario for gas breakdown mitigation. Corresponding parameters (cf.\ Table~\ref{tab:surface}) are used to evaluate the fitting formula due to Young, \cite{young_dissipation_1957} as well as Lye and Dekker, \cite{lye_theory_1957}
\begin{align}
\delta(E) &= \frac{\delta_\text{m}}{1-e^{-x_\text{m}}} \left(\frac{E}{E_\text{m}}\right)^{1-k} \left( 1 - \exp\left[ -x_\text{m} \left( \frac{E}{E_\text{m}} \right)^k \right] \right).
\label{eq:SEE}
\end{align}
$E_\text{m}$ defines the energy of maximal electron emission, whereas $\delta_\text{m}$ specifies its value. $x_\text{m}$ and $k$ are dimensionless fitting coefficients. All listed parameters are material dependent.

\begin{table*}[t!]
\centering
\begin{tabularx}{0.7\textwidth}{ X X X }
& Cu & SS \\
\hline
\hline
$Z$ & 29 & 25.8 \\
\hline
$\eta$ at $E=1$\,keV & 0.333 & 0.317 \\
\hline
$\delta_\text{m}$ & 1.3 & 1.22 \\
\hline
$E_\text{m}$ (eV) & 600 & 400 \\
\hline
$x_\text{m}$ & 2.013 & 1.931 \\
\hline
$k$ & 1.45 & 1.487 \\
\hline
\hline
\end{tabularx}
\caption{Parameters used for the evaluation of surface coefficients using Equations~\eqref{eq:SEE} and \eqref{eq:reflection}. Cu parameters from  \cite{kollath_sekundarelektronen-emission_1956,tolias_secondary_2014}. For SS effective values have been calculated by weighting the atomic results assuming atomic fractions 70\% Fe, 20\% Cr, 10\% Ni. \cite{janev_atomic_1991}}
\label{tab:surface}
\end{table*}

\begin{figure}[b]
\includegraphics[width=8cm]{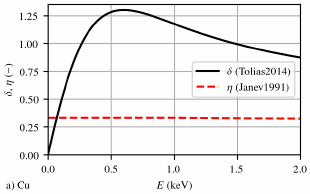}
\includegraphics[width=8cm]{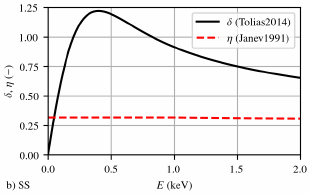}
\caption{e-SEE coefficient $\delta$ and reflection coefficient $\eta$ as a function of energy $E$ for a) Cu and b) SS. Curves are evaluated based on Equations~\eqref{eq:SEE} and \eqref{eq:reflection} using the values from Table~\ref{tab:surface}.}
\label{fig:surface}
\end{figure}

The process of electron reflection has been found to predominantly depend on atomic number $Z$.\cite{janev_atomic_1991,lye_theory_1957,hunger_measurements_1979} Following Hunger and Küchler,\cite{hunger_measurements_1979} it may be evaluated based on the proposed fitting formula
\begin{align}
\eta(E,Z) &= \left(E/\text{keV}\right)^{m(Z)} e^{c(Z)}, \qquad E \text{ in keV}
\label{eq:reflection} \\
m(Z) &= 0.1382-0.9211^{-Z}, \nonumber \\
e^{c(Z)} &= 0.1904 - 0.2236 \ln(Z) + 0.1292 \ln^2(Z) - 0.01491 \ln^3(Z). \nonumber
\end{align}
Whereas the dependence $m(Z)$ governs an energy power law, $c(Z)$ scales the total electron reflection. Both are dimensionless functions of the atomic number $Z$.

While data for Cu has been proposed, \cite{janev_atomic_1991,lye_theory_1957,hunger_measurements_1979} the uncertainty of the available data suggests that atomic fractions 70\% Fe, 20\% Cr, 10\% Ni provide a reasonable estimate for SS. These were used accordingly in the following. Surface coefficients for Cu and SS are depicted in Figure~\ref{fig:surface}. The parameters used to evaluate Equations~\eqref{eq:SEE} and \eqref{eq:reflection} are listed in Table~\ref{tab:surface}. While the e-SEE coefficient $\delta$ initially increases to $\delta_\text{m}$ with incident energy, it steadily drops for energies $E>E_\text{m}$. In contrast, the electron reflection coefficient $\eta$ is nearly constant for the relevant energies.

The process of ion-induced electron emission (i-SEE) is substantially weaker than electron-induced electron emission at ion bombardment energies $E_\text{i}$ in the eV to keV range.\cite{brown_basic_1967,janev_atomic_1991} For the present investigation, two cases need to be distinguished: (i) Ions created in volume ionization processes are not significantly heated by the RF electric field, due to their large mass and inertia. Without a bias electric field and in the absence of a fully established plasma (and corresponding boundary sheaths), ions are close to thermal equilibrium with the gas background. Their approximate average energy is correspondingly low, $E_\text{i} \lesssim 50$\,meV. (ii) With a bias electric field, the ion energy is on the order of the bias voltage, $E_\text{i} \approx e V_\text{bias} \lesssim 2$\,keV. Ions are removed from the waveguide volume within approximately the ion sweep time $\tau_\text{sweep} \approx 570$\,ns, much longer than the fast ECR heating dynamics, as estimated for $^1$H$_2$ following Equation~\eqref{eq:sweeptime}. In both cases (i) and (ii), the ion-induced electron emission process is in the potential and kinetic emission transition regime (eV to keV range) and can be consequently neglected, as reasoned by an emission coefficient of $\gamma \lesssim 0.1$.\cite{large_secondary_1962,svensson_electron_1982,zalm_ion-induced_1985,szapiro_electron_1988} Moreover, to incorporate i-SEE into the model and assess its influence on gas breakdown would require to also simulate the ion dynamics. The expected small influence does not justify these additional, significant computational costs.

Electrons emitted from the surface in the simulation are assumed to have a Maxwell-Boltzmann distribution with $k_\text{B} T_\text{e} = 2.6$\,eV (approximating fractional energy input from incident electrons). In the absence of more reliable data, electrons reflected from the surface are divided into 90\% diffuse and 10\% specularly reflected contributions, approximating measured emitted electron energy spectra. \cite{kollath_sekundarelektronen-emission_1956,janev_atomic_1991} The diffuse fraction is re-emitted identical to the primary emitted secondaries.

\subsection{Collision Processes}
\label{ssec:collisions}
Collisional interactions included in the calculation are reduced to elastic scattering and direct electron impact ionization to the singly ionized state. The cross sections have been obtained using LXCat from the IST Lisbon and Biagi database. \cite{pancheshnyi_lxcat_2012,pitchford_lxcat_2017,biagi_lxcat_nodate,alves_ist-lisbon_2014,alves_lxcat_nodate} Argon, as well as molecular hydrogen, nitrogen, and oxygen were obtained and implemented. \cite{alves_ist-lisbon_2014,alves_lxcat_nodate,phelps_span_2008,wingerden_elastic_1977,rapp_cross_1965,rapp_total_1965,gorse_progress_1987,tawara_cross_1990,simko_computer_1997,biagi_lxcat_nodate} In the absence of reliable cross section data for hydrogen isotopes (deuterium and tritium),\cite{korolov_breakdown_2015} their cross sections are approximated using hydrogen $^1$H$_2$ cross sections (which are correspondingly used throughout).

While elastic scattering cross sections reveal peculiar features associated with their species' atomic structure, the more important ionization cross sections consistently follow a general trend. Starting from the ionization threshold $E_\text{iz}$, they reveal a steep rise followed by a maximum and a subsequent decline.

\begin{figure}[t!]
\includegraphics[width=8cm]{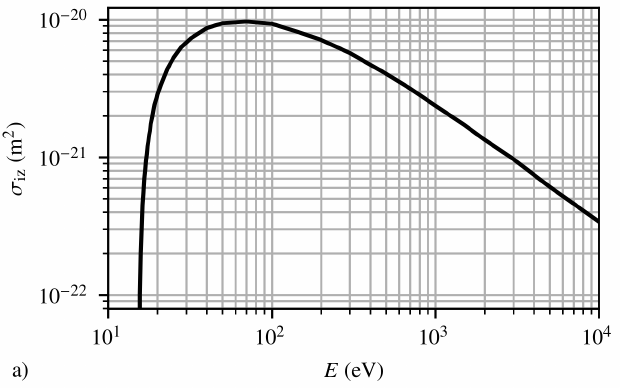}
\includegraphics[width=8cm]{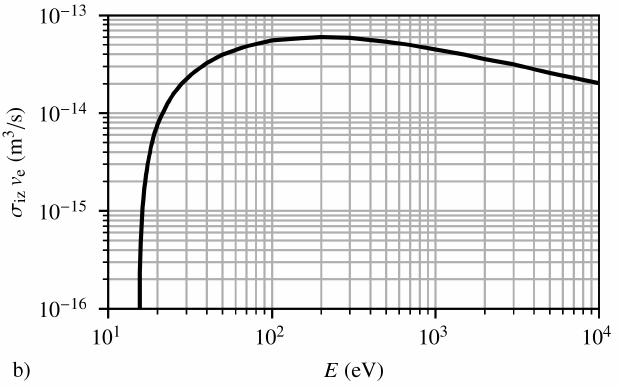}
\caption{a) Electron impact ionization collision cross section $\sigma_\text{iz}$ and b) product $\sigma_\text{iz}(v_\text{e}) v_\text{e}$  for $^1$H$_2$ plotted over energy to illustrate their characteristic dependence.}
\label{fig:sigma}
\end{figure}

As depicted in Figure~\ref{fig:sigma}a), the electron impact ionization collision cross section for $^1$H$_2$ reaches a maximum value of $\sigma_\text{iz} \approx 10^{-20}\,\text{m}^2$ at an energy of approximately $E \approx 70$\,eV. In contrast, the product $\sigma_\text{iz}(v_\text{e}) v_\text{e}$ -- which determines the ionization rate -- is peaked at $E \approx 285$\,eV. This kinetic energy corresponds to an electron velocity of approximately $v_\text{e} \approx 10^7$\,m/s. As illustrated in Figure~\ref{fig:sigma}b), $\sigma_\text{iz}(v_\text{e}) v_\text{e}$ varies less than 10\,\% from its maximum for energies in the range $E \approx 70 \text{~to~} 700$\,eV. This energy range determines the optimum window for the volume contribution to an ionization avalanche. The cross sections of Ar, N$_2$, and O$_2$ are slightly higher but in the same order of magnitude (not shown). Equation~\eqref{eq:meancollisiontime} was evaluated using these values to obtain Table~\ref{tab:collisiontime}. Given no precise gas temperature specifications, in all cases $T=400$\,K is assumed (both for estimates and simulations).

One of the constituents of the ITER fuel is Tritium ($^3$H$_2$). Tritium is subject to radioactive $\beta$-decay with a lifetime of approximately 12.32\,years. It correspondingly acts as a constant electron source (average electron energy of 5.7\,keV), with a source rate on the order of $10^{11}\,\text{m}^{-3}\text{s}^{-1}$ (assuming a pressure of 1\,Pa and a tritium fraction of 0.5). Although these $\beta$-electrons have a substantial chance of subsequently undergoing an electron-impact ionization collision ($\sigma_\text{iz}(v_\text{e}) v_\text{e} \approx 2.5 \cdot 10^{-14}\,\text{m}^{3}/\text{s}$), the total source rate is estimated to be much smaller than a conservatively approximated electron source rate due to thermal seed electrons and electron-impact ionization on the order of $10^{17}\,\text{m}^{-3}\text{s}^{-1}$ (using an initial electron density $n_\text{e} = 10^{10}\,\text{m}^{-3}$ and $\sigma_\text{iz}(v_\text{e}) v_\text{e} \approx 5 \cdot 10^{-14}\,\text{m}^{3}/\text{s}$). Consequently, even if tritium was used in the modeling, this process would be of subordinate importance for the present study. Note, however, that Tritium $\beta$-decay will be a constant source of initial seed electrons for the initiation of the breakdown process. This also means that it is not possible to completely deplete the resonant volume free electrons.

\section{Results and discussion}
\subsection{Validation with results of Moeller}
\label{ssec:Moeller}

\begin{table}[b!]
\centering
\begin{tabularx}{0.5\textwidth}{ X X }
$p$ (Pa) & minimum $V_\text{bias}$ (kV) \\
\hline
\hline
0.01 & not mentioned ($\leq 1$)$^*$ \\
\hline
0.08 & 1 \\
\hline
0.133 & 2.3 \\
\hline
\hline
\end{tabularx}
\caption{Collection of hydrogen pressures where breakdown was reported and minimum bias voltages required to mitigate breakdown via the $\vec{E}_\text{bias} \times \vec{B}$ drift as enlisted in references \cite{moeller_avoidance_1987,[{}][{, U.S. Patent 4,687,616, August 18, 1987.}]moeller_method_1987}. *Estimated from higher pressure case.}
\label{tab:moellerresults}
\end{table}

\begin{figure}[b!]
\includegraphics[width=8cm]{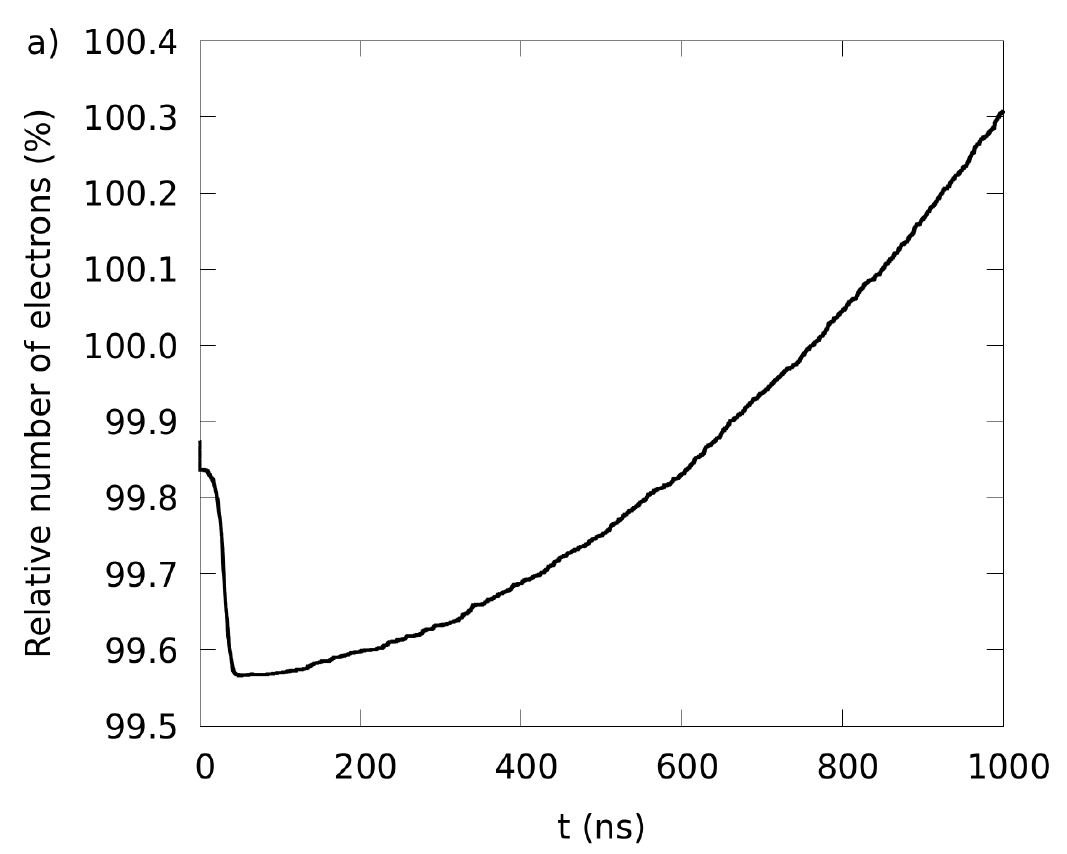}
\includegraphics[width=8cm]{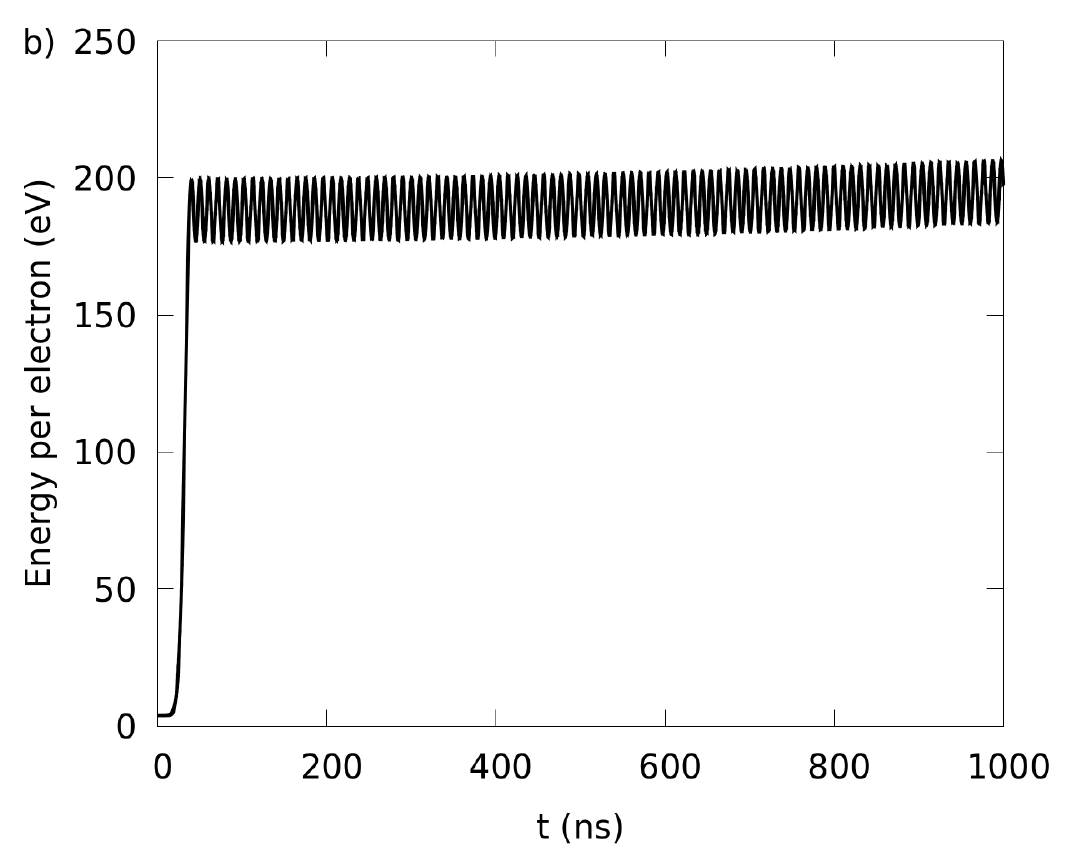}
\caption{Moeller case (2D, $E_\text{rf} \perp E_\text{bias} \perp B$, $p=0.01$\,Pa H$_2$, $E_0=80$\,kV/m, $B=2.14$\,T, without bias voltage). a) Number of electrons and b) average energy per electron plotted over time.}
\label{fig:Moeller0-01}
\end{figure}

Moeller reports on observations of ECR breakdown for a number of specific cases as summarized in Table~\ref{tab:moellerresults}. In the following, simulation results for these cases are presented. The simulated configuration is distinct in the sense that the axial magnetic field allows for transport in the axial direction (not resolved in the simulation), but effectively inhibits transport in the radial and azimuthal direction (except for $\vec{E}_\text{bias}\times\vec{B}$ drift contributions, due to the bias electric field). The losses to the walls are correspondingly small. In addition, the resonant condition is satisfied in the entire internal length of the solenoid (in contrast to the ITER case with a small resonant region). Hence, gas breakdown can be observed down to a very low pressure of $p=0.01$\,Pa (the lowest pressure achievable with Moeller's equipment;\cite{moeller_avoidance_1987,[{}][{, U.S. Patent 4,687,616, August 18, 1987.}]moeller_method_1987} not necessarily the pressure where breakdown will occur). For this case, the evolution of the total number of electrons, as well as the average energy per electron is depicted in Figure~\ref{fig:Moeller0-01}. Following an initial rise of the pulse and an accompanying loss of electrons out of the resonant volume ($<40$\,ns), a phase of electron heating and collisional relaxation develops into an exponential rise of the number of electrons. As can be seen, the timescale of this breakdown is governed by the mean collision time on the order of $\tau_\text{c} = 5.5$\,$\mu$s with a correspondingly slow rise of the electron density (net increase of about 120 electrons or 0.8\,\% in 950\,ns). Evidently, the average energy per electron remains rather stable at $E \approx 190$\,eV after the initial rise of the pulse (close to the optimum ionization energy). As previously discussed, intermediate frequency oscillations with time period $T \approx 9$\,ns are observed. A comparison with the simulation pulse rise time $\sigma = 10$\,ns underlines the rapid excitation of the system as root cause. This `ringing' is argued to be negligible for the gas breakdown dynamics, due to its small magnitude with respect to the total average energy per electron.

\begin{figure}[b!]
\includegraphics[width=8cm]{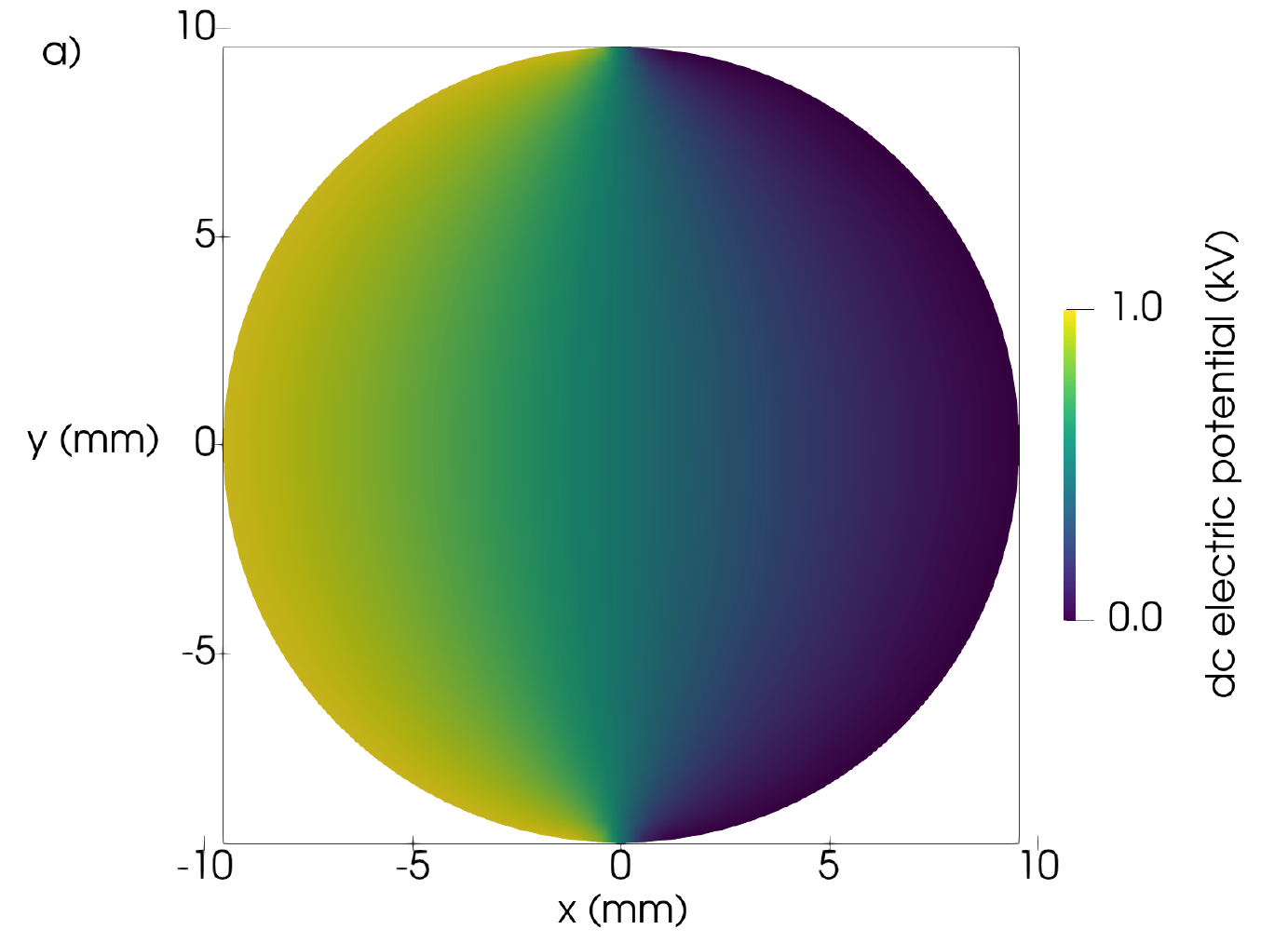}
\includegraphics[width=8cm]{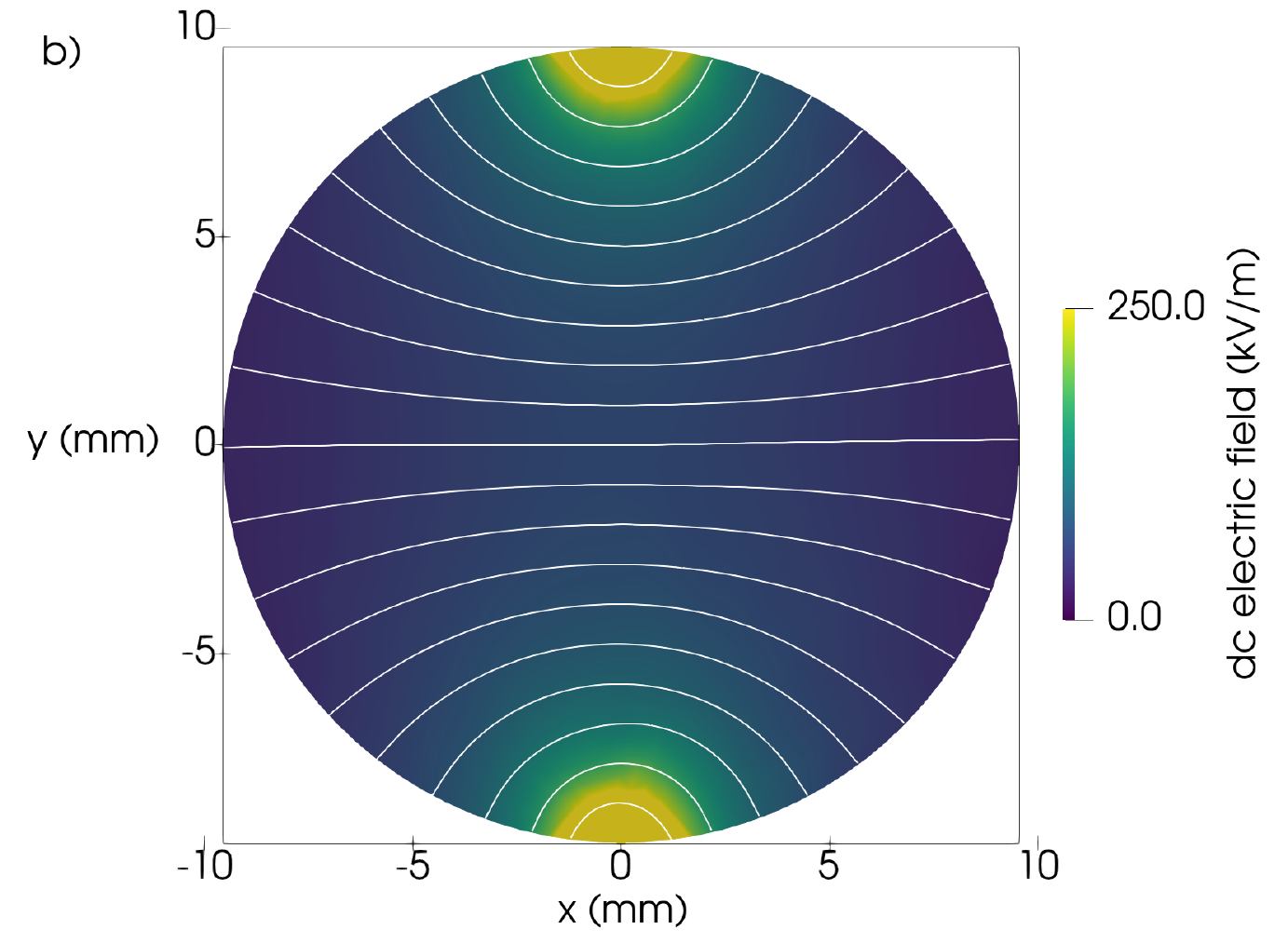}
\caption{`Moeller' case (2D, $E_\text{rf} \perp E_\text{bias} \perp B$, $V_\text{bias}=1$\,kV). a) Bias electric potential and b) bias electric field distributions plotted over the simulation domain. White lines indicate the curvature of the field lines. The data range of the electric field represented by false-colors is limited to 250\,kV/m to depict the distribution in the most relevant central region of the waveguide. It is out of this bound at the top/bottom gaps.}
\label{fig:Moellermitigation}
\end{figure}

Breakdown has also been reported for an increased pressure of $p=0.08$\,Pa and successful mitigation has been observed using a bias voltage of $V_\text{bias} = 1$\,kV. This result is reproduced by the simulation, using the bias electric field presented in Figure~\ref{fig:Moellermitigation}. The latter has been consistently calculated in the simulation domain, imposing the respective boundary electric potential. While the bias electric potential and electric field are symmetric with respect to the $x$ axis, the circular geometry enforces a corresponding curvature of the electric field lines. This effect is of relevance for electron removal, due to an $\vec{E}_\text{bias}\times\vec{B}$ drift, which scales in magnitude proportional to the local fields and is directed in the local perpendicular direction. The bias electric field in the center of the waveguide is systematically larger than the one dimensional approximation, $E_\text{bias} \approx 67\,\text{kV/m} > V_\text{bias}/D \approx 52\,\text{kV/m}$.

\begin{figure}[t!]
\includegraphics[width=8cm]{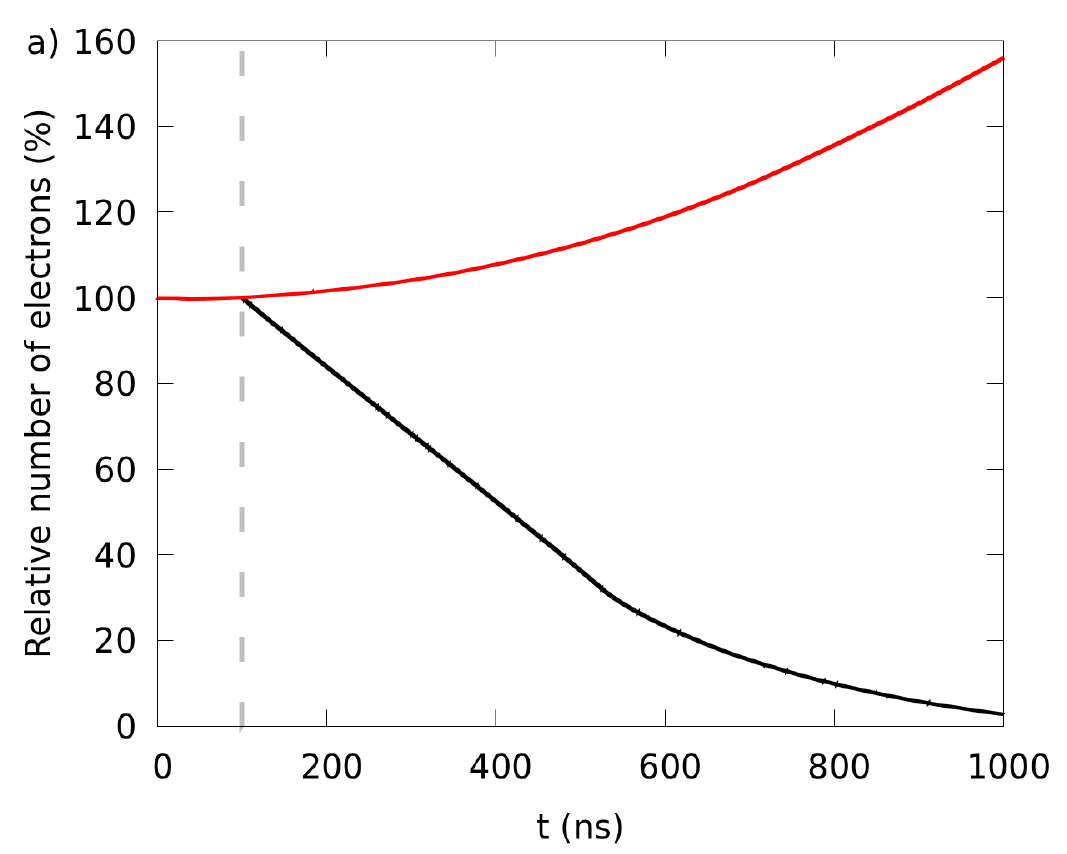}
\includegraphics[width=8cm]{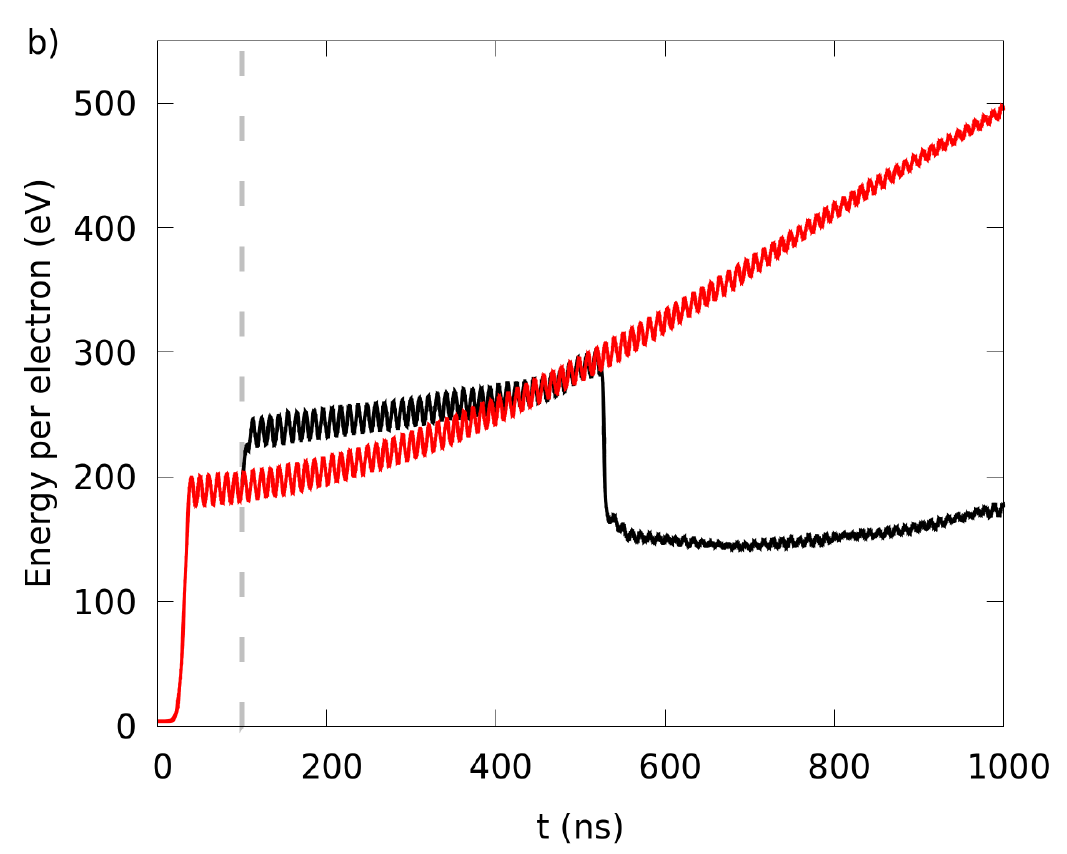}
\caption{Moeller case (2D, $E_\text{rf} \perp E_\text{bias} \perp B$, $p=0.08$\,Pa H$_2$, $E_0=80$\,kV/m, $B=2.14$\,T, $V_\text{bias}=1$\,kV). a) Number of electrons and b) average energy per electron plotted over time. Results without (red) and with (black) bias voltage applied after $t=100$\,ns.}
\label{fig:Moeller0-08}
\end{figure}

Depicted in Figure~\ref{fig:Moeller0-08} are the number of electrons within the simulation domain and their average energy per electron for a pressure of $p=0.08$\,Pa. Shown in red is the evolution without bias voltage, while the black line gives the evolution with a bias voltage of $V_\text{bias}=1$\,kV applied after $t=100$\,ns. For the former case without mitigation, an exponential increase in number of electrons is again observed after a short transient. The breakdown timescale is reduced compared to the case with $p=0.01$\,Pa, due to the reduced mean collision time, $\tau_\text{c}(p=0.08\,\text{Pa}) \approx 0.7\,\mu\text{s} \ll \tau_\text{c}(p=0.01\,\text{Pa}) \approx 5.5\,\mu\text{s}$. The average energy per electron initially increases sharply with the pulse rise and more steadily during the breakdown (due to continued heating of the confined electrons by the RF electric field).

\begin{figure}[t!]
\includegraphics[width=8cm]{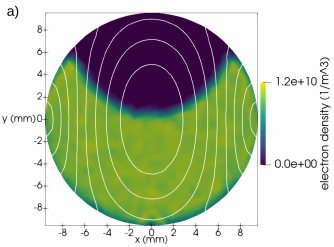}
\includegraphics[width=8cm]{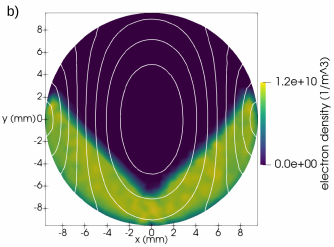}
\caption{Moeller case (2D, $E_\text{rf} \perp E_\text{bias} \perp B$, $p=0.08$\,Pa H$_2$, $E_0=80$\,kV/m, $B=2.14$\,T, $V_\text{bias}=1$\,kV). Electron density distributions over waveguide cross section, during electron removal process following $\vec{E}\times\vec{B}$ drift at a) $t=300$\,ns and b) $t=500$\,ns. Indicated by white contours is the TE$_{11}$ RF electric field magnitude (cf.\ Figure~\ref{fig:modes}).}
\label{fig:Moeller0-08electrons}
\end{figure}

For the mitigation case, depicted by the black lines in Figure~\ref{fig:Moeller0-08}, a rise in electron energy is noticeable within approximately 10\,ns after switching on the bias electric field, associated with an initial distortion of the electron dynamics. The decay of the number of electrons proceeds on the order of $\tau_\text{sweep} \approx 0.8\,\mu$s, caused by the $\vec{E}_\text{bias}\times\vec{B}$ drift. Reflection and emission of electrons when reaching the adjacent walls have an additional contribution. A sharp drop in electron energy is observed at $t \approx 500$\,ns. It is associated with the removal of electrons from the high RF electric field region in the waveguide center. To illustrate this, two snapshots of the electron density are plotted over the simulation domain at $t=300$\,ns and $t=500$\,ns as shown in Figure~\ref{fig:Moeller0-08electrons}. With the expected $\vec{E}_\text{bias}\times\vec{B}$ drift and direction, electrons are removed from the central waveguide region (white isocurves indicating the RF electric field magnitude), substantially reducing ECR heating. The kink in the slope of the number of electrons stems from these electron removal dynamics, dictated by the bias electric field and magnetic field distributions.

\begin{figure}[t!]
\includegraphics[width=8cm]{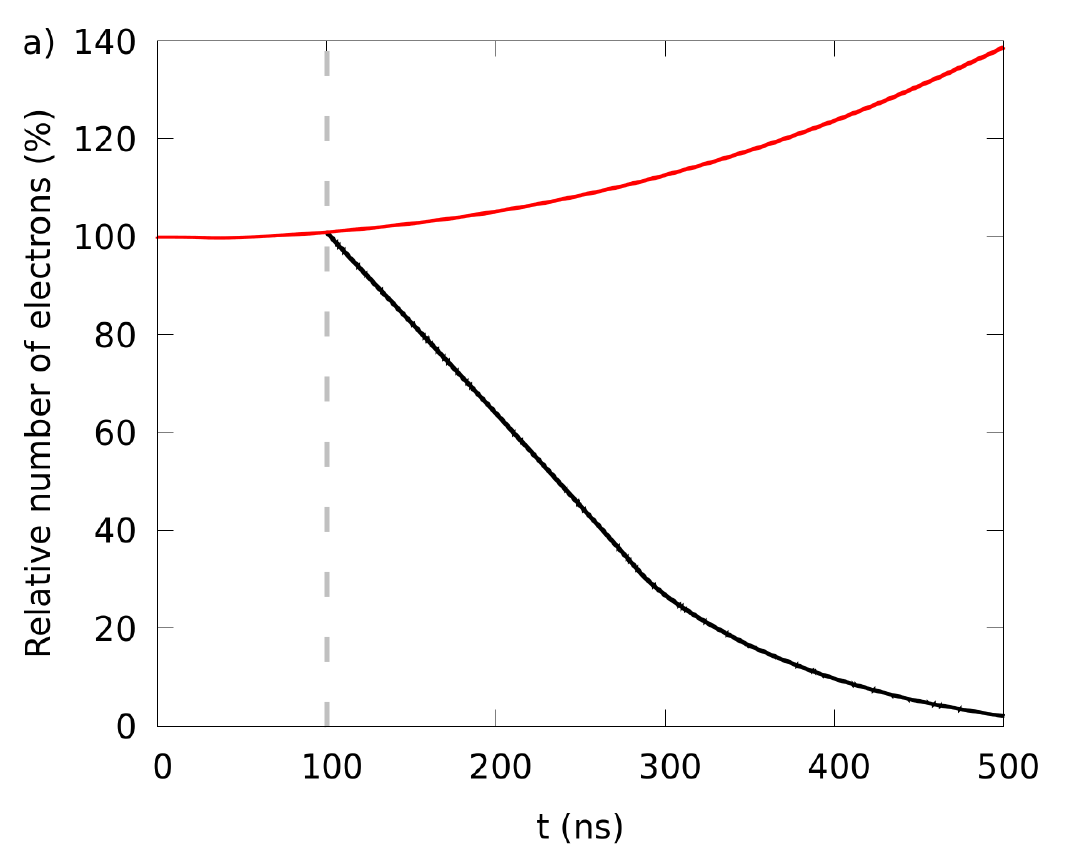}
\includegraphics[width=8cm]{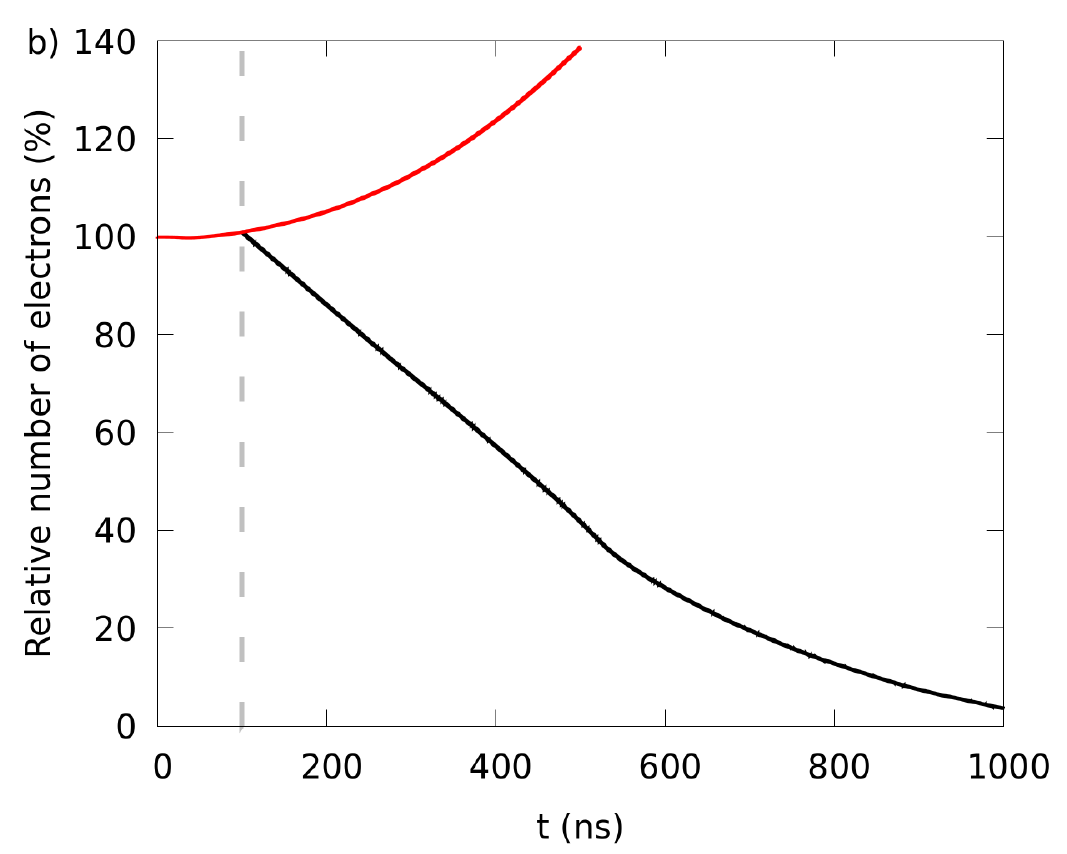}
\caption{Moeller case (2D, $E_\text{rf} \perp E_\text{bias} \perp B$, $p=0.133$\,Pa H$_2$, $E_0=80$\,kV/m, $B=2.14$\,T). Number of electrons with a) $V_\text{bias}=2.3$\,kV and b) $V_\text{bias}=1$\,kV (*) plotted over time. Results without (red) and with (black) bias voltage applied after $t=100$\,ns. [(*)$V_\text{bias}=1$\,kV was not sufficient to suppress gas breakdown in experiments, possibly due to outgassing from the walls.\cite{moeller_high_1987}]}
\label{fig:Moeller0-133_2-3kV_1kV}
\end{figure}

Moeller has reported gas breakdown and successful mitigation for a further increased pressure of $p=0.133$\,Pa, where he found a bias voltage of $V_\text{bias}=1$\,kV to be insufficient for mitigation, but mitigation was successful at $V_\text{bias}=2.3$\,kV. Simulation results for these cases are presented in Figure~\ref{fig:Moeller0-133_2-3kV_1kV} with bias voltages $V_\text{bias}=2.3$\,kV and $V_\text{bias}=1$\,kV, switched on after $t=100$\,ns. The red and black lines denote the evolution without and with mitigation, respectively. For both applied bias voltages, the number of electrons proceeds similar to the low pressure case. Without mitigation an exponential increase in the number of electrons is observed at the timescale of the mean collision time $\tau_\text{c} \approx 425$\,ns. When using a bias voltage $V_\text{bias}=2.3$\,kV (Figure~\ref{fig:Moeller0-133_2-3kV_1kV}a), electron removal following the $\vec{E}_\text{bias}\times\vec{B}$ drift proceeds more than twice as fast, $\tau_\text{sweep} \approx 350$\,ns, compared to the case with $V_\text{bias} = 1$\,kV and $\tau_\text{sweep} \approx 800$\,ns. The ratio of the sweep times is consistent with the inverse ratio of the corresponding bias voltages and can clearly be attributed to the $\vec{E}_\text{bias}\times\vec{B}$ transport mechanism.

Using a reduced bias voltage of $V_\text{bias}=1$\,kV (Figure~\ref{fig:Moeller0-133_2-3kV_1kV}b) reveals that even in this case -- in contrast to the observation by Moeller -- mitigation is effective and gas breakdown is disrupted. The principle dynamics and transport mechanisms are identical (breakdown and ECR electron heating without mitigation; removal of electrons from the high RF field region with mitigation). Electron removal takes place with the previously reported sweep time of $\tau_\text{sweep} \approx 800$\,ns. This appears to be sufficiently fast in comparison with a mean collision time $\tau_\text{c} \approx 425$\,ns: A ratio $\tau_\text{sweep} / \tau_\text{c} \approx 1$ means that an electron experiences approximately a single collision encounter during sweep out. In contrast, to initiate a substantial ionization avalanche, $\tau_\text{sweep} / \tau_\text{c} > 1$ would be required.

\begin{figure}[b!]
\includegraphics[width=8cm]{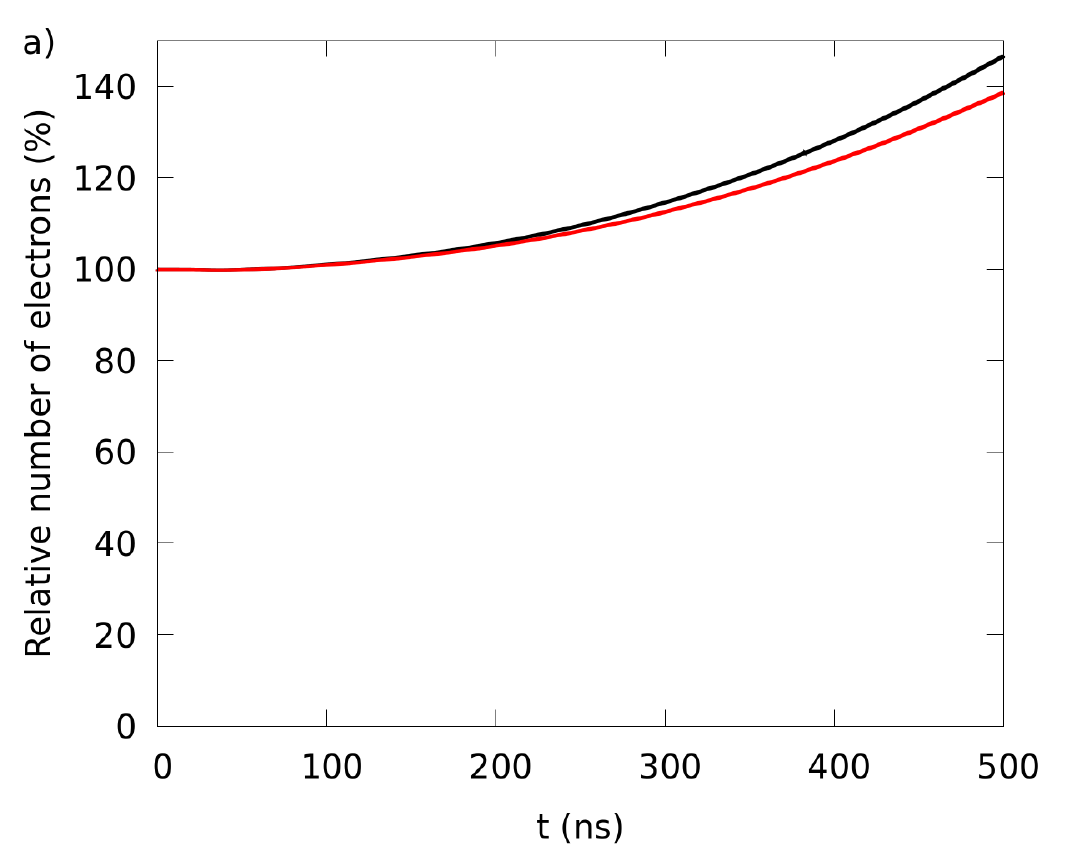}
\includegraphics[width=8cm]{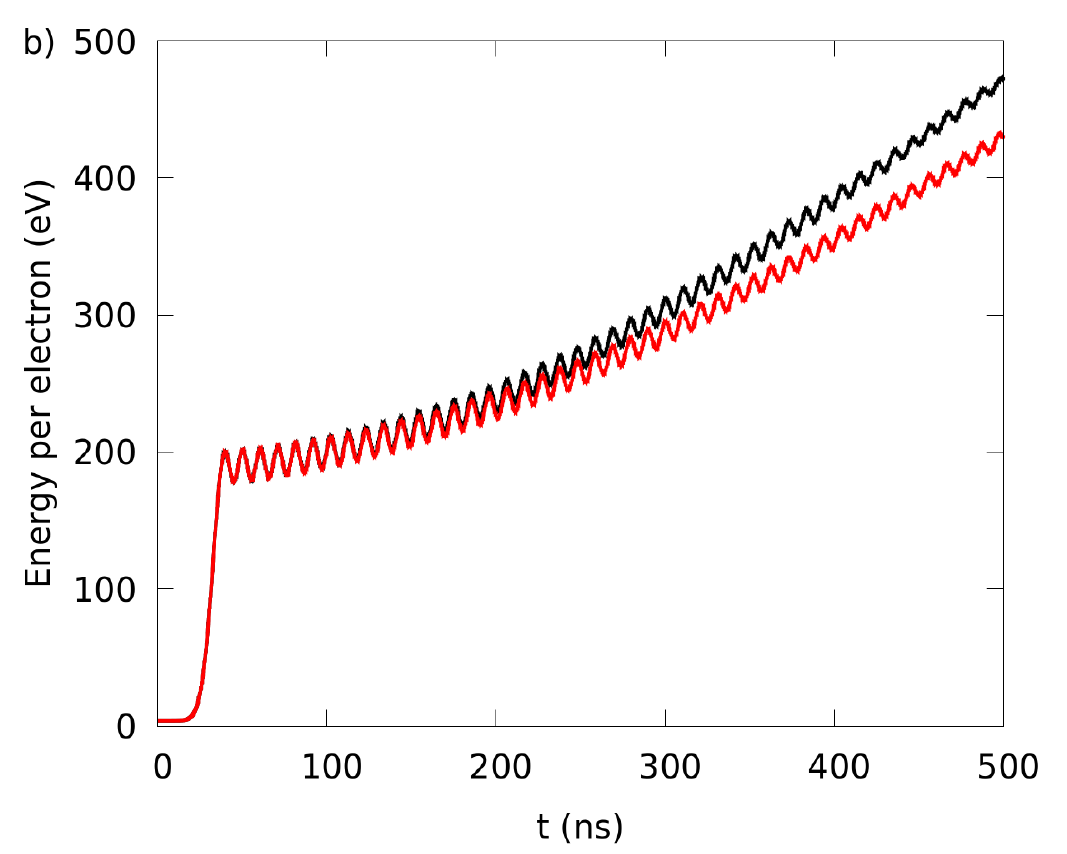}
\caption{Moeller case (2D, $E_\text{rf} \perp E_\text{bias} \perp B$, $p=0.133$\,Pa, $E_0=80$\,kV/m, $B=2.14$\,T, without bias voltage). a) Number of electrons and b) average energy per electron plotted over time. Results for pure $^1$H$_2$ (red) and $^1$H$_2$ + 10\% synthetic air impurity (black).}
\label{fig:Moeller0-133_Air}
\end{figure}

A number of aspects can be raised to explain the apparent discrepancy between simulations and experimental observations (in presumed order of importance): (i) Moeller has mentioned that conditioning of the waveguide surfaces had a significant impact on the gas breakdown behavior. \cite{moeller_high_1987} In particular, he has mentioned the influence of outgassing from the SS waveguide walls, which probably assisted the breakdown phenomenon (cf.\ subsequent paragraph). (ii) The simplifying assumption of a two dimensional geometry with a homogeneous magnetic field is not an exact representation of the setup which has been used experimentally.\cite{moeller_avoidance_1987,[{}][{, U.S. Patent 4,687,616, August 18, 1987.}]moeller_method_1987} Possible improvements of the model are uncertain, however, as only insufficient details on the geometry and the magnetic field setup are documented. (iii) The surface coefficients (e-SEE, electron reflection) and the reduced set of collision cross sections are subject to uncertainty as well. Improvements of the model are impossible in the absence of more reliable data.

To obtain an estimate of the influence of outgassing, simulations were performed assuming a fraction of 10\% of synthetic air (78\% N$_2$, 21\% O$_2$, 1\% Ar) as an impurity and otherwise unaltered parameters (i.e., a maintained total pressure $p=0.133$\,Pa). The corresponding number of electrons and the average energy per electron are presented in Figure~\ref{fig:Moeller0-133_Air}. It is found that, after an initial relaxation of the system, ECR heating and breakdown evolution are decisively different. The lower ionization thresholds and larger cross sections for the introduced gas impurity enhance breakdown. This is reasoned by the decreased mean collision time and is observed despite the impurity's minor concentration. At $t=500$\,ns, the relative increase in the number of electrons is more than 10\% larger compared to the case without impurity. Notably, breakdown follows the expected exponential dependence leading to an even more pronounced effect on the later evolution. Electron heating also appears to proceed more efficiently due to the more local energy conversion. Due to a shorter mean free path and an inhibited transport, electrons remain in the high RF electric field region for a longer period of time. It should be noted that outgassing from the walls in the experiments is merely an addition to the present gas, not a substitute (varied total pressure). Consequently, a synergistic effect of a lower ionization threshold paired with an increased gas pressure would be expected for the situation reported by Moeller.

\clearpage

\subsection{Simulation prediction for ITER CTS}
\label{ssec:ITER}
The ITER CTS scenario differs in three main aspects from the `Moeller' case: (i) waveguide inner diameter $D=88.9$\,mm (i.e., lower peak RF electric field strength $E_0$); (ii) axially varying magnetic field with only about 35\% axial component in the resonant region, with $B \approx 2.14$\,T; (iii) CuCrZr ITER-grade alloy waveguide surfaces (approximated by Cu).

In the following, two and three dimensional simulation results are presented, each depicting a specific aspect of the gas breakdown phenomenon. Initially, two dimensional simulation results are discussed, focusing on gas breakdown with a magnetic field solely in the transversal waveguide plane (i.e., directed toward the bounding waveguide walls). Subsequently, three dimensional simulation results highlight the influence of an axial magnetic field contribution. Finally, two particular situations are analyzed related to variations in the magnetic field. The cases to be considered are summarized in Table~\ref{tab:ITERresults}.

\begin{table}[b!]
\centering
\begin{tabularx}{0.5\textwidth}{ L{0.5} L{1.9} L{0.7} L{0.9} }
Dim. & Configuration & $p$ (Pa) & $V_\text{bias}$ (kV) \\
\hline
\hline
2D & $E_\text{rf} \perp E_\text{bias} \parallel B$ & 1 & 1 \\
\hline
2D & $E_\text{rf} \perp E_\text{bias} \parallel B$ & 10 & 1 \\
\hline
2D & $E_\text{rf} \perp E_\text{bias} \parallel B$ & 0.3 & 1 \\
\hline
2D & $E_\text{rf} \perp E_\text{bias} \parallel B$ & 20 & 1 and 2 \\
\hline
3D & $E_\text{rf} \perp E_\text{bias} \parallel B_\text{trans.}$ & 1 & 1 \\
\hline
3D & $E_\text{rf} \parallel E_\text{bias} \parallel B_\text{trans.}$ & 1 & 1 \\
\hline
3D & $E_\text{rf} \parallel E_\text{bias} \perp B_\text{trans.}$ & 1 & 1 \\
\hline
3D & $E_\text{rf} \perp E_\text{bias} \parallel B_\text{trans.}$ & 5 & 1 \\
\hline
\hline
\end{tabularx}
\caption{Collection of ITER CTS cases considered in the proceeding discussion.}
\label{tab:ITERresults}
\end{table}

\subsubsection{Two dimensional}
\begin{figure}[t]
\includegraphics[width=8cm]{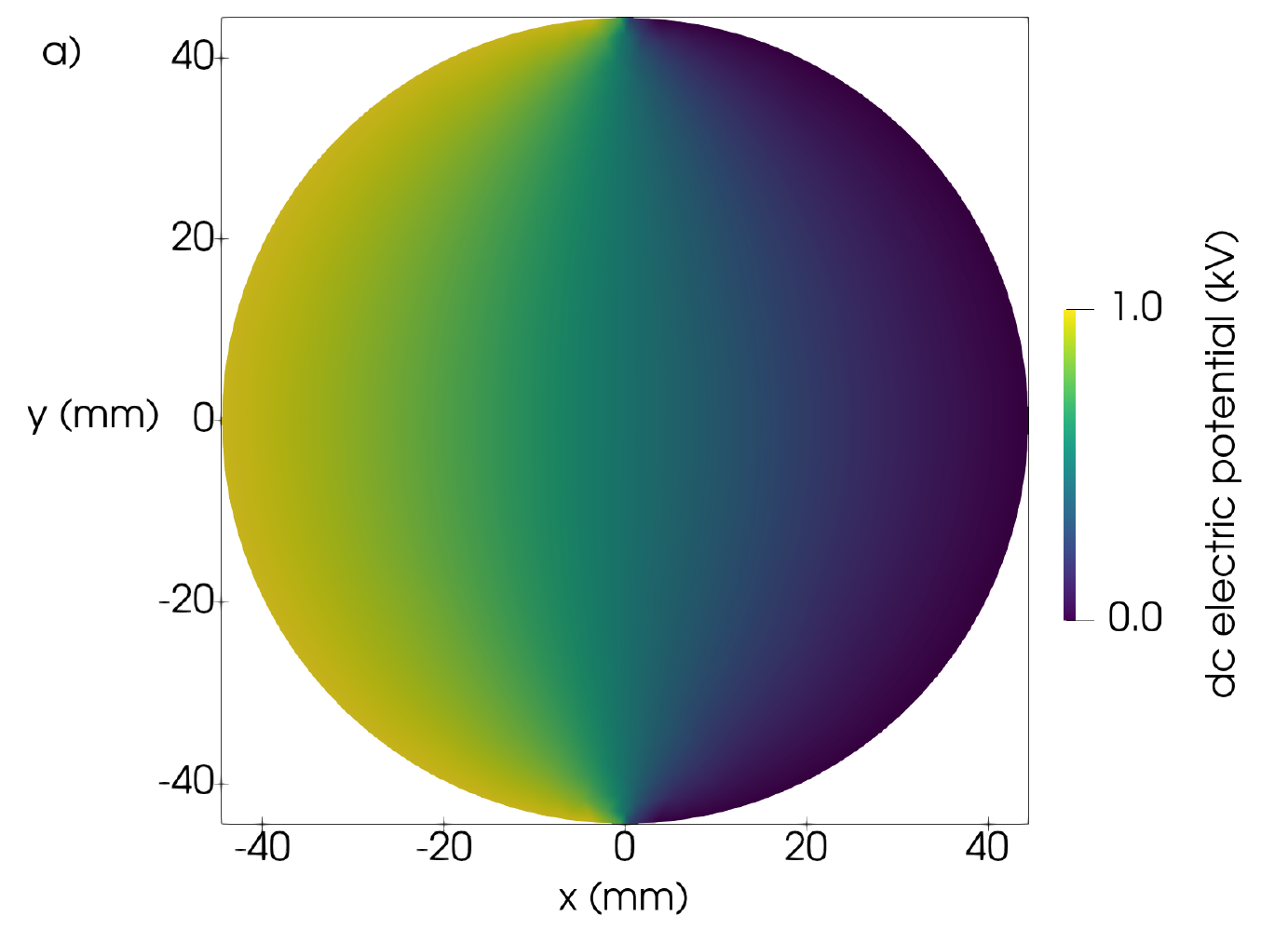}
\includegraphics[width=8cm]{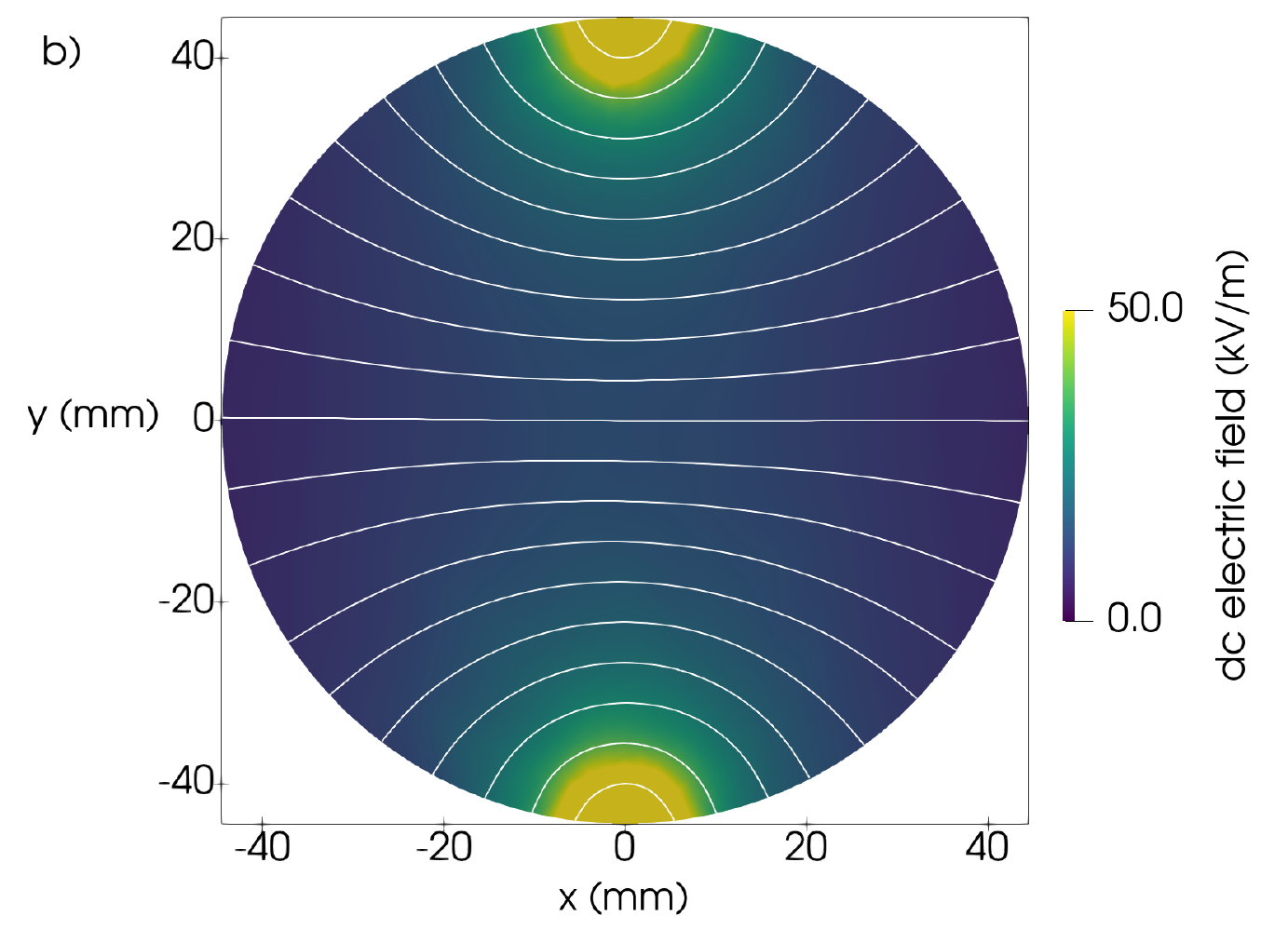}
\caption{ITER CTS case (2D, $E_\text{rf} \perp E_\text{bias} \parallel B$, $V_\text{bias}=1$\,kV). a) Bias electric potential and b) bias electric field distributions plotted over the simulation domain. White lines indicate the curvature of the field lines. The data range of the electric field represented by false-colors is limited to 50\,kV/m to depict the distribution in the most relevant central region of the waveguide. It is out of this bound at the top/bottom gaps.}
\label{fig:ITERmitigation}
\end{figure}

A bias voltage of $V_\text{bias}=1$\,kV is initially specified for ITER CTS and was correspondingly used in the proceeding analysis once mitigation was switched on. It was later increased to 2\,kV, following the findings in this study. The corresponding bias electric potential and field are shown in Figure~\ref{fig:ITERmitigation}. The bias electric field in the center of the waveguide is again systematically larger ($\approx 28$\,\%) compared to a one dimensional approximation due to the circular geometry, $E_\text{bias} \approx 14\,\text{kV/m} > V_\text{bias}/D \approx 11\,\text{kV/m}$. Consistent with the scaling in waveguide diameter, the maximum bias electric field is approximately 5 times smaller than for the `Moeller' case (cf.\ Figure~\ref{fig:Moellermitigation}). In the $\vec{E}_\text{bias}\times\vec{B}$ configuration, this results in a correspondingly slowed down removal of electrons. However, by aligning the SBWG halves appropriately with the magnetic field in the resonant region (i.e., $\vec{E}_\text{bias} \parallel \vec{B}$), electron removal can proceed along magnetic field lines, resulting in short electron sweep times on the order of $\tau_\text{sweep} \approx 10$\,ns (cf.\ Section~\ref{ssec:SBWG}).

\begin{figure}[b!]
\includegraphics[width=8cm]{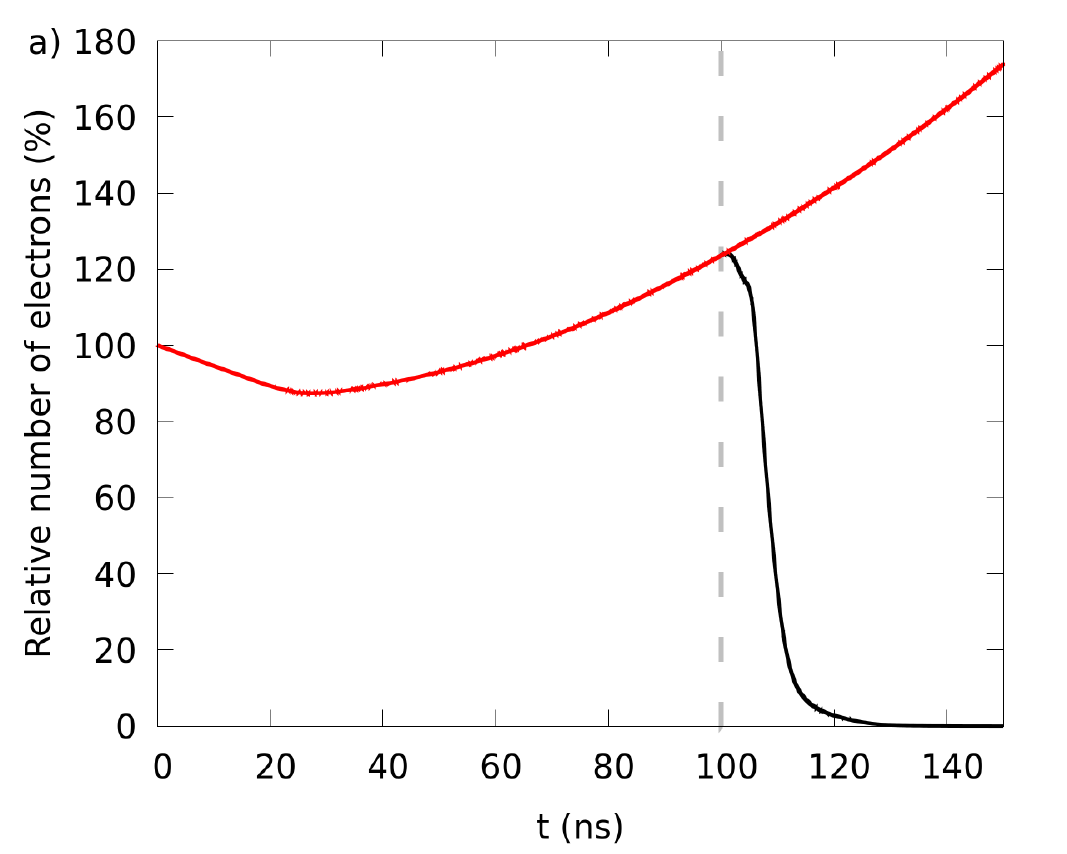}
\includegraphics[width=8cm]{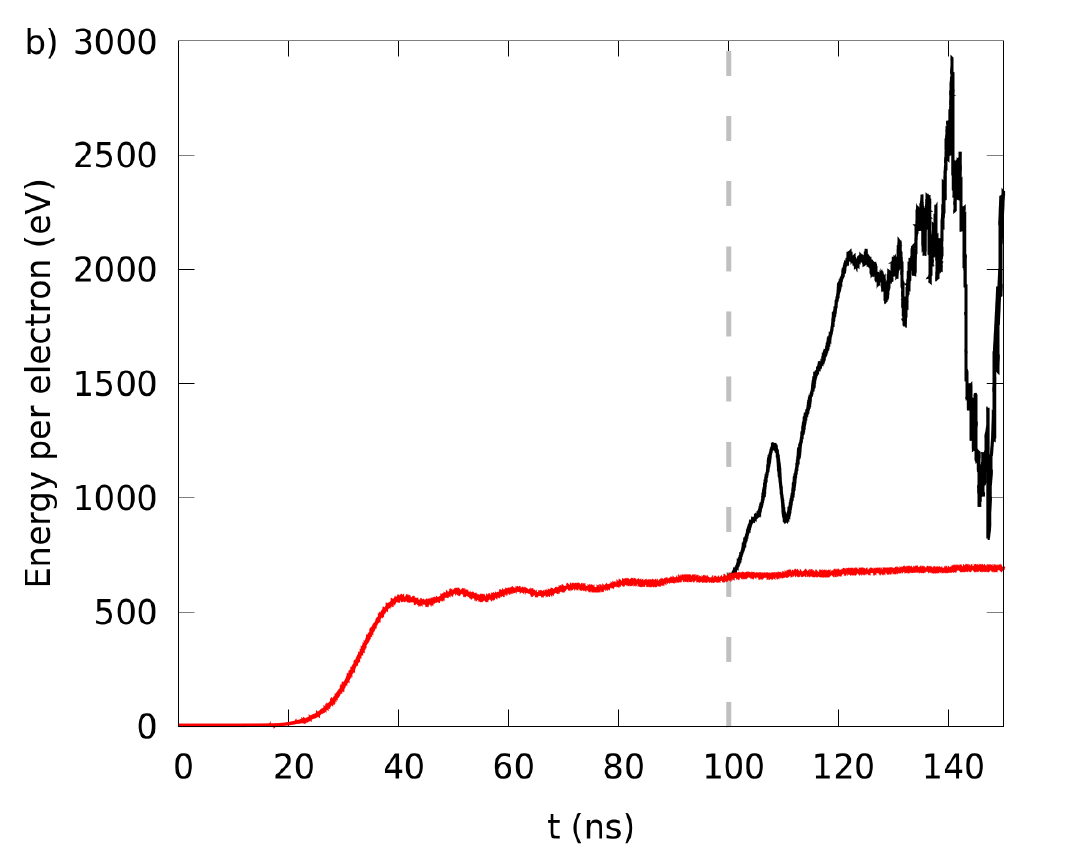}
\caption{ITER CTS case (2D, $E_\text{rf} \perp E_\text{bias} \parallel B$, $p=1$\,Pa H$_2$, $E_0=150$\,kV/m, $B=2.14$\,T, $V_\text{bias}=1$\,kV). a) Number of electrons and b) average energy per electron plotted over time. Results without (red) and with (black) bias voltage applied after $t=100$\,ns.}
\label{fig:Iter1}
\end{figure}

\begin{figure}[t!]
\includegraphics[width=8cm]{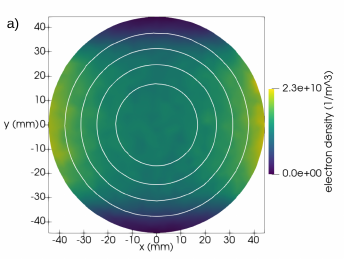}
\includegraphics[width=8cm]{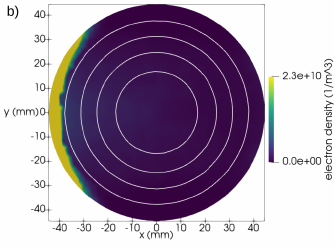}
\caption{ITER CTS case (2D, $E_\text{rf} \perp E_\text{bias} \parallel B$, $p=1$\,Pa H$_2$, $E_0=150$\,kV/m, $B=2.14$\,T, $V_\text{bias}=1$\,kV). Electron density distributions plotted over waveguide cross section during electron removal process, following acceleration due to $\vec{E}_\text{bias}$ at $t=100$\,ns (left) and $t=110$\,ns (right). Indicated by contours is the LP$_{01}$/HE$_{11}$ RF electric field magnitude (cf.\ Figure~\ref{fig:modes}).}
\label{fig:ITER1electrons}
\end{figure}

As a reference case for ITER CTS gas breakdown and mitigation simulations, a hydrogen pressure of $p=1$\,Pa was used. In the two dimensional representation, the magnetic field points solely in the $x$ direction. The axial magnetic field component (along the waveguide) is set to zero. A transversal magnetic field magnitude of $B=2.14$\,T is used to maintain ECR conditions. In Figure~\ref{fig:Iter1} the evolution of the number of electrons and the average energy per electron is shown. Without mitigation, breakdown is observed at the timescale of the mean collision time $\tau_\text{c} \approx 55$\,ns, following the initial onset of the RF modulated pulse. ECR heating proceeds accordingly, increasing above $E \approx 500$\,eV for $t \gtrsim 40$\,ns. When a bias voltage $V_\text{bias}=1$\,kV is applied after $t=100$\,ns, the expected behavior of a rapid depletion of electrons within 15\,ns is observed. This is accompanied by a steep increase in electron energy after switching on the bias voltage. It is reasoned by the acceleration of electrons due to the bias electric field. For the case with mitigation (black line), the average energy per electron after $t \approx 110$\,ns is subject to substantial statistical fluctuations, due to the low number of electrons involved and should be considered with caution.

\begin{figure}[t!]
\includegraphics[width=8cm]{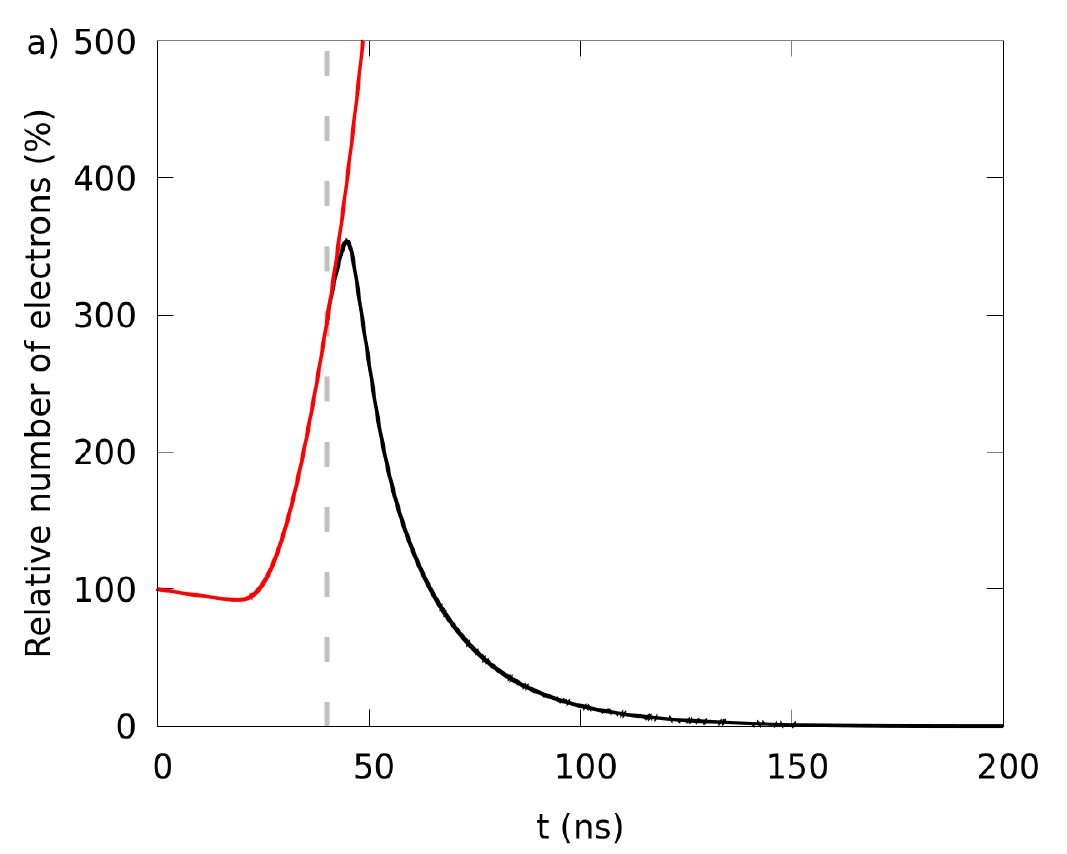}
\includegraphics[width=8cm]{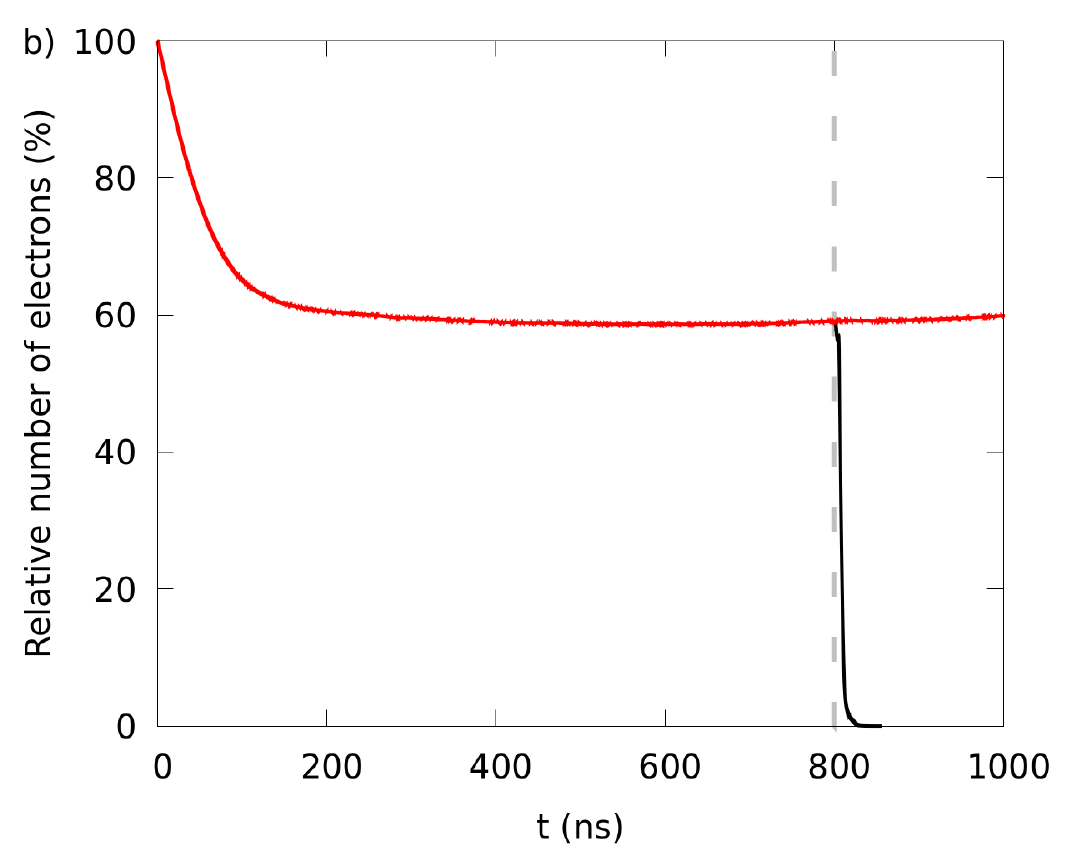}
\caption{ITER CTS case (2D, $E_\text{rf} \perp E_\text{bias} \parallel B$, H$_2$, $E_0=150$\,kV/m, $B=2.14$\,T, $V_\text{bias}=1$\,kV). Number of electrons for a) $p=10$\,Pa and b) $p=0.3$\,Pa plotted over time. Results without (red) and with (black) bias voltage applied after $t=40$\,ns and $t=800$\,ns, respectively.}
\label{fig:Iter6_0-3}
\end{figure}

The dynamics of electron removal can be understood from the spatial distributions of the electron density right before the bias voltage is switched on and 10\,ns after. Figure~\ref{fig:ITER1electrons} shows corresponding electron density profiles plotted over the simulated domain. Notably, even before active mitigation the electrons diffusively distribute along the magnetic field lines. The density maxima close to the wall stem from the electrons' wall interactions and subsequent reflection or secondary electron emission. This effect is correspondingly enhanced with a bias electric field which promotes the electron flux toward the wall. That is, electrons are rapidly removed from the central region of high RF electric field strength (indicated by white isocurves), diminishing ECR heating, but at the same time may accumulate close to the wall until breakdown is finally disrupted. Consequently, wall processes have a noticeable influence on the total electron count (slow absorption), but little influence on the removal of electrons from the high RF electric field region and hence the mitigation dynamics.

A different characteristic can be observed in the two dimensional simulations with a pressure of $p=10$\,Pa, as demonstrated in Figure~\ref{fig:Iter6_0-3}a). The dynamics of gas breakdown proceed faster due to the reduced mean collision time -- on the order of $\tau_\text{c} \approx 5.5$\,ns. Using a bias voltage of $V_\text{bias}=1$\,kV corresponding to $\tau_\text{sweep} \approx 10$\,ns for $t>40$\,ns, this is associated with a prolonged effective removal of electrons. Ionization in the volume is enhanced (black line) and removal is achieved only with an effective time constant $\tau \gtrsim 25\,\text{ns} \gg \tau_\text{sweep} \approx 10$\,ns. It is reasoned by a shortened inelastic collision timescale and a corresponding contribution of the bias electric field in energy input and sustaining breakdown. Note that $\tau_\text{sweep}$ only approximates an upper bound, whereas the actual sweep time is presumably shorter. For higher pressures above $p \gtrsim 10$\,Pa, a hybrid DC-ECR breakdown may occur. In this situation, the SBWG mitigation procedure requires an increased bias voltage, and thus electric field, to reduce $\tau_\text{sweep}$ and remove electrons prior their ionization collisions with the gas background.

At the opposite pressure end, breakdown can be observed for pressures as low as $p=0.3$\,Pa in the two dimensional simulations. As depicted in Figure~\ref{fig:Iter6_0-3}b), the dynamics slow down significantly to approximately a few times $\tau_\text{c} \approx 183$\,ns, in line with the previous reasoning on the mean collision time. The principle breakdown and mitigation dynamics consistently remain. For pressures below $p \lesssim 0.3$\,Pa, no breakdown is observed in the simulations despite high electron energies (not shown). Electron multiplication, due to ionization collisions, is too slow to compensate the diffusion loss to the walls. No breakdown is expected once these losses dominate. This can be estimated to occur when $\tau_\text{c} \approx 275$\,ns, corresponding to a pressure of about $p \approx 0.2$\,Pa. Therefore, mitigation using a bias voltage $V_\text{bias}=1$\,kV is inherently sufficient for all pressures $p \lesssim 10$\,Pa, given the two dimensional setup ($\vec{E}_\text{bias} \parallel \vec{B}$).

\begin{figure}[b!]
\includegraphics[width=8cm]{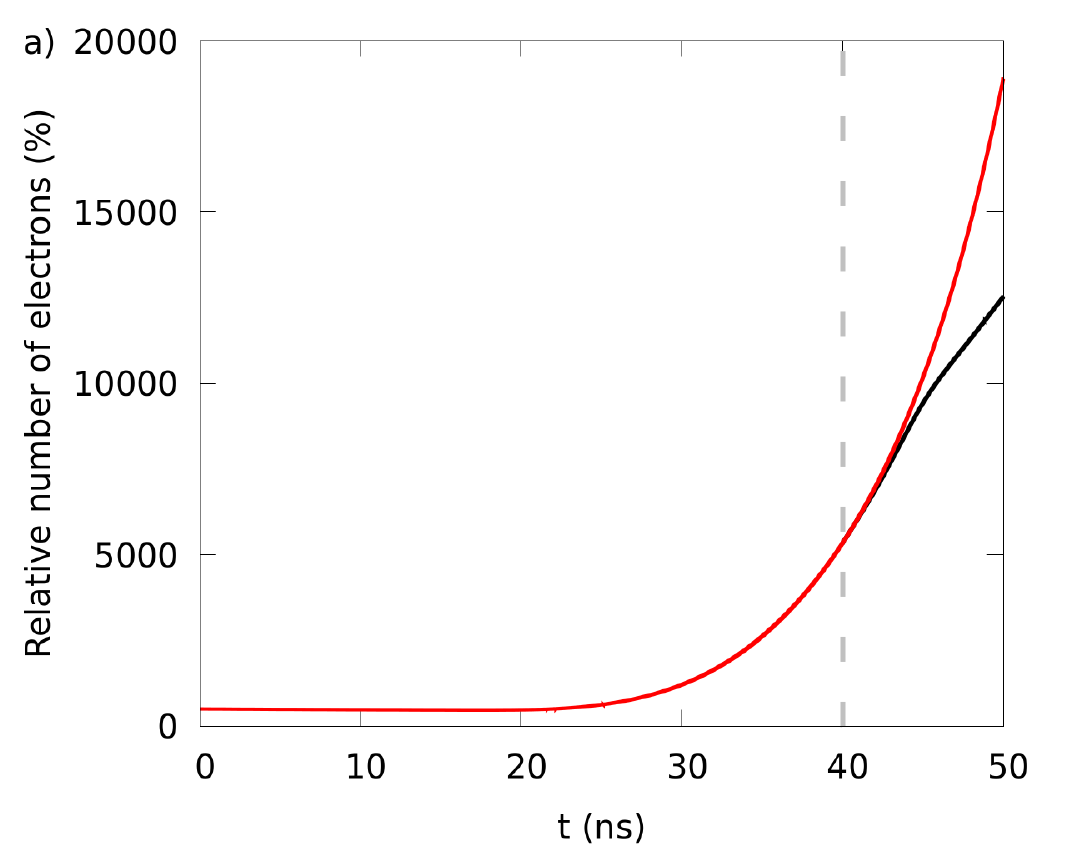}
\includegraphics[width=8cm]{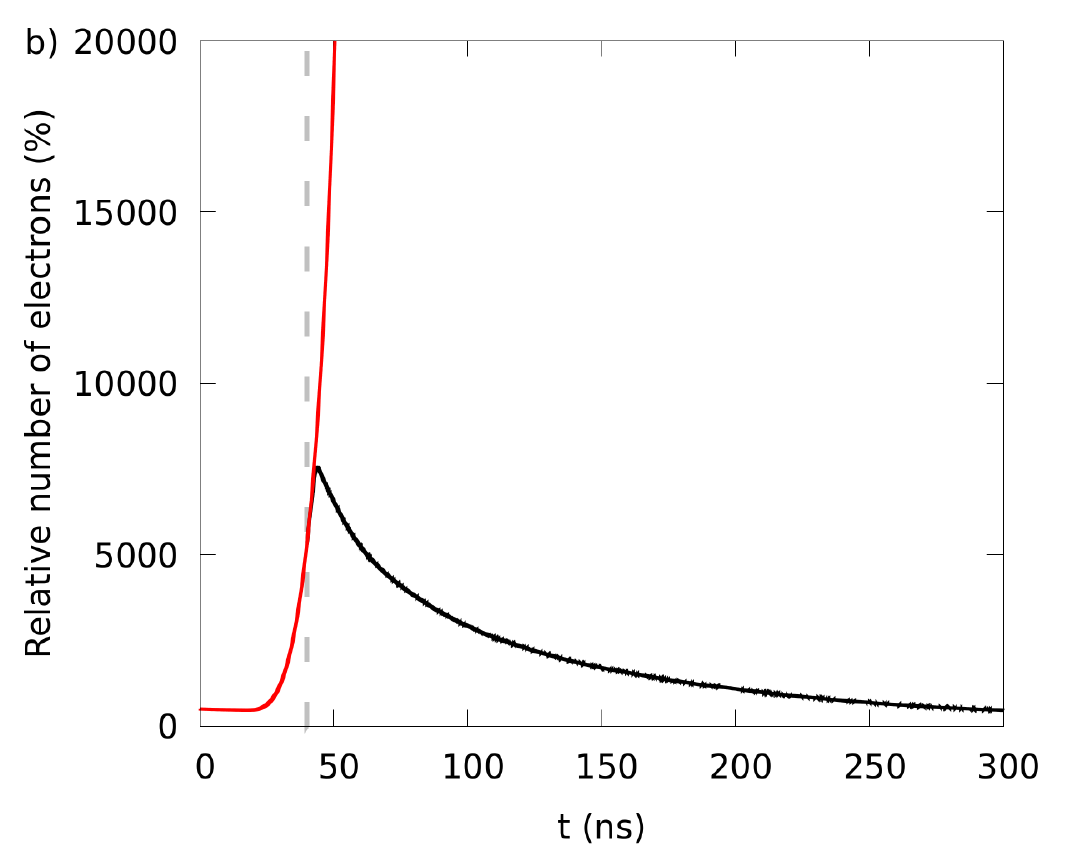}
\caption{ITER CTS case (2D, $E_\text{rf} \perp E_\text{bias} \parallel B_\text{trans.}$, $p=20$\,Pa H$_2$, $E_0=150$\,kV/m, $B=2.14$\,T). Number of electrons for a) $V_\text{bias}=1$\,kV and b) $V_\text{bias}=2$\,kV plotted over time. Results without (red) and with (black) bias voltage applied after $t=40$\,ns.}
\label{fig:Iter10_2kV}
\end{figure}

Mitigation requires an increased bias voltage for pressures above $p \gtrsim 10$\,Pa. To illustrate this, simulation results for a pressure $p=20$\,Pa and bias voltages $V_\text{bias} = 1$\,kV and $V_\text{bias} = 2$\,kV are depicted in Figure~\ref{fig:Iter10_2kV}. As apparent from the graphs, the original bias voltage $V_\text{bias} = 1$\,kV  is uncertain to successfully mitigate breakdown. Electrons are accelerated and removed from the resonant region. However, breakdown continues to proceed with a slowed down transient. Due to the exponential dynamics, the simulations were only conducted within a feasible computational run-time. A doubled bias voltage is required and sufficient to suppress breakdown at least up to this pressure. This is reasoned by the significantly faster breakdown dynamics compared to the previous cases, with a timescale on the order of the mean collision time $\tau_\text{c} \approx 2.8$\,ns. Following Equation~\eqref{eq:sweeptime}, a sweep time $\tau_\text{sweep} \approx 10$\,ns corresponding to $V_\text{bias} = 1$\,kV is too long to inhibit an ionization avalanche (at least up to a feasible simulation time). A value of $\tau_\text{sweep} \approx 6.7$\,ns for $V_\text{bias} = 2$\,kV is of the same order as $\tau_\text{c}$. Hence, with a prolonged effective time constant $\tau \gtrsim \text{65\,ns} \gg \tau_\text{sweep} \approx 6.7$\,ns and a similar reasoning as for $p=10$\,Pa, the balance between electron multiplication through ionization and removal from the central high RF electric field region is sufficient for effective mitigation. Again the actual sweep time is presumably shorter than its upper bound $\tau_\text{sweep}$. Notably, the cause for the further prolonged effective time constant for mitigation is in the different approximate scaling relations of the mean collision time with pressure $\tau_\text{c} \propto p^{-1}$ and the sweep time with bias voltage $\tau_\text{sweep} \propto V_\text{bias}^{-0.5}$, following Equations~\eqref{eq:meancollisionrate} to \eqref{eq:sweeptime}. A doubled pressure reduces the mean collision time more strongly than a doubled bias voltage reduces the sweep time. The ratio of the time constants $\tau_\text{c}/\tau_\text{sweep}$ specifies the point of break even.

Variations of the magnetic field additionally influence the gas breakdown dynamics. By maintaining an excitation frequency $f=60$\,GHz and varying the magnetic field magnitude to $B=0$\,T and $B=1.07$\,T (i.e., a classical RF heating regime, or a 2nd harmonic ECR heating regime may be established). It was found in both cases that even an elevated RF electric field strength of $E_0 = 1$\,MV/m was insufficient to sustain the initial number of electrons within the waveguide. The electron populations decay to zero at the diffusion timescale $\tau_\text{diff} \approx 300$\,ns for $p=1$\,Pa (not shown). These observations are in line with results by Aanesland et al.,\cite{aanesland_electron_2003} who report on the small influence of 2nd harmonic ECR heating.

A peculiar difference in the `Moeller' and the two dimensional ITER CTS scenario is that due to acceleration and removal of electrons along magnetic field lines, electrons are rapidly removed from the center of the waveguide, where ECR heating is strongest (within $\tau \lesssim 20$\,ns; cf.\ Figures~\ref{fig:modes} and \ref{fig:ITER1electrons}). In contrast, however, they accumulate temporarily close to the absorbing (and partially emitting) walls until they are finally removed due to the continued drag toward the wall (see left wall in Figure~\ref{fig:ITER1electrons}b). The mitigation scheme in the `Moeller' case, in contrast, relies on the significantly slower $\vec{E}_\text{bias}\times\vec{B}$ drift, which does not lead to emphasized accumulation in front of the wall (cf.\ Figure~\ref{fig:Moeller0-08electrons}b).

\subsubsection{Three dimensional}

\begin{figure*}[t!]
\includegraphics[width=8cm]{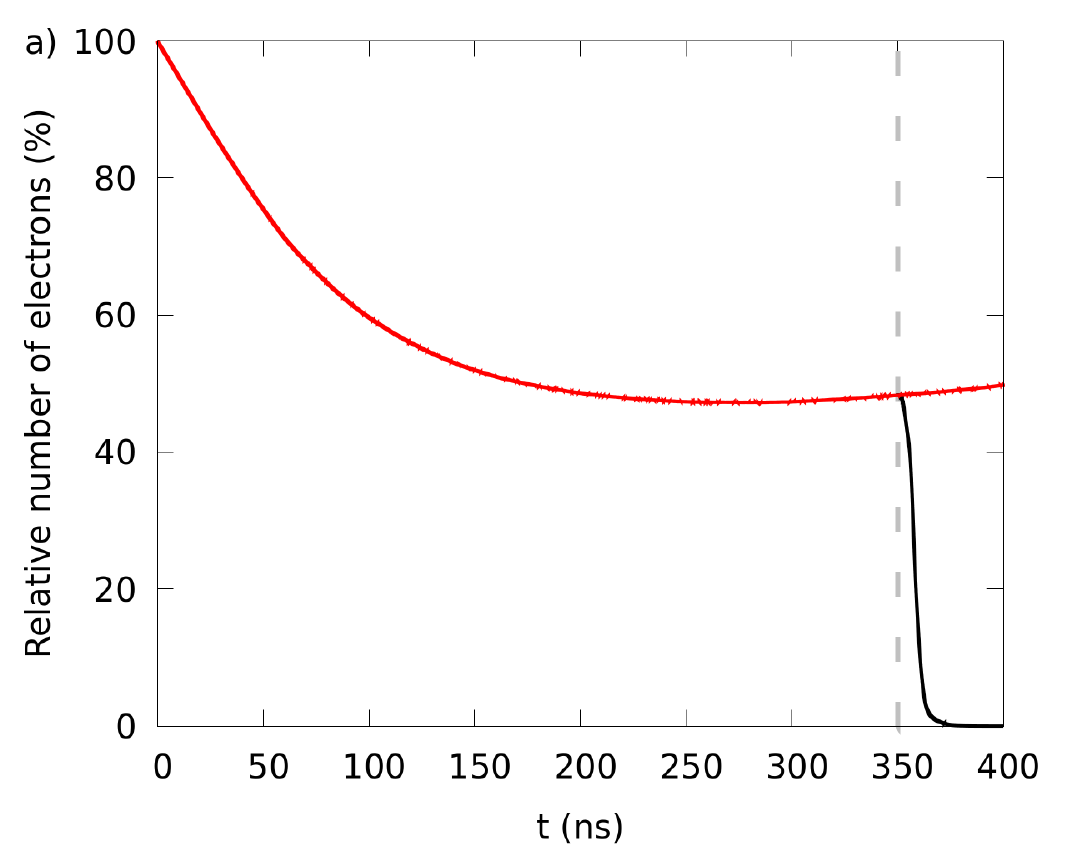}
\includegraphics[width=8cm]{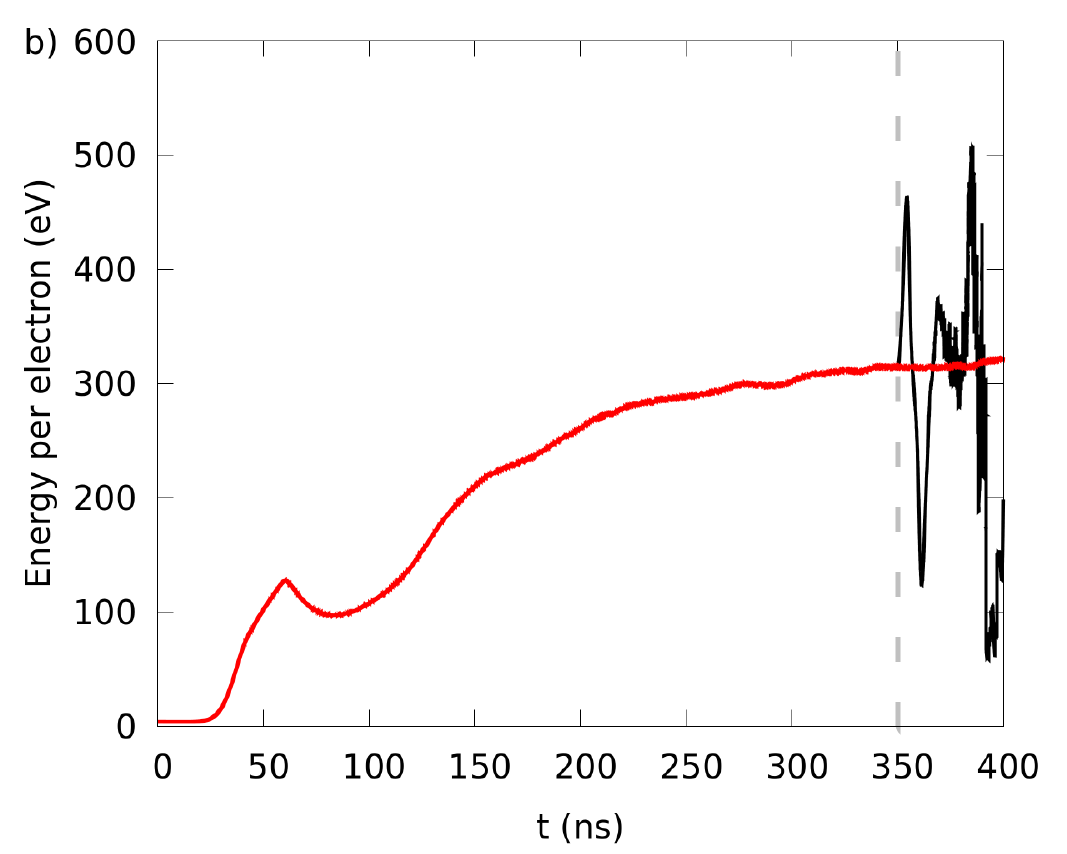}
\caption{ITER CTS case (3D, $E_\text{rf} \perp E_\text{bias} \parallel B_\text{trans.}$, $p=1$\,Pa H$_2$, $E_0=35$\,kV/m, $B=2.14$\,T, $V_\text{bias}=1$\,kV). a) Number of electrons and b) average energy per electron plotted over time. Results without (red) and with (black) bias voltage applied after $t=350$\,ns.}
\label{fig:Iter1_3D}
\end{figure*}

To verify the mechanisms involved in gas breakdown for ITER CTS, also three dimensional simulations were performed for a section of the waveguide of length $L=80$\,mm. The main difference compared to the two dimensional case is posed by an axial magnetic field component, which is expected to reduce the efficacy of ECR heating, due to a smaller RF electric field component $\vec{E}_\text{rf}$ perpendicular to the magnetic field $\vec{B}$. Due to the computational effort involved in these simulations, only a number of representative cases were performed and are depicted. The conceptual similarity of the two and three dimensional cases is illustrated.

\begin{figure*}[b!]
\includegraphics[width=8cm]{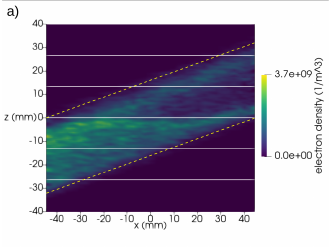}
\includegraphics[width=8cm]{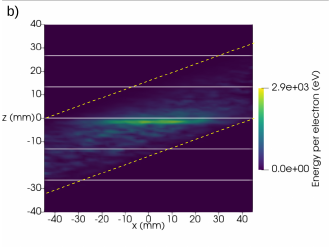}
\caption{ITER CTS case (3D, $E_\text{rf} \perp E_\text{bias} \parallel B_\text{trans.}$, $p=1$\,Pa H$_2$, $E_0=35$\,kV/m, $B=2.14$\,T, without bias voltage). a) Electron density and b) local energy per electron plotted over the $x-z$ plane at $t=350$\,ns. White solid contours indicate isocurves of the magnetic field magnitude. Yellow dashed lines indicate the magnetic field direction.}
\label{fig:Iter1_3D_electrons1}
\end{figure*}

The number of electrons and the average energy per electron for the three dimensional case at a pressure of $p=1$\,Pa and $E_\text{rf} \perp E_\text{bias} \parallel B_\text{trans.}$ (where $B_\text{trans.}$ indicates the magnetic field component in the transveral waveguide plane) are presented in Figure~\ref{fig:Iter1_3D}. Gas breakdown is observed following an initial heating and relaxation phase after approximately $t=200$\,ns. The slow breakdown dynamics -- compared to the two dimensional case of Figure~\ref{fig:Iter1} -- may be attributed to less efficient ECR heating at an identical RF electric field magnitude. Electrons are heated in the central high RF electric field region of the waveguide. This is depicted in Figure~\ref{fig:Iter1_3D_electrons1}, where a) the instantaneous electron density and b) their kinetic energy per electron is plotted over a cross sectional cut through the $x-z$ plane at $t=350$\,ns. While electrons distribute along the magnetic field lines, ECR heating is effective only in the narrow resonant zone where $B \approx 2.14$\,T. As electron transport is bound to the oblique magnetic field lines, however, electrons diffuse out of the resonant zone at the respective angle. Subsequently, they gain kinetic energy only when passing through the resonant region and remain off-resonance along their remaining trajectory along $\vec{B}$. This leads to a large discrepancy between the local and the average energy per electron: local 3\,keV, global average 300\,eV (cf.\ Figures~\ref{fig:Iter1_3D}b and \ref{fig:Iter1_3D_electrons1}). It further appears to influence the electron source/loss balance at the surfaces, specifically through e-SEE. With a maximum emission yield $\delta_\text{m} = 1.3$ at $E_\text{m} = 600$\,eV for Cu, less energetic electrons which are heated in the off-central region are more likely to cause secondary electron emission. Being projected onto the magnetic field lines, this causes an increased electron density along the edges of the resonant region (best visible close to the right-hand boundary at $z \approx 0 \text{ and } 25$\,mm depicted in Figure~\ref{fig:Iter1_3D_electrons1}a).

An additional asymmetry is found in the electron density below ($z<0$\,mm) and above ($z>0$\,mm) the ECR location. Despite the statistical fluctuations apparent in Figure~\ref{fig:Iter1_3D_electrons1}, clearly more electrons populate the volume below the resonance condition, where the magnetic field is smaller and the RF beam originates from. This is a known phenomenon which can be understood from the gradient in $\vec{B}$ approximated by the given field structure. \cite{lieberman_principles_2005} Electrons experience a net drift toward regions of smaller magnetic field, while generally being constrained to their respective field lines. Note that to some extent, the electron and gas breakdown dynamics may be affected by the limited statistics of the three dimensional simulation (i.e., fewer particles per computational cell $\approx 20$).
%This is visible from the statistical fluctuations apparent to Figure~\ref{fig:Iter1_3D_electrons1}.

\begin{figure*}[t!]
\includegraphics[width=8cm]{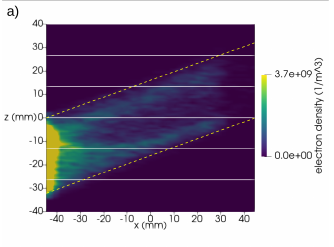}
\includegraphics[width=8cm]{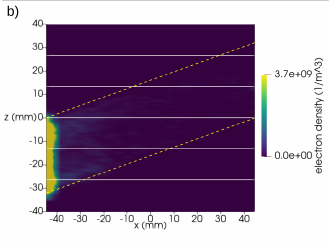}
\caption{ITER CTS case (3D, $E_\text{rf} \perp E_\text{bias} \parallel B_\text{trans.}$, $p=1$\,Pa H$_2$, $E_0=35$\,kV/m, $B=2.14$\,T, $V_\text{bias}=1$\,kV). Electron density plotted over the $x-z$ plane after the bias voltage has been switched for a) $t=355$\,ns and b) $t=360$\,ns. White solid contours indicate the magnitude of the magnetic field. Yellow dashed lines indicate the magnetic field direction.}
\label{fig:Iter1_3D_electrons2}
\end{figure*}

The mechanism of electron removal subsequent to switching on the bias voltage $V_\text{bias}=1$\,kV, at time $t=350$\,ns, is illustrated in Figures~\ref{fig:Iter1_3D} and \ref{fig:Iter1_3D_electrons2}. Transport and removal of electrons proceeds mainly in the direction of the magnetic field, whereas an $\vec{E}_\text{bias}\times\vec{B}$ drift contribution is estimated to be orders of magnitude smaller. With an angle $\angle(\vec{E}_\text{bias},\vec{B}) \approx 20 ^\circ$, the effective bias electric field component parallel to $\vec{B}$ is slightly reduced to approximately 94\,\% of the 2D case. Irrespective of this reduction, also in this case electrons are rapidly removed from the high RF electric field region within a sweep time $\tau_\text{sweep} \approx 10$\,ns. By following the oblique magnetic field lines, electrons are additionally drawn out of the resonant zone with $B \approx 2.14$\,T, further reducing energy input from the RF electric field. The number of electrons decays quickly after a short phase of accumulation at the wall (cf.\ Figure~\ref{fig:Iter1_3D_electrons2}).

\begin{figure*}[t!]
\includegraphics[width=8cm]{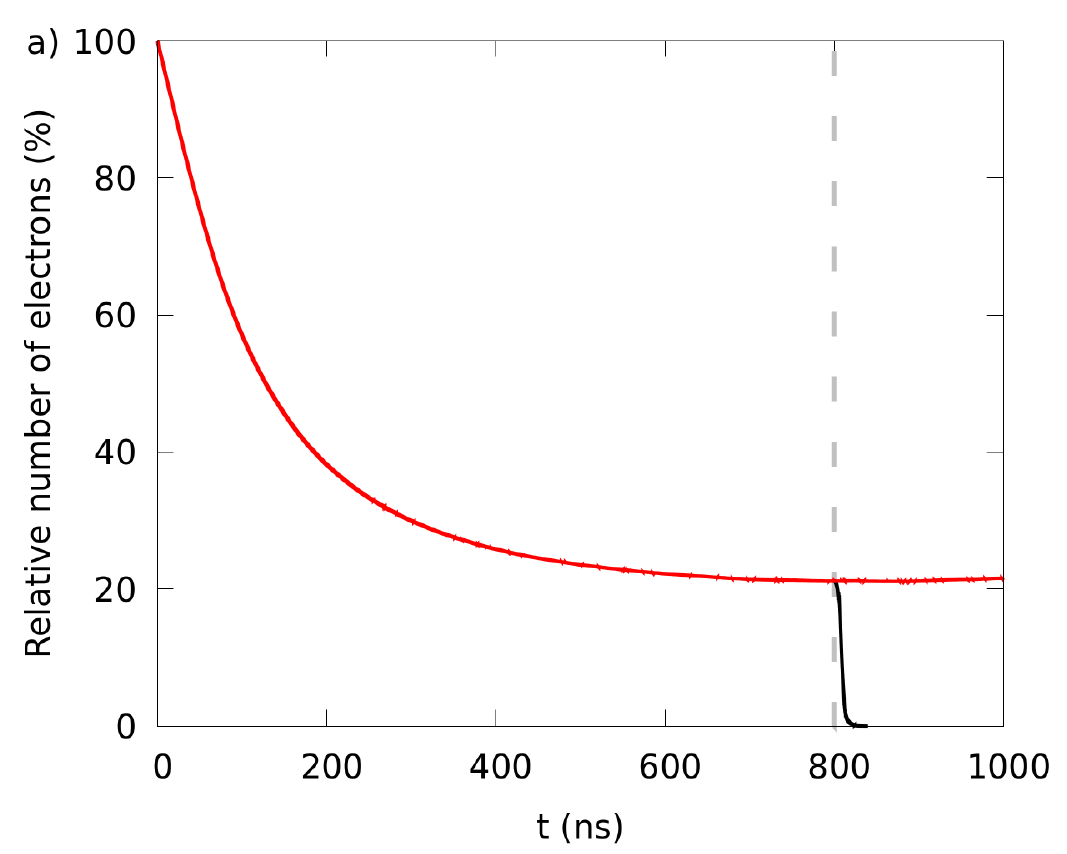}
\includegraphics[width=8cm]{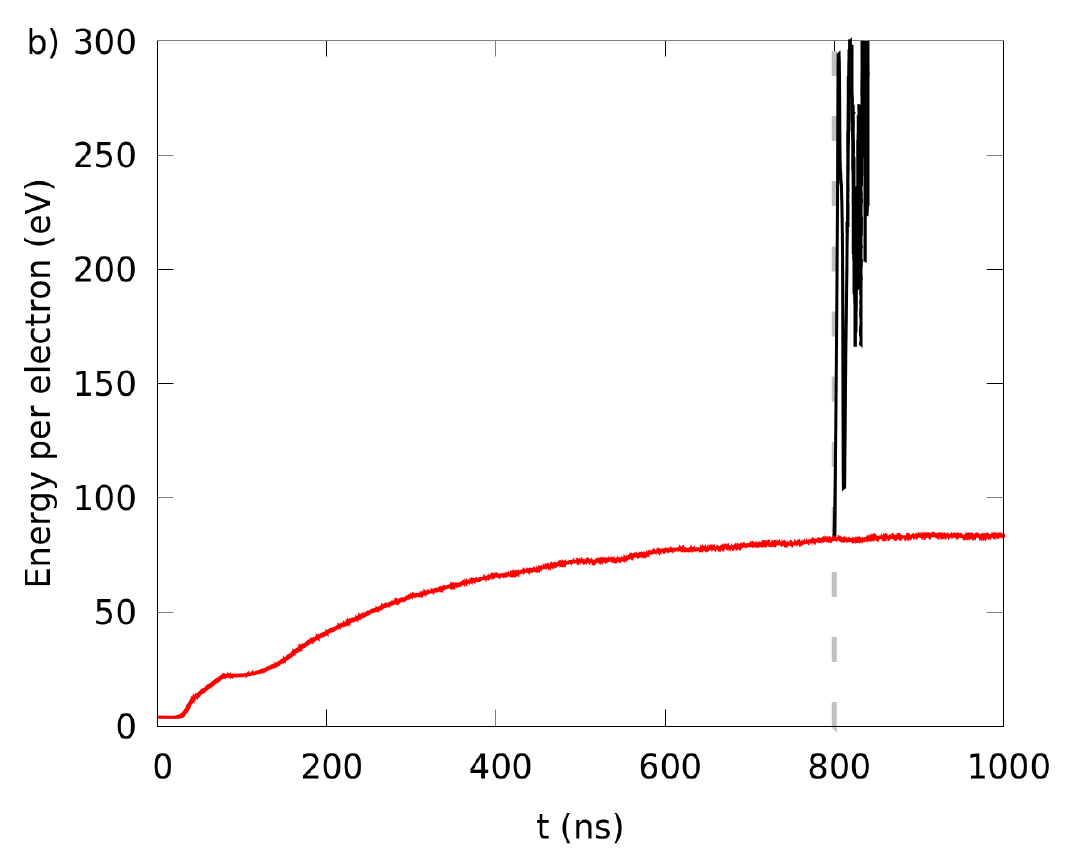}
\caption{ITER CTS case (3D, $E_\text{rf} \parallel E_\text{bias} \parallel B_\text{trans.}$, $p=1$\,Pa H$_2$, $E_0=35$\,kV/m, $B=2.14$\,T, $V_\text{bias}=1$\,kV). a) Number of electrons and b) average energy per electron plotted over time. Results for setup with all transversal field components approximately parallel without (red) and with (black) bias voltage applied after $t=800$\,ns.}
\label{fig:Iter1_3D_allParallel}
\end{figure*}

To obtain further insight into the ECR heating and breakdown dynamics in the three dimensional situation, an altered scenario was simulated, where also the RF electric field points in the $x$ direction, $\vec{E}_\text{rf} \parallel \vec{E}_\text{bias} \parallel \vec{B}_\text{trans.}$. ECR heating in this case is due only to the RF electric field contribution $\vec{E}_\text{rf}$ perpendicular to $\vec{B}$, which is reduced to approximately 34\,\% of the 2D case, as estimated from the axial magnetic field contribution. For otherwise unaltered parameters the results are shown in Figure~\ref{fig:Iter1_3D_allParallel}. Gas breakdown is observed after $t \approx 800$\,ns. The pre-breakdown duration is three times longer, compared to the $\vec{E}_\text{rf} \perp \vec{E}_\text{bias} \parallel \vec{B}_\text{trans.}$ situation, and stems from the slower ECR heating dynamics (in line with the previously estimated RF electric field contribution). Eventually electrons obtain sufficient energy ($E \gtrsim 70$\,eV) to establish an ECR breakdown. Electron removal after the bias voltage $V_\text{bias}=1$\,kV is switched on proceeds analogously to the $\vec{E}_\text{rf} \perp \vec{E}_\text{bias} \parallel \vec{B}_\text{trans.}$ case.

\begin{figure*}[t!]
\includegraphics[width=8cm]{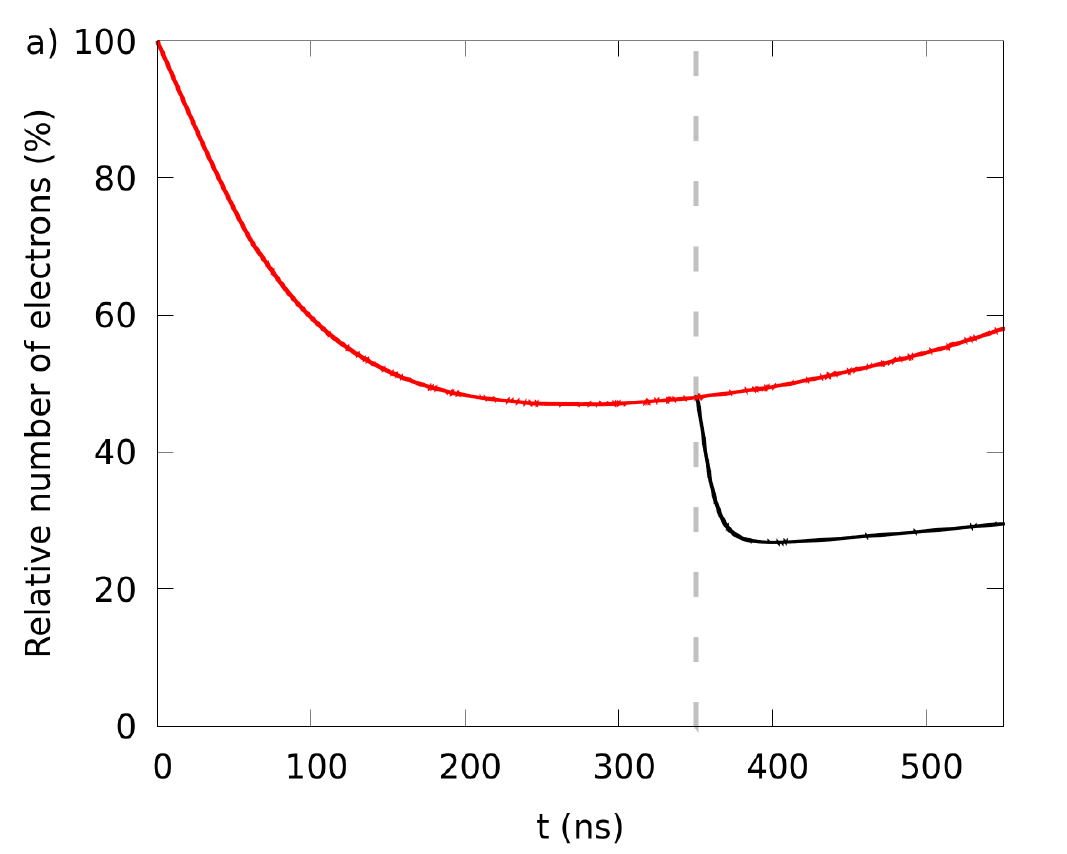}
\includegraphics[width=8cm]{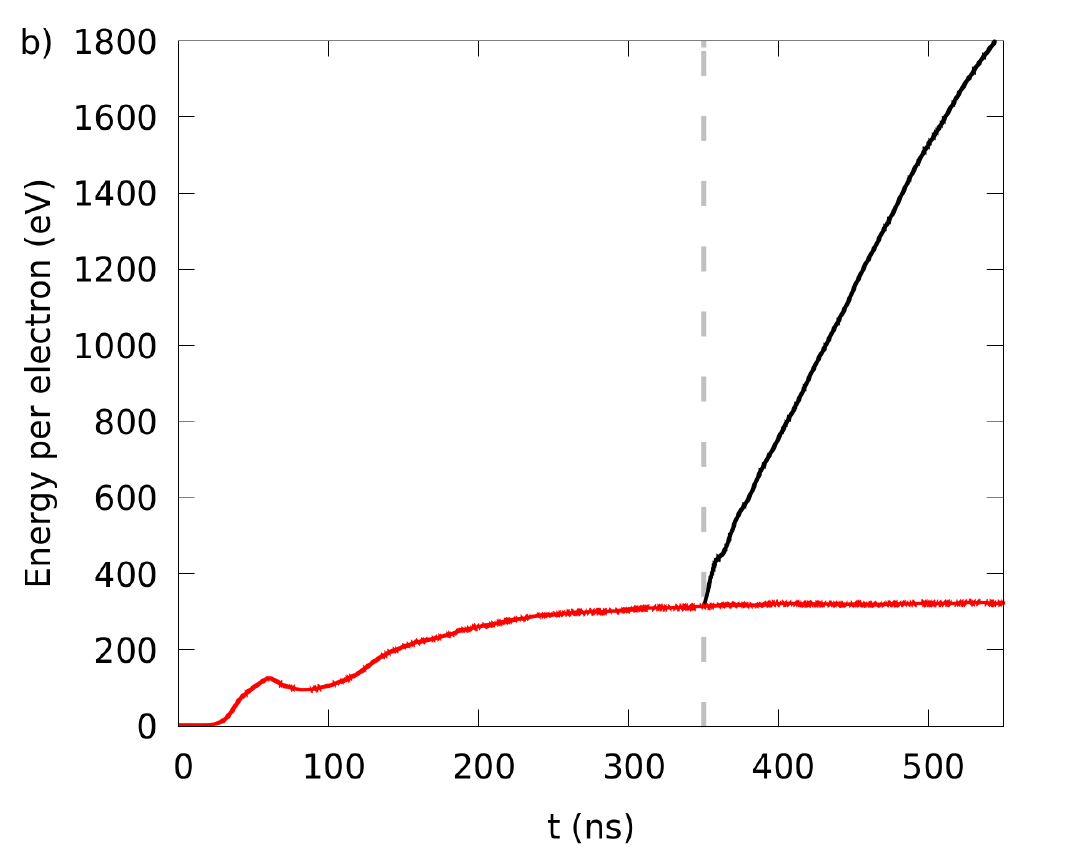}
\caption{ITER CTS case (3D, $E_\text{rf} \parallel E_\text{bias} \perp B_\text{trans.}$, $p=1$\,Pa H$_2$, $E_0=35$\,kV/m, $B=2.14$\,T, $V_\text{bias}=1$\,kV). a) Number of electrons and b) and energy per electron plotted over time. Results for setup with bias electric field approximately perpendicular to the magnetic field without (red) and with (black) bias voltage applied after $t=350$\,ns.}
\label{fig:Iter1_3D_EParallel}
\end{figure*}

Yet another field arrangement was considered to investigate the effectiveness of mitigation when the bias electric field is perpendicular to the transversal component of the magnetic field (i.e., $\vec{E}_\text{bias}$ mainly in the $y$ direction and $\vec{E}_\text{rf} \parallel \vec{E}_\text{bias} \perp \vec{B}_\text{trans.}$). As depicted in Figure~\ref{fig:Iter1_3D_EParallel}, about half of the electrons present in the waveguide are initially lost to the walls within $\tau \approx 20$\,ns. This can be understood from the circumstance that after the bias voltage is switched on, electrons are accelerated by the bias electric field component parallel to the magnetic field ($\vec{E}_\text{bias} \parallel \vec{B}$). Due to the inherent symmetry of the circular SBWG, however, this bias electric field component accelerates electrons toward the wall following the parallel projection, $\vec{E}_\text{bias} \cdot \vec{B}/B$, only in one half of the waveguide (i.e., on one side of the longitudinal waveguide split; cf.\ Figure~\ref{fig:ITERmitigation}). In the other half of the waveguide, $\vec{E}_\text{bias} \cdot \vec{B}/B$ has opposite sign, pinching electrons into the volume of this waveguide half. This pinching is associated with a corresponding continuous energy gain due to the fundamental bias electric field acceleration (see Figure~\ref{fig:Iter1_3D_EParallel}b). In addition, the constricted electrons are concentrated close to the waveguide center with a high RF electric field. This leads to a continued gas breakdown in this half of the waveguide after $t \gtrsim 380$\,ns. The proceeding longtime breakdown dynamics for times $t \gtrsim 400$\,ns are governed by a balance between ionization processes in the pinched waveguide half and transport out of the resonant region following an $\vec{E}_\text{bias} \times \vec{B}$ drift. As this drift is orders of magnitude slower than the $\vec{E}_\text{bias} \parallel \vec{B}$ removal (cf.\ Section~\ref{ssec:Moeller}), the subsequent dynamics cannot be feasibly resolved with the present three dimensional simulation. In addition, a simulation of the referenced effect is problematic for the considered three dimensional setup, because the corresponding drift is directed to the wall only for a sufficiently long waveguide section. As the currently simulated waveguide section is comparably short ($L=8$\,cm), and the simulation assumes periodic boundary conditions at the top and bottom, this transport mechanism can only insufficiently be reproduced by the model. It is, however, expected to be effective in a `non-simplified' ITER CTS setup.

\begin{figure*}[t!]
\includegraphics[width=8cm]{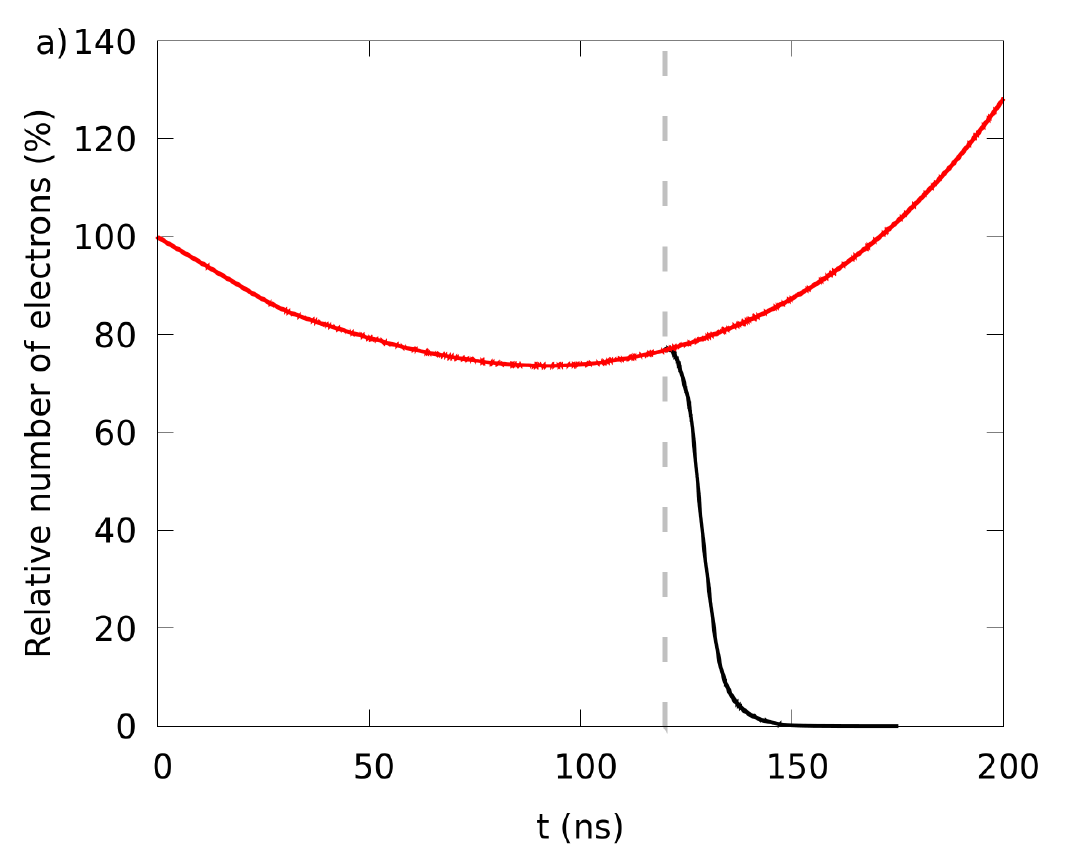}
\includegraphics[width=8cm]{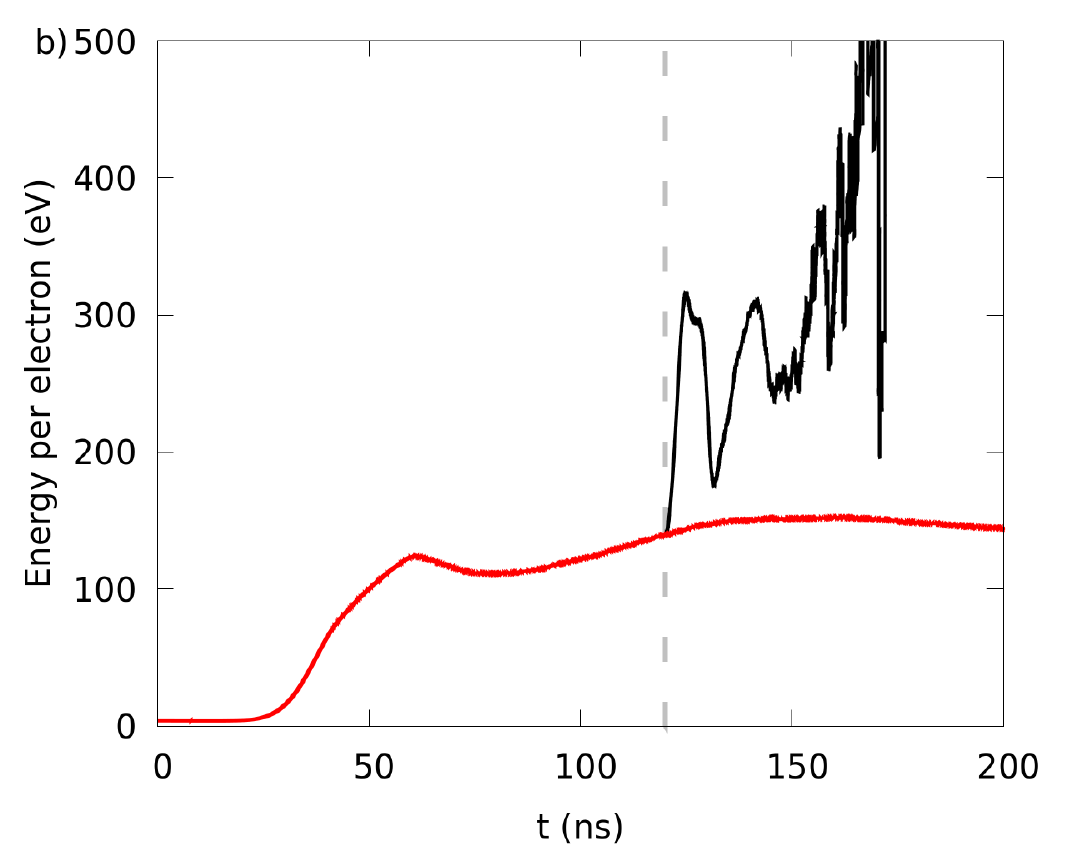}
\caption{ITER CTS case (3D, $E_\text{rf} \perp E_\text{bias} \parallel B_\text{trans.}$, $p=5$\,Pa H$_2$, $E_0=35$\,kV/m, $B=2.14$\,T, $V_\text{bias}=1$\,kV). a) Number of electrons and b) average energy per electron plotted over time. Results without (red) and with (black) bias voltage applied after $t=120$\,ns.}
\label{fig:Iter5_3D}
\end{figure*}

A final three dimensional study concerns a high pressure case $p=5$\,Pa with $\vec{E}_\text{rf} \perp \vec{E}_\text{bias} \parallel \vec{B}_\text{trans.}$. As seen from Figure~\ref{fig:Iter5_3D}, gas breakdown proceeds significantly faster compared to the $p=1$\,Pa case, on the order of the mean collision time $\tau_\text{c} \approx 11$\,ns. The principle ECR heating and breakdown dynamics remain similar, however. It can be seen from the case with a bias voltage $V_\text{bias}=1$\,kV, switched on after $t=120$\,ns (black line), that the SBWG mitigation scheme is also effectively interrupting gas breakdown at this pressure. The mitigation timescale is consistently on the order of the electron sweep time $\tau_\text{sweep} \approx 10$\,ns.

From the preceding analysis it is argued that despite the different model assumptions for the two and three dimensional ITER CTS setup, the principle ECR heating dynamics, gas breakdown, as well as a conceptually similar electron removal mechanism is involved. While the two dimensional situation systematically does not account for any processes related to an axial magnetic field component, it provides a meaningful approximation for the prediction of gas breakdown and mitigation effectiveness. At the same time -- due to the reduced computational cost -- it allows for more detailed studies concerning the specifying parameters, providing a reliable understanding of the inherent processes. In fact, the two dimensional situation merely represents a worst case scenario of the three dimensional case concerning ECR heating, since the RF electric field and the magnetic field are aligned most ideally concerning the ECR condition. Notably, however, the three dimensional scenario also shows physical aspects, which are not present in the simplified two dimensional situation. This especially relates to the peculiar combination of electron transport to the walls, their interaction with the respective surfaces, and their accumulation due to the magnetic field structure and strength. Most apparently this manifests in an altered optimum RF electric field determined to be almost an order of magnitude smaller in the three dimensional case.

\section{Conclusions}

The scope of this work was to investigate ECR heating and the proposed SBWG mitigation scheme for the ITER CTS diagnostic. The results were divided in two main sections:

(i) In Section~\ref{ssec:Moeller}, the simulation scheme was compared against experimental reference data available from Moeller et al. \cite{moeller_avoidance_1987,[{}][{, U.S. Patent 4,687,616, August 18, 1987.}]moeller_method_1987,dellis_experience_1987,moeller_high_1987,moeller_trip_1987} It was shown that ECR-assisted gas breakdown dynamics reported from experiments could be reproduced with our model for breakdown and the Monte Carlo electron simulations. The effectiveness of the SBWG mitigation scheme -- as also reported by Moeller -- was subsequently shown by the simulations. A critical assessment was provided regarding the discrepancy of the lower mitigation bias voltage limit for a pressure of $p=0.133$\,Pa. An analysis of the underlying cause was conducted. It was argued that the uncertainty in outgassing from the SS waveguide walls in the experiments is the most probable contribution to which the deviations are attributed. This is also corroborated by the commissioning phase at the tokamak DITE, where a gyrotron 'conditioning' of the waveguides was necessary, before significant power could be transmitted through the ECR for significantly lengths of time. \cite{moeller_high_1987} The proposed simulation procedure is finally argued to be sufficiently accurate for a reliable prediction of the phenomena expected for the ITER CTS diagnostic system.

(ii) In Section~\ref{ssec:ITER}, the ITER CTS scenario was initially investigated with respect to the fundamental physical processes taking place during gas breakdown and mitigation. Pressure limits (cf.\ following paragraph) were determined for a reduced two dimensional situation, whereas they were hypothesized to provide a reliable measure also for the more realistic three dimensional situation. This three dimensional scenario was subsequently investigated for a number of representative cases to manifest the following aspects:
\begin{enumerate}
\item The similarity in ECR breakdown dynamics for the simulations for 2D and 3D ITER CTS scenarios was initially elaborated.
\item The similarity of SBWG mitigation for 2D and 3D ITER CTS scenarios, with a bias electric field predominantly parallel to the transversal magnetic field component, was laid out thereafter. Notably, for both relevant three dimensional cases ($\vec{E}_\text{rf} \perp \vec{E}_\text{bias} \parallel \vec{B}_\text{trans.}$ and $\vec{E}_\text{rf} \parallel \vec{E}_\text{bias} \parallel \vec{B}_\text{trans.}$), the effectiveness of SBWG mitigation could be demonstrated.
\item The physical processes and the mitigation effectiveness was investigated for a less ideal 3D case with $\vec{E}_\text{bias} \perp \vec{B}_\text{trans.}$, which is conceptually related to the `Moeller' scenario. The attention was drawn to the peculiar effect of electron `pinching' and continued ECR heating in one half of the SBWG.
\end{enumerate}

To conclude, the pressure limits for gas breakdown and mitigation for ITER CTS -- as predicted by the simulation, within the limits of the model assumptions and uncertainties -- are: ECR breakdown is observed down to a hydrogen pressure of $p=0.3$\,Pa when no bias voltage is applied as mitigation action, reasoned by the circumstance that for smaller pressures the diffusion losses to the walls dominate the particle balance. The diffusion loss time is on the order of the mean collision time $\tau_\text{c} \approx 300$\,ns for this pressure and the given setup. An upper hydrogen pressure limit where breakdown mitigation in the ECR heating regime is shown to be effective is $p=10$\,Pa for a bias voltage of $V_\text{bias} = 1$\,kV. The effectiveness of the SBWG mitigation scheme with a bias voltage $V_\text{bias}=1$\,kV is uncertain for higher pressures $p \gtrsim 10$\,Pa, due to a too long electron sweep time, $\tau_\text{sweep} \approx 10$\,ns, compared to the mean collision time $\tau_\text{c} \lesssim \tau_\text{sweep}$. For an increased bias voltage of $V_\text{bias} = 2$\,kV, effective mitigation is demonstrated for a pressure up to $p = 20$\,Pa with a similar reasoning (note the different scaling of the mean collision time and the electron sweep time with pressure and bias voltage). Finally, for two peculiar situations with pure RF heating ($B=0$\,T), and 2nd harmonic ECR heating ($B=1.07$\,T), insufficient electron heating is found even for $E_0 = 1\,\text{MV/m} > 670$\,kV/m (the maximum RF electric field expected in the ITER CTS diagnostic).

It is worth noting that -- given the corresponding parameters -- the procedure and the model used in this work can in principle be extended to accommodate various gas mixtures or wall materials  (e.g., including the reaction processes for tritium decay). Therefore, it could prove useful for future studies of different types of gas breakdown.

\clearpage

\section*{Acknowledgment}
The work leading to this publication has been funded partially by Fusion for Energy under the Framework Partnership Agreement F4E-FPA-393. This publication reflects the views only of the authors, and Fusion for Energy cannot be held responsible for any use, which may be made of the information contained therein. The authors would also like to thank Charles Moeller for considerable input regarding the integration of SBWGs into existing tokamaks.

\section*{Data availability}
The data that support the findings of this study are available from the corresponding author upon reasonable request.

\section*{ORCiD IDs}
\noindent
Jan Trieschmann: \url{https://orcid.org/0000-0001-9136-8019}\\
Axel Wright Larsen: \url{https://orcid.org/0000-0002-7837-717X}\\
Thomas Mussenbrock: \url{https://orcid.org/0000-0001-6445-4990}\\
Søren Bang Korsholm: \url{https://orcid.org/0000-0001-7160-8361}

\bibliography{references}

\begin{thebibliography}{65}%
\makeatletter
\providecommand \@ifxundefined [1]{%
 \@ifx{#1\undefined}
}%
\providecommand \@ifnum [1]{%
 \ifnum #1\expandafter \@firstoftwo
 \else \expandafter \@secondoftwo
 \fi
}%
\providecommand \@ifx [1]{%
 \ifx #1\expandafter \@firstoftwo
 \else \expandafter \@secondoftwo
 \fi
}%
\providecommand \natexlab [1]{#1}%
\providecommand \enquote  [1]{``#1''}%
\providecommand \bibnamefont  [1]{#1}%
\providecommand \bibfnamefont [1]{#1}%
\providecommand \citenamefont [1]{#1}%
\providecommand \href@noop [0]{\@secondoftwo}%
\providecommand \href [0]{\begingroup \@sanitize@url \@href}%
\providecommand \@href[1]{\@@startlink{#1}\@@href}%
\providecommand \@@href[1]{\endgroup#1\@@endlink}%
\providecommand \@sanitize@url [0]{\catcode `\\12\catcode `\$12\catcode
  `\&12\catcode `\#12\catcode `\^12\catcode `\_12\catcode `\%12\relax}%
\providecommand \@@startlink[1]{}%
\providecommand \@@endlink[0]{}%
\providecommand \url  [0]{\begingroup\@sanitize@url \@url }%
\providecommand \@url [1]{\endgroup\@href {#1}{\urlprefix }}%
\providecommand \urlprefix  [0]{URL }%
\providecommand \Eprint [0]{\href }%
\providecommand \doibase [0]{http://dx.doi.org/}%
\providecommand \selectlanguage [0]{\@gobble}%
\providecommand \bibinfo  [0]{\@secondoftwo}%
\providecommand \bibfield  [0]{\@secondoftwo}%
\providecommand \translation [1]{[#1]}%
\providecommand \BibitemOpen [0]{}%
\providecommand \bibitemStop [0]{}%
\providecommand \bibitemNoStop [0]{.\EOS\space}%
\providecommand \EOS [0]{\spacefactor3000\relax}%
\providecommand \BibitemShut  [1]{\csname bibitem#1\endcsname}%
\let\auto@bib@innerbib\@empty
%</preamble>
\bibitem [{\citenamefont {Korsholm}\ \emph {et~al.}(2019)\citenamefont
  {Korsholm}, \citenamefont {Gon{\c c}alves}, \citenamefont {Gutierrez},
  \citenamefont {Henriques}, \citenamefont {Infante}, \citenamefont {Jensen},
  \citenamefont {Jessen}, \citenamefont {Klinkby}, \citenamefont {Larsen},
  \citenamefont {Leipold}, \citenamefont {Lopes}, \citenamefont {Luis},
  \citenamefont {Naulin}, \citenamefont {Nielsen}, \citenamefont {Nonb{\o}l},
  \citenamefont {Rasmussen}, \citenamefont {Salewski}, \citenamefont {Stejner},
  \citenamefont {Taormina}, \citenamefont {Vale}, \citenamefont {Vidal},
  \citenamefont {Sanchez}, \citenamefont {Ballester},\ and\ \citenamefont
  {Udintsev}}]{korsholm_design_2019}%
  \BibitemOpen
  \bibfield  {author} {\bibinfo {author} {\bibfnamefont {S.~B.}\ \bibnamefont
  {Korsholm}}, \bibinfo {author} {\bibfnamefont {B.}~\bibnamefont {Gon{\c
  c}alves}}, \bibinfo {author} {\bibfnamefont {H.~E.}\ \bibnamefont
  {Gutierrez}}, \bibinfo {author} {\bibfnamefont {E.}~\bibnamefont
  {Henriques}}, \bibinfo {author} {\bibfnamefont {V.}~\bibnamefont {Infante}},
  \bibinfo {author} {\bibfnamefont {T.}~\bibnamefont {Jensen}}, \bibinfo
  {author} {\bibfnamefont {M.}~\bibnamefont {Jessen}}, \bibinfo {author}
  {\bibfnamefont {E.~B.}\ \bibnamefont {Klinkby}}, \bibinfo {author}
  {\bibfnamefont {A.~W.}\ \bibnamefont {Larsen}}, \bibinfo {author}
  {\bibfnamefont {F.}~\bibnamefont {Leipold}}, \bibinfo {author} {\bibfnamefont
  {A.}~\bibnamefont {Lopes}}, \bibinfo {author} {\bibfnamefont
  {R.}~\bibnamefont {Luis}}, \bibinfo {author} {\bibfnamefont {V.}~\bibnamefont
  {Naulin}}, \bibinfo {author} {\bibfnamefont {S.~K.}\ \bibnamefont {Nielsen}},
  \bibinfo {author} {\bibfnamefont {E.}~\bibnamefont {Nonb{\o}l}}, \bibinfo
  {author} {\bibfnamefont {J.}~\bibnamefont {Rasmussen}}, \bibinfo {author}
  {\bibfnamefont {M.}~\bibnamefont {Salewski}}, \bibinfo {author}
  {\bibfnamefont {M.}~\bibnamefont {Stejner}}, \bibinfo {author} {\bibfnamefont
  {A.}~\bibnamefont {Taormina}}, \bibinfo {author} {\bibfnamefont
  {A.}~\bibnamefont {Vale}}, \bibinfo {author} {\bibfnamefont {C.}~\bibnamefont
  {Vidal}}, \bibinfo {author} {\bibfnamefont {L.}~\bibnamefont {Sanchez}},
  \bibinfo {author} {\bibfnamefont {R.~M.}\ \bibnamefont {Ballester}}, \ and\
  \bibinfo {author} {\bibfnamefont {V.}~\bibnamefont {Udintsev}},\ }\href
  {\doibase 10.1051/epjconf/201920303002} {\bibfield  {journal} {\bibinfo
  {journal} {EPJ Web of Conferences}\ }\textbf {\bibinfo {volume} {203}},\
  \bibinfo {pages} {03002} (\bibinfo {year} {2019})}\BibitemShut {NoStop}%
\bibitem [{\citenamefont {Salewski}\ \emph {et~al.}(2018)\citenamefont
  {Salewski}, \citenamefont {Nocente}, \citenamefont {Madsen}, \citenamefont
  {Abramovic}, \citenamefont {Fitzgerald}, \citenamefont {Gorini},
  \citenamefont {Hansen}, \citenamefont {Heidbrink}, \citenamefont {Jacobsen},
  \citenamefont {Jensen}, \citenamefont {Kiptily}, \citenamefont {Klinkby},
  \citenamefont {Korsholm}, \citenamefont {{Kurki-Suonio}}, \citenamefont
  {Larsen}, \citenamefont {Leipold}, \citenamefont {Moseev}, \citenamefont
  {Nielsen}, \citenamefont {Pinches}, \citenamefont {Rasmussen}, \citenamefont
  {Rebai}, \citenamefont {Schneider}, \citenamefont {Shevelev}, \citenamefont
  {Sipil{\"a}}, \citenamefont {Stejner},\ and\ \citenamefont
  {Tardocchi}}]{salewski_alpha-particle_2018}%
  \BibitemOpen
  \bibfield  {author} {\bibinfo {author} {\bibfnamefont {M.}~\bibnamefont
  {Salewski}}, \bibinfo {author} {\bibfnamefont {M.}~\bibnamefont {Nocente}},
  \bibinfo {author} {\bibfnamefont {B.}~\bibnamefont {Madsen}}, \bibinfo
  {author} {\bibfnamefont {I.}~\bibnamefont {Abramovic}}, \bibinfo {author}
  {\bibfnamefont {M.}~\bibnamefont {Fitzgerald}}, \bibinfo {author}
  {\bibfnamefont {G.}~\bibnamefont {Gorini}}, \bibinfo {author} {\bibfnamefont
  {P.~C.}\ \bibnamefont {Hansen}}, \bibinfo {author} {\bibfnamefont {W.~W.}\
  \bibnamefont {Heidbrink}}, \bibinfo {author} {\bibfnamefont {A.~S.}\
  \bibnamefont {Jacobsen}}, \bibinfo {author} {\bibfnamefont {T.}~\bibnamefont
  {Jensen}}, \bibinfo {author} {\bibfnamefont {V.~G.}\ \bibnamefont {Kiptily}},
  \bibinfo {author} {\bibfnamefont {E.~B.}\ \bibnamefont {Klinkby}}, \bibinfo
  {author} {\bibfnamefont {S.~B.}\ \bibnamefont {Korsholm}}, \bibinfo {author}
  {\bibfnamefont {T.}~\bibnamefont {{Kurki-Suonio}}}, \bibinfo {author}
  {\bibfnamefont {A.~W.}\ \bibnamefont {Larsen}}, \bibinfo {author}
  {\bibfnamefont {F.}~\bibnamefont {Leipold}}, \bibinfo {author} {\bibfnamefont
  {D.}~\bibnamefont {Moseev}}, \bibinfo {author} {\bibfnamefont {S.~K.}\
  \bibnamefont {Nielsen}}, \bibinfo {author} {\bibfnamefont {S.~D.}\
  \bibnamefont {Pinches}}, \bibinfo {author} {\bibfnamefont {J.}~\bibnamefont
  {Rasmussen}}, \bibinfo {author} {\bibfnamefont {M.}~\bibnamefont {Rebai}},
  \bibinfo {author} {\bibfnamefont {M.}~\bibnamefont {Schneider}}, \bibinfo
  {author} {\bibfnamefont {A.}~\bibnamefont {Shevelev}}, \bibinfo {author}
  {\bibfnamefont {S.}~\bibnamefont {Sipil{\"a}}}, \bibinfo {author}
  {\bibfnamefont {M.}~\bibnamefont {Stejner}}, \ and\ \bibinfo {author}
  {\bibfnamefont {M.}~\bibnamefont {Tardocchi}},\ }\href {\doibase
  10.1088/1741-4326/aace05} {\bibfield  {journal} {\bibinfo  {journal} {Nuclear
  Fusion}\ }\textbf {\bibinfo {volume} {58}},\ \bibinfo {pages} {096019}
  (\bibinfo {year} {2018})}\BibitemShut {NoStop}%
\bibitem [{\citenamefont {Paschen}(1889)}]{paschen_uber_1889}%
  \BibitemOpen
  \bibfield  {author} {\bibinfo {author} {\bibfnamefont {F.}~\bibnamefont
  {Paschen}},\ }\href {\doibase 10.1002/andp.18892730505} {\bibfield  {journal}
  {\bibinfo  {journal} {Annalen der Physik}\ }\textbf {\bibinfo {volume}
  {273}},\ \bibinfo {pages} {69} (\bibinfo {year} {1889})}\BibitemShut
  {NoStop}%
\bibitem [{\citenamefont {Townsend}(1910)}]{townsend_theory_1910}%
  \BibitemOpen
  \bibfield  {author} {\bibinfo {author} {\bibfnamefont {J.~S.}\ \bibnamefont
  {Townsend}},\ }\href@noop {} {\emph {\bibinfo {title} {Theory of
  {{Ionization}} of {{Gases}} by {{Collision}}}}}\ (\bibinfo  {publisher}
  {{Constable \& Company}},\ \bibinfo {address} {{London, UK}},\ \bibinfo
  {year} {1910})\BibitemShut {NoStop}%
\bibitem [{\citenamefont {Lieberman}\ and\ \citenamefont
  {Lichtenberg}(2005)}]{lieberman_principles_2005}%
  \BibitemOpen
  \bibfield  {author} {\bibinfo {author} {\bibfnamefont {M.~A.}\ \bibnamefont
  {Lieberman}}\ and\ \bibinfo {author} {\bibfnamefont {A.~J.}\ \bibnamefont
  {Lichtenberg}},\ }\href@noop {} {\emph {\bibinfo {title} {Principles of
  {{Plasma Discharges}} and {{Materials Processing}}}}},\ \bibinfo {edition}
  {2nd}\ ed.\ (\bibinfo  {publisher} {{Wiley}},\ \bibinfo {address} {{Hoboken,
  USA}},\ \bibinfo {year} {2005})\BibitemShut {NoStop}%
\bibitem [{\citenamefont {MacDonald}\ and\ \citenamefont
  {Brown}(1949)}]{macdonald_high_1949-1}%
  \BibitemOpen
  \bibfield  {author} {\bibinfo {author} {\bibfnamefont {A.~D.}\ \bibnamefont
  {MacDonald}}\ and\ \bibinfo {author} {\bibfnamefont {S.~C.}\ \bibnamefont
  {Brown}},\ }\href {\doibase 10.1103/PhysRev.76.1634} {\bibfield  {journal}
  {\bibinfo  {journal} {Physical Review}\ }\textbf {\bibinfo {volume} {76}},\
  \bibinfo {pages} {1634} (\bibinfo {year} {1949})}\BibitemShut {NoStop}%
\bibitem [{\citenamefont {Lax}, \citenamefont {Allis},\ and\ \citenamefont
  {Brown}(1950)}]{lax_effect_1950}%
  \BibitemOpen
  \bibfield  {author} {\bibinfo {author} {\bibfnamefont {B.}~\bibnamefont
  {Lax}}, \bibinfo {author} {\bibfnamefont {W.~P.}\ \bibnamefont {Allis}}, \
  and\ \bibinfo {author} {\bibfnamefont {S.~C.}\ \bibnamefont {Brown}},\ }\href
  {\doibase 10.1063/1.1699594} {\bibfield  {journal} {\bibinfo  {journal}
  {Journal of Applied Physics}\ }\textbf {\bibinfo {volume} {21}},\ \bibinfo
  {pages} {1297} (\bibinfo {year} {1950})}\BibitemShut {NoStop}%
\bibitem [{\citenamefont {Lax}\ and\ \citenamefont
  {Cohn}(1973)}]{lax_cyclotron_1973}%
  \BibitemOpen
  \bibfield  {author} {\bibinfo {author} {\bibfnamefont {B.}~\bibnamefont
  {Lax}}\ and\ \bibinfo {author} {\bibfnamefont {D.~R.}\ \bibnamefont {Cohn}},\
  }\href {\doibase 10.1063/1.1654920} {\bibfield  {journal} {\bibinfo
  {journal} {Applied Physics Letters}\ }\textbf {\bibinfo {volume} {23}},\
  \bibinfo {pages} {363} (\bibinfo {year} {1973})}\BibitemShut {NoStop}%
\bibitem [{\citenamefont {Bornatici}\ \emph {et~al.}(1983)\citenamefont
  {Bornatici}, \citenamefont {Cano}, \citenamefont {Barbieri},\ and\
  \citenamefont {Engelmann}}]{bornatici_electron_1983}%
  \BibitemOpen
  \bibfield  {author} {\bibinfo {author} {\bibfnamefont {M.}~\bibnamefont
  {Bornatici}}, \bibinfo {author} {\bibfnamefont {R.}~\bibnamefont {Cano}},
  \bibinfo {author} {\bibfnamefont {O.~D.}\ \bibnamefont {Barbieri}}, \ and\
  \bibinfo {author} {\bibfnamefont {F.}~\bibnamefont {Engelmann}},\ }\href
  {\doibase 10.1088/0029-5515/23/9/005} {\bibfield  {journal} {\bibinfo
  {journal} {Nuclear Fusion}\ }\textbf {\bibinfo {volume} {23}},\ \bibinfo
  {pages} {1153} (\bibinfo {year} {1983})}\BibitemShut {NoStop}%
\bibitem [{\citenamefont {Strauss}\ \emph {et~al.}(2019)\citenamefont
  {Strauss}, \citenamefont {Aiello}, \citenamefont {Bertizzolo}, \citenamefont
  {Bruschi}, \citenamefont {Casal}, \citenamefont {Chavan}, \citenamefont
  {Farina}, \citenamefont {Figini}, \citenamefont {Gagliardi}, \citenamefont
  {Goodman}, \citenamefont {Grossetti}, \citenamefont {Heemskerk},
  \citenamefont {Henderson}, \citenamefont {Kasparek}, \citenamefont {Koning},
  \citenamefont {Landis}, \citenamefont {Leichtle}, \citenamefont {Meier},
  \citenamefont {Moro}, \citenamefont {Nowak}, \citenamefont {Pacheco},
  \citenamefont {Platania}, \citenamefont {Plaum}, \citenamefont {Poli},
  \citenamefont {Ramseyer}, \citenamefont {Ronden}, \citenamefont {Saibene},
  \citenamefont {{M{\'a}s-Sanchez}}, \citenamefont {Santos~Silva},
  \citenamefont {Sauter}, \citenamefont {Scherer}, \citenamefont {Schreck},
  \citenamefont {Sozzi}, \citenamefont {Spaeh}, \citenamefont {Vagnoni},
  \citenamefont {Vaccaro},\ and\ \citenamefont
  {Weinhorst}}]{strauss_nearing_2019}%
  \BibitemOpen
  \bibfield  {author} {\bibinfo {author} {\bibfnamefont {D.}~\bibnamefont
  {Strauss}}, \bibinfo {author} {\bibfnamefont {G.}~\bibnamefont {Aiello}},
  \bibinfo {author} {\bibfnamefont {R.}~\bibnamefont {Bertizzolo}}, \bibinfo
  {author} {\bibfnamefont {A.}~\bibnamefont {Bruschi}}, \bibinfo {author}
  {\bibfnamefont {N.}~\bibnamefont {Casal}}, \bibinfo {author} {\bibfnamefont
  {R.}~\bibnamefont {Chavan}}, \bibinfo {author} {\bibfnamefont
  {D.}~\bibnamefont {Farina}}, \bibinfo {author} {\bibfnamefont
  {L.}~\bibnamefont {Figini}}, \bibinfo {author} {\bibfnamefont
  {M.}~\bibnamefont {Gagliardi}}, \bibinfo {author} {\bibfnamefont {T.~P.}\
  \bibnamefont {Goodman}}, \bibinfo {author} {\bibfnamefont {G.}~\bibnamefont
  {Grossetti}}, \bibinfo {author} {\bibfnamefont {C.}~\bibnamefont
  {Heemskerk}}, \bibinfo {author} {\bibfnamefont {M.~A.}\ \bibnamefont
  {Henderson}}, \bibinfo {author} {\bibfnamefont {W.}~\bibnamefont {Kasparek}},
  \bibinfo {author} {\bibfnamefont {J.}~\bibnamefont {Koning}}, \bibinfo
  {author} {\bibfnamefont {J.~D.}\ \bibnamefont {Landis}}, \bibinfo {author}
  {\bibfnamefont {D.}~\bibnamefont {Leichtle}}, \bibinfo {author}
  {\bibfnamefont {A.}~\bibnamefont {Meier}}, \bibinfo {author} {\bibfnamefont
  {A.}~\bibnamefont {Moro}}, \bibinfo {author} {\bibfnamefont {S.}~\bibnamefont
  {Nowak}}, \bibinfo {author} {\bibfnamefont {J.}~\bibnamefont {Pacheco}},
  \bibinfo {author} {\bibfnamefont {P.}~\bibnamefont {Platania}}, \bibinfo
  {author} {\bibfnamefont {B.}~\bibnamefont {Plaum}}, \bibinfo {author}
  {\bibfnamefont {E.}~\bibnamefont {Poli}}, \bibinfo {author} {\bibfnamefont
  {F.}~\bibnamefont {Ramseyer}}, \bibinfo {author} {\bibfnamefont
  {D.}~\bibnamefont {Ronden}}, \bibinfo {author} {\bibfnamefont
  {G.}~\bibnamefont {Saibene}}, \bibinfo {author} {\bibfnamefont
  {A.}~\bibnamefont {{M{\'a}s-Sanchez}}}, \bibinfo {author} {\bibfnamefont
  {P.}~\bibnamefont {Santos~Silva}}, \bibinfo {author} {\bibfnamefont
  {O.}~\bibnamefont {Sauter}}, \bibinfo {author} {\bibfnamefont
  {T.}~\bibnamefont {Scherer}}, \bibinfo {author} {\bibfnamefont
  {S.}~\bibnamefont {Schreck}}, \bibinfo {author} {\bibfnamefont
  {C.}~\bibnamefont {Sozzi}}, \bibinfo {author} {\bibfnamefont
  {P.}~\bibnamefont {Spaeh}}, \bibinfo {author} {\bibfnamefont
  {M.}~\bibnamefont {Vagnoni}}, \bibinfo {author} {\bibfnamefont
  {A.}~\bibnamefont {Vaccaro}}, \ and\ \bibinfo {author} {\bibfnamefont
  {B.}~\bibnamefont {Weinhorst}},\ }\href {\doibase
  10.1016/j.fusengdes.2018.11.013} {\bibfield  {journal} {\bibinfo  {journal}
  {Fusion Engineering and Design}\ }\textbf {\bibinfo {volume} {146}},\
  \bibinfo {pages} {23} (\bibinfo {year} {2019})}\BibitemShut {NoStop}%
\bibitem [{\citenamefont {Li}\ \emph {et~al.}(2012)\citenamefont {Li},
  \citenamefont {Teunissen}, \citenamefont {Nool}, \citenamefont
  {Hundsdorfer},\ and\ \citenamefont {Ebert}}]{li_comparison_2012}%
  \BibitemOpen
  \bibfield  {author} {\bibinfo {author} {\bibfnamefont {C.}~\bibnamefont
  {Li}}, \bibinfo {author} {\bibfnamefont {J.}~\bibnamefont {Teunissen}},
  \bibinfo {author} {\bibfnamefont {M.}~\bibnamefont {Nool}}, \bibinfo {author}
  {\bibfnamefont {W.}~\bibnamefont {Hundsdorfer}}, \ and\ \bibinfo {author}
  {\bibfnamefont {U.}~\bibnamefont {Ebert}},\ }\href {\doibase
  10.1088/0963-0252/21/5/055019} {\bibfield  {journal} {\bibinfo  {journal}
  {Plasma Sources Science and Technology}\ }\textbf {\bibinfo {volume} {21}},\
  \bibinfo {pages} {055019} (\bibinfo {year} {2012})}\BibitemShut {NoStop}%
\bibitem [{\citenamefont {Teunissen}\ and\ \citenamefont
  {Ebert}(2016)}]{teunissen_3d_2016}%
  \BibitemOpen
  \bibfield  {author} {\bibinfo {author} {\bibfnamefont {J.}~\bibnamefont
  {Teunissen}}\ and\ \bibinfo {author} {\bibfnamefont {U.}~\bibnamefont
  {Ebert}},\ }\href {\doibase 10.1088/0963-0252/25/4/044005} {\bibfield
  {journal} {\bibinfo  {journal} {Plasma Sources Science and Technology}\
  }\textbf {\bibinfo {volume} {25}},\ \bibinfo {pages} {044005} (\bibinfo
  {year} {2016})}\BibitemShut {NoStop}%
\bibitem [{\citenamefont {Mao}\ \emph {et~al.}(2020)\citenamefont {Mao},
  \citenamefont {Li}, \citenamefont {Ye},\ and\ \citenamefont
  {He}}]{mao_monte_2020}%
  \BibitemOpen
  \bibfield  {author} {\bibinfo {author} {\bibfnamefont {Z.}~\bibnamefont
  {Mao}}, \bibinfo {author} {\bibfnamefont {Y.}~\bibnamefont {Li}}, \bibinfo
  {author} {\bibfnamefont {M.}~\bibnamefont {Ye}}, \ and\ \bibinfo {author}
  {\bibfnamefont {Y.}~\bibnamefont {He}},\ }\href {\doibase 10.1063/5.0010169}
  {\bibfield  {journal} {\bibinfo  {journal} {Physics of Plasmas}\ }\textbf
  {\bibinfo {volume} {27}},\ \bibinfo {pages} {093502} (\bibinfo {year}
  {2020})}\BibitemShut {NoStop}%
\bibitem [{\citenamefont
  {Moeller}(1987{\natexlab{a}})}]{moeller_avoidance_1987}%
  \BibitemOpen
  \bibfield  {author} {\bibinfo {author} {\bibfnamefont {C.~P.}\ \bibnamefont
  {Moeller}},\ }\href@noop {} {\enquote {\bibinfo {title} {Avoidance of
  cyclotron breakdown in partially evacuated waveguides},}\ }\bibinfo {type}
  {Tech. Rep.}\ \bibinfo {number} {GA-A--18836}\ (\bibinfo  {institution} {{GA
  Technologies}},\ \bibinfo {year} {1987})\BibitemShut {NoStop}%
\bibitem [{\citenamefont {Moeller}(1987{\natexlab{b}})}]{moeller_method_1987}%
  \BibitemOpen
  \bibfield  {author} {\bibinfo {author} {\bibfnamefont {C.~P.}\ \bibnamefont
  {Moeller}},\ }\href@noop {} {\enquote {\bibinfo {title} {Method and apparatus
  for preventing cyclotron breakdown in partially evacuated waveguide},}\ }
  (\bibinfo {year} {1987}{\natexlab{b}})\BibitemShut {NoStop}%
\bibitem [{\citenamefont {Dellis}\ \emph {et~al.}(1987)\citenamefont {Dellis},
  \citenamefont {Alcock}, \citenamefont {Ainsworth}, \citenamefont {Collins},
  \citenamefont {Fielding}, \citenamefont {Hugill}, \citenamefont {Johnson},
  \citenamefont {Riviere},\ and\ \citenamefont
  {Moeller}}]{dellis_experience_1987}%
  \BibitemOpen
  \bibfield  {author} {\bibinfo {author} {\bibfnamefont {A.~N.}\ \bibnamefont
  {Dellis}}, \bibinfo {author} {\bibfnamefont {M.~W.}\ \bibnamefont {Alcock}},
  \bibinfo {author} {\bibfnamefont {N.~R.~G.}\ \bibnamefont {Ainsworth}},
  \bibinfo {author} {\bibfnamefont {P.~R.}\ \bibnamefont {Collins}}, \bibinfo
  {author} {\bibfnamefont {S.~J.}\ \bibnamefont {Fielding}}, \bibinfo {author}
  {\bibfnamefont {J.}~\bibnamefont {Hugill}}, \bibinfo {author} {\bibfnamefont
  {P.~C.}\ \bibnamefont {Johnson}}, \bibinfo {author} {\bibfnamefont {A.~C.}\
  \bibnamefont {Riviere}}, \ and\ \bibinfo {author} {\bibfnamefont {C.~P.}\
  \bibnamefont {Moeller}},\ }in\ \href@noop {} {\emph {\bibinfo {booktitle}
  {Proceedings of the 6th {{Joint Workshop}} on {{Electron Cyclotron Emission}}
  ({{ECE}}) and {{Electron Cyclotron Resonance Heating}} ({{ECRH}})}}}\
  (\bibinfo {address} {{Oxford, UK}},\ \bibinfo {year} {1987})\ p.\ \bibinfo
  {pages} {247}\BibitemShut {NoStop}%
\bibitem [{\citenamefont {Moeller}(1987{\natexlab{c}})}]{moeller_trip_1987}%
  \BibitemOpen
  \bibfield  {author} {\bibinfo {author} {\bibfnamefont {C.~P.}\ \bibnamefont
  {Moeller}},\ }\href@noop {} {\enquote {\bibinfo {title} {Trip {{Report}}:
  {{Trip}} to {{Culham Laboratory}}},}\ }\bibinfo {type} {Tech. Rep.}\ \bibinfo
  {number} {GA-D--18928}\ (\bibinfo  {institution} {{GA Technologies}},\
  \bibinfo {year} {1987})\BibitemShut {NoStop}%
\bibitem [{\citenamefont {Moeller}\ \emph {et~al.}(1987)\citenamefont
  {Moeller}, \citenamefont {Prater}, \citenamefont {Riviere}, \citenamefont
  {Ainsworth}, \citenamefont {Dellis},\ and\ \citenamefont
  {Johnson}}]{moeller_high_1987}%
  \BibitemOpen
  \bibfield  {author} {\bibinfo {author} {\bibfnamefont {C.~P.}\ \bibnamefont
  {Moeller}}, \bibinfo {author} {\bibfnamefont {R.}~\bibnamefont {Prater}},
  \bibinfo {author} {\bibfnamefont {A.~C.}\ \bibnamefont {Riviere}}, \bibinfo
  {author} {\bibfnamefont {N.~R.~G.}\ \bibnamefont {Ainsworth}}, \bibinfo
  {author} {\bibfnamefont {A.~N.}\ \bibnamefont {Dellis}}, \ and\ \bibinfo
  {author} {\bibfnamefont {P.~C.}\ \bibnamefont {Johnson}},\ }in\ \href@noop {}
  {\emph {\bibinfo {booktitle} {Proceedings of the 6th {{Joint Workshop}} on
  {{Electron Cyclotron Emission}} ({{ECE}}) and {{Electron Cyclotron Resonance
  Heating}} ({{ECRH}})}}}\ (\bibinfo {address} {{Oxford, UK}},\ \bibinfo {year}
  {1987})\ p.\ \bibinfo {pages} {355}\BibitemShut {NoStop}%
\bibitem [{\citenamefont {Larsen}\ \emph {et~al.}(2019)\citenamefont {Larsen},
  \citenamefont {Korsholm}, \citenamefont {Gon{\c c}alves}, \citenamefont
  {Gutierrez}, \citenamefont {Henriques}, \citenamefont {Infante},
  \citenamefont {Jensen}, \citenamefont {Jessen}, \citenamefont {Klinkby},
  \citenamefont {Nonb{\o}l}, \citenamefont {Luis}, \citenamefont {Vale},
  \citenamefont {Lopes}, \citenamefont {Naulin}, \citenamefont {Nielsen},
  \citenamefont {Salewski}, \citenamefont {Rasmussen}, \citenamefont
  {Taormina}, \citenamefont {M{\o}lls{\o}e}, \citenamefont {Mussenbrock},\ and\
  \citenamefont {Trieschmann}}]{larsen_mitigation_2019}%
  \BibitemOpen
  \bibfield  {author} {\bibinfo {author} {\bibfnamefont {A.~W.}\ \bibnamefont
  {Larsen}}, \bibinfo {author} {\bibfnamefont {S.~B.}\ \bibnamefont
  {Korsholm}}, \bibinfo {author} {\bibfnamefont {B.}~\bibnamefont {Gon{\c
  c}alves}}, \bibinfo {author} {\bibfnamefont {H.~E.}\ \bibnamefont
  {Gutierrez}}, \bibinfo {author} {\bibfnamefont {E.}~\bibnamefont
  {Henriques}}, \bibinfo {author} {\bibfnamefont {V.}~\bibnamefont {Infante}},
  \bibinfo {author} {\bibfnamefont {T.}~\bibnamefont {Jensen}}, \bibinfo
  {author} {\bibfnamefont {M.}~\bibnamefont {Jessen}}, \bibinfo {author}
  {\bibfnamefont {E.~B.}\ \bibnamefont {Klinkby}}, \bibinfo {author}
  {\bibfnamefont {E.}~\bibnamefont {Nonb{\o}l}}, \bibinfo {author}
  {\bibfnamefont {R.}~\bibnamefont {Luis}}, \bibinfo {author} {\bibfnamefont
  {A.}~\bibnamefont {Vale}}, \bibinfo {author} {\bibfnamefont {A.}~\bibnamefont
  {Lopes}}, \bibinfo {author} {\bibfnamefont {V.}~\bibnamefont {Naulin}},
  \bibinfo {author} {\bibfnamefont {S.~K.}\ \bibnamefont {Nielsen}}, \bibinfo
  {author} {\bibfnamefont {M.}~\bibnamefont {Salewski}}, \bibinfo {author}
  {\bibfnamefont {J.}~\bibnamefont {Rasmussen}}, \bibinfo {author}
  {\bibfnamefont {A.}~\bibnamefont {Taormina}}, \bibinfo {author}
  {\bibfnamefont {C.}~\bibnamefont {M{\o}lls{\o}e}}, \bibinfo {author}
  {\bibfnamefont {T.}~\bibnamefont {Mussenbrock}}, \ and\ \bibinfo {author}
  {\bibfnamefont {J.}~\bibnamefont {Trieschmann}},\ }\href {\doibase
  10.1088/1748-0221/14/11/C11009} {\bibfield  {journal} {\bibinfo  {journal}
  {Journal of Instrumentation}\ }\textbf {\bibinfo {volume} {14}},\ \bibinfo
  {pages} {C11009} (\bibinfo {year} {2019})}\BibitemShut {NoStop}%
\bibitem [{\citenamefont {Gould}(1956)}]{gould_handbook_1956}%
  \BibitemOpen
  \bibfield  {author} {\bibinfo {author} {\bibfnamefont {L.}~\bibnamefont
  {Gould}},\ }\href@noop {} {\emph {\bibinfo {title} {Handbook on Breakdown of
  Air in Waveguide Systems}}}\ (\bibinfo  {publisher} {{Microwave
  Associates}},\ \bibinfo {address} {{Boston, USA}},\ \bibinfo {year}
  {1956})\BibitemShut {NoStop}%
\bibitem [{\citenamefont {Balanis}(1989)}]{balanis_advanced_1989}%
  \BibitemOpen
  \bibfield  {author} {\bibinfo {author} {\bibfnamefont {C.~A.}\ \bibnamefont
  {Balanis}},\ }\href@noop {} {\emph {\bibinfo {title} {Advanced {{Engineering
  Electromagnetics}}}}}\ (\bibinfo  {publisher} {{Wiley}},\ \bibinfo {address}
  {{New York, USA}},\ \bibinfo {year} {1989})\BibitemShut {NoStop}%
\bibitem [{\citenamefont {Kowalski}\ \emph {et~al.}(2010)\citenamefont
  {Kowalski}, \citenamefont {Tax}, \citenamefont {Shapiro}, \citenamefont
  {Sirigiri}, \citenamefont {Temkin}, \citenamefont {Bigelow},\ and\
  \citenamefont {Rasmussen}}]{kowalski_linearly_2010}%
  \BibitemOpen
  \bibfield  {author} {\bibinfo {author} {\bibfnamefont {E.~J.}\ \bibnamefont
  {Kowalski}}, \bibinfo {author} {\bibfnamefont {D.~S.}\ \bibnamefont {Tax}},
  \bibinfo {author} {\bibfnamefont {M.~A.}\ \bibnamefont {Shapiro}}, \bibinfo
  {author} {\bibfnamefont {J.~R.}\ \bibnamefont {Sirigiri}}, \bibinfo {author}
  {\bibfnamefont {R.~J.}\ \bibnamefont {Temkin}}, \bibinfo {author}
  {\bibfnamefont {T.~S.}\ \bibnamefont {Bigelow}}, \ and\ \bibinfo {author}
  {\bibfnamefont {D.~A.}\ \bibnamefont {Rasmussen}},\ }\href {\doibase
  10.1109/TMTT.2010.2078972} {\bibfield  {journal} {\bibinfo  {journal} {IEEE
  Transactions on Microwave Theory and Techniques}\ }\textbf {\bibinfo {volume}
  {58}},\ \bibinfo {pages} {2772} (\bibinfo {year} {2010})}\BibitemShut
  {NoStop}%
\bibitem [{noa()}]{noauthor_iter_d_q2j6me_nodate}%
  \BibitemOpen
  \href@noop {} {\enquote {\bibinfo {title} {{{ITER}}\_{{D}}\_{{Q2J6ME}},
  {{ITER Static Flux Density EPP}} ({{Coil}} + {{Plasma}}) v. 2.0},}\ }\bibinfo
  {type} {Private Communication}\BibitemShut {NoStop}%
\bibitem [{\citenamefont {Kartikeyan}, \citenamefont {Borie},\ and\
  \citenamefont {Thumm}(2004)}]{kartikeyan_gyrotrons_2004}%
  \BibitemOpen
  \bibfield  {author} {\bibinfo {author} {\bibfnamefont {M.~V.}\ \bibnamefont
  {Kartikeyan}}, \bibinfo {author} {\bibfnamefont {E.}~\bibnamefont {Borie}}, \
  and\ \bibinfo {author} {\bibfnamefont {M.}~\bibnamefont {Thumm}},\
  }\href@noop {} {\emph {\bibinfo {title} {Gyrotrons: {{High}}-{{Power
  Microwave}} and {{Millimeter Wave Technology}}}}},\ Advanced {{Texts}} in
  {{Physics}}\ (\bibinfo  {publisher} {{Springer}},\ \bibinfo {address}
  {{Berlin, Germany}},\ \bibinfo {year} {2004})\BibitemShut {NoStop}%
\bibitem [{\citenamefont {Weller}, \citenamefont {Greenshields},\ and\
  \citenamefont {Janssens}(2020)}]{weller_openfoam_2020}%
  \BibitemOpen
  \bibfield  {author} {\bibinfo {author} {\bibfnamefont {H.~G.}\ \bibnamefont
  {Weller}}, \bibinfo {author} {\bibfnamefont {C.~J.}\ \bibnamefont
  {Greenshields}}, \ and\ \bibinfo {author} {\bibfnamefont {M.}~\bibnamefont
  {Janssens}},\ }\href@noop {} {\enquote {\bibinfo {title} {{{OpenFOAM}},
  {{www.openfoam.org}}},}\ }\bibinfo {type} {Developement Version}\ (\bibinfo
  {year} {2020})\BibitemShut {NoStop}%
\bibitem [{\citenamefont {Scanlon}\ \emph {et~al.}(2010)\citenamefont
  {Scanlon}, \citenamefont {Roohi}, \citenamefont {White}, \citenamefont
  {Darbandi},\ and\ \citenamefont {Reese}}]{scanlon_open_2010}%
  \BibitemOpen
  \bibfield  {author} {\bibinfo {author} {\bibfnamefont {T.~J.}\ \bibnamefont
  {Scanlon}}, \bibinfo {author} {\bibfnamefont {E.}~\bibnamefont {Roohi}},
  \bibinfo {author} {\bibfnamefont {C.}~\bibnamefont {White}}, \bibinfo
  {author} {\bibfnamefont {M.}~\bibnamefont {Darbandi}}, \ and\ \bibinfo
  {author} {\bibfnamefont {J.~M.}\ \bibnamefont {Reese}},\ }\href {\doibase
  10.1016/j.compfluid.2010.07.014} {\bibfield  {journal} {\bibinfo  {journal}
  {Computers and Fluids}\ }\textbf {\bibinfo {volume} {39}},\ \bibinfo {pages}
  {2078} (\bibinfo {year} {2010})}\BibitemShut {NoStop}%
\bibitem [{\citenamefont {Bobzin}\ \emph {et~al.}(2013)\citenamefont {Bobzin},
  \citenamefont {Brinkmann}, \citenamefont {Mussenbrock}, \citenamefont
  {Bagcivan}, \citenamefont {Brugnara}, \citenamefont {Sch{\"a}fer},\ and\
  \citenamefont {Trieschmann}}]{bobzin_continuum_2013}%
  \BibitemOpen
  \bibfield  {author} {\bibinfo {author} {\bibfnamefont {K.}~\bibnamefont
  {Bobzin}}, \bibinfo {author} {\bibfnamefont {R.~P.}\ \bibnamefont
  {Brinkmann}}, \bibinfo {author} {\bibfnamefont {T.}~\bibnamefont
  {Mussenbrock}}, \bibinfo {author} {\bibfnamefont {N.}~\bibnamefont
  {Bagcivan}}, \bibinfo {author} {\bibfnamefont {R.~H.}\ \bibnamefont
  {Brugnara}}, \bibinfo {author} {\bibfnamefont {M.}~\bibnamefont
  {Sch{\"a}fer}}, \ and\ \bibinfo {author} {\bibfnamefont {J.}~\bibnamefont
  {Trieschmann}},\ }\href {\doibase 10.1016/j.surfcoat.2013.08.018} {\bibfield
  {journal} {\bibinfo  {journal} {Surface and Coatings Technology}\ }\textbf
  {\bibinfo {volume} {237}},\ \bibinfo {pages} {176} (\bibinfo {year}
  {2013})}\BibitemShut {NoStop}%
\bibitem [{\citenamefont {Trieschmann}\ and\ \citenamefont
  {Mussenbrock}(2015)}]{trieschmann_transport_2015}%
  \BibitemOpen
  \bibfield  {author} {\bibinfo {author} {\bibfnamefont {J.}~\bibnamefont
  {Trieschmann}}\ and\ \bibinfo {author} {\bibfnamefont {T.}~\bibnamefont
  {Mussenbrock}},\ }\href {\doibase 10.1063/1.4926878} {\bibfield  {journal}
  {\bibinfo  {journal} {Journal of Applied Physics}\ }\textbf {\bibinfo
  {volume} {118}},\ \bibinfo {pages} {033302} (\bibinfo {year}
  {2015})}\BibitemShut {NoStop}%
\bibitem [{\citenamefont {Trieschmann}(2018)}]{trieschmann_neutral_2018}%
  \BibitemOpen
  \bibfield  {author} {\bibinfo {author} {\bibfnamefont {J.}~\bibnamefont
  {Trieschmann}},\ }\href {\doibase 10.1002/ctpp.201700062} {\bibfield
  {journal} {\bibinfo  {journal} {Contributions to Plasma Physics}\ }\textbf
  {\bibinfo {volume} {58}},\ \bibinfo {pages} {394} (\bibinfo {year}
  {2018})}\BibitemShut {NoStop}%
\bibitem [{\citenamefont {Layes}\ \emph {et~al.}(2017)\citenamefont {Layes},
  \citenamefont {Monje}, \citenamefont {Corbella}, \citenamefont {Trieschmann},
  \citenamefont {{de los Arcos}},\ and\ \citenamefont {{von
  Keudell}}}]{layes_species_2017}%
  \BibitemOpen
  \bibfield  {author} {\bibinfo {author} {\bibfnamefont {V.}~\bibnamefont
  {Layes}}, \bibinfo {author} {\bibfnamefont {S.}~\bibnamefont {Monje}},
  \bibinfo {author} {\bibfnamefont {C.}~\bibnamefont {Corbella}}, \bibinfo
  {author} {\bibfnamefont {J.}~\bibnamefont {Trieschmann}}, \bibinfo {author}
  {\bibfnamefont {T.}~\bibnamefont {{de los Arcos}}}, \ and\ \bibinfo {author}
  {\bibfnamefont {A.}~\bibnamefont {{von Keudell}}},\ }\href {\doibase
  10.1063/1.4976999} {\bibfield  {journal} {\bibinfo  {journal} {Applied
  Physics Letters}\ }\textbf {\bibinfo {volume} {110}},\ \bibinfo {pages}
  {081603} (\bibinfo {year} {2017})}\BibitemShut {NoStop}%
\bibitem [{\citenamefont {Trieschmann}\ \emph {et~al.}(2018)\citenamefont
  {Trieschmann}, \citenamefont {Ries}, \citenamefont {Bibinov}, \citenamefont
  {Awakowicz}, \citenamefont {Mr{\'a}z}, \citenamefont {Schneider},\ and\
  \citenamefont {Mussenbrock}}]{trieschmann_combined_2018}%
  \BibitemOpen
  \bibfield  {author} {\bibinfo {author} {\bibfnamefont {J.}~\bibnamefont
  {Trieschmann}}, \bibinfo {author} {\bibfnamefont {S.}~\bibnamefont {Ries}},
  \bibinfo {author} {\bibfnamefont {N.}~\bibnamefont {Bibinov}}, \bibinfo
  {author} {\bibfnamefont {P.}~\bibnamefont {Awakowicz}}, \bibinfo {author}
  {\bibfnamefont {S.}~\bibnamefont {Mr{\'a}z}}, \bibinfo {author}
  {\bibfnamefont {J.~M.}\ \bibnamefont {Schneider}}, \ and\ \bibinfo {author}
  {\bibfnamefont {T.}~\bibnamefont {Mussenbrock}},\ }\href {\doibase
  10.1088/1361-6595/aac23e} {\bibfield  {journal} {\bibinfo  {journal} {Plasma
  Sources Science and Technology}\ }\textbf {\bibinfo {volume} {27}},\ \bibinfo
  {pages} {054003} (\bibinfo {year} {2018})}\BibitemShut {NoStop}%
\bibitem [{\citenamefont {Kirchheim}\ \emph {et~al.}(2019)\citenamefont
  {Kirchheim}, \citenamefont {Wilski}, \citenamefont {Jaritz}, \citenamefont
  {Mitschker}, \citenamefont {Oberberg}, \citenamefont {Trieschmann},
  \citenamefont {Banko}, \citenamefont {Brochhagen}, \citenamefont
  {Schreckenberg}, \citenamefont {Hopmann}, \citenamefont {B{\"o}ke},
  \citenamefont {Benedikt}, \citenamefont {{de los Arcos}}, \citenamefont
  {Grundmeier}, \citenamefont {Grochla}, \citenamefont {Ludwig}, \citenamefont
  {Mussenbrock}, \citenamefont {Brinkmann}, \citenamefont {Awakowicz},\ and\
  \citenamefont {Dahlmann}}]{kirchheim_improved_2019}%
  \BibitemOpen
  \bibfield  {author} {\bibinfo {author} {\bibfnamefont {D.}~\bibnamefont
  {Kirchheim}}, \bibinfo {author} {\bibfnamefont {S.}~\bibnamefont {Wilski}},
  \bibinfo {author} {\bibfnamefont {M.}~\bibnamefont {Jaritz}}, \bibinfo
  {author} {\bibfnamefont {F.}~\bibnamefont {Mitschker}}, \bibinfo {author}
  {\bibfnamefont {M.}~\bibnamefont {Oberberg}}, \bibinfo {author}
  {\bibfnamefont {J.}~\bibnamefont {Trieschmann}}, \bibinfo {author}
  {\bibfnamefont {L.}~\bibnamefont {Banko}}, \bibinfo {author} {\bibfnamefont
  {M.}~\bibnamefont {Brochhagen}}, \bibinfo {author} {\bibfnamefont
  {R.}~\bibnamefont {Schreckenberg}}, \bibinfo {author} {\bibfnamefont
  {C.}~\bibnamefont {Hopmann}}, \bibinfo {author} {\bibfnamefont
  {M.}~\bibnamefont {B{\"o}ke}}, \bibinfo {author} {\bibfnamefont
  {J.}~\bibnamefont {Benedikt}}, \bibinfo {author} {\bibfnamefont
  {T.}~\bibnamefont {{de los Arcos}}}, \bibinfo {author} {\bibfnamefont
  {G.}~\bibnamefont {Grundmeier}}, \bibinfo {author} {\bibfnamefont
  {D.}~\bibnamefont {Grochla}}, \bibinfo {author} {\bibfnamefont
  {A.}~\bibnamefont {Ludwig}}, \bibinfo {author} {\bibfnamefont
  {T.}~\bibnamefont {Mussenbrock}}, \bibinfo {author} {\bibfnamefont {R.~P.}\
  \bibnamefont {Brinkmann}}, \bibinfo {author} {\bibfnamefont {P.}~\bibnamefont
  {Awakowicz}}, \ and\ \bibinfo {author} {\bibfnamefont {R.}~\bibnamefont
  {Dahlmann}},\ }\href {\doibase 10.1007/s11998-018-0138-4} {\bibfield
  {journal} {\bibinfo  {journal} {Journal of Coatings Technology and Research}\
  }\textbf {\bibinfo {volume} {16}},\ \bibinfo {pages} {573} (\bibinfo {year}
  {2019})}\BibitemShut {NoStop}%
\bibitem [{\citenamefont {Turner}\ \emph {et~al.}(2013)\citenamefont {Turner},
  \citenamefont {Derzsi}, \citenamefont {Donk{\'o}}, \citenamefont {Eremin},
  \citenamefont {Kelly}, \citenamefont {Lafleur},\ and\ \citenamefont
  {Mussenbrock}}]{turner_simulation_2013}%
  \BibitemOpen
  \bibfield  {author} {\bibinfo {author} {\bibfnamefont {M.~M.}\ \bibnamefont
  {Turner}}, \bibinfo {author} {\bibfnamefont {A.}~\bibnamefont {Derzsi}},
  \bibinfo {author} {\bibfnamefont {Z.}~\bibnamefont {Donk{\'o}}}, \bibinfo
  {author} {\bibfnamefont {D.}~\bibnamefont {Eremin}}, \bibinfo {author}
  {\bibfnamefont {S.~J.}\ \bibnamefont {Kelly}}, \bibinfo {author}
  {\bibfnamefont {T.}~\bibnamefont {Lafleur}}, \ and\ \bibinfo {author}
  {\bibfnamefont {T.}~\bibnamefont {Mussenbrock}},\ }\href {\doibase
  10.1063/1.4775084} {\bibfield  {journal} {\bibinfo  {journal} {Physics of
  Plasmas}\ }\textbf {\bibinfo {volume} {20}},\ \bibinfo {pages} {013507}
  (\bibinfo {year} {2013})}\BibitemShut {NoStop}%
\bibitem [{\citenamefont {Trieschmann}(2017)}]{trieschmann_particle_2017}%
  \BibitemOpen
  \bibfield  {author} {\bibinfo {author} {\bibfnamefont {J.}~\bibnamefont
  {Trieschmann}},\ }\emph {\bibinfo {title} {Particle Transport in
  Technological Plasmas}},\ \href@noop {} {\bibinfo {type} {{{PhD Thesis}}}},\
  \bibinfo  {school} {Ruhr-Universit\"at Bochum}, \bibinfo {address} {{Bochum,
  Germany}} (\bibinfo {year} {2017})\BibitemShut {NoStop}%
\bibitem [{\citenamefont {Boris}(1970)}]{boris_relativistic_1970}%
  \BibitemOpen
  \bibfield  {author} {\bibinfo {author} {\bibfnamefont {J.}~\bibnamefont
  {Boris}},\ }in\ \href@noop {} {\emph {\bibinfo {booktitle} {Proceedings of
  4th {{Conference}} on {{Numerical Simulation}} of {{Plasmas}}}}}\ (\bibinfo
  {publisher} {{Naval Research Laboratory}},\ \bibinfo {address} {{Washington
  DC, USA}},\ \bibinfo {year} {1970})\ p.~\bibinfo {pages} {3}\BibitemShut
  {NoStop}%
\bibitem [{\citenamefont {Zenitani}\ and\ \citenamefont
  {Umeda}(2018)}]{zenitani_boris_2018}%
  \BibitemOpen
  \bibfield  {author} {\bibinfo {author} {\bibfnamefont {S.}~\bibnamefont
  {Zenitani}}\ and\ \bibinfo {author} {\bibfnamefont {T.}~\bibnamefont
  {Umeda}},\ }\href {\doibase 10.1063/1.5051077} {\bibfield  {journal}
  {\bibinfo  {journal} {Physics of Plasmas}\ }\textbf {\bibinfo {volume}
  {25}},\ \bibinfo {pages} {112110} (\bibinfo {year} {2018})}\BibitemShut
  {NoStop}%
\bibitem [{\citenamefont {Bird}(1994)}]{bird_molecular_1994}%
  \BibitemOpen
  \bibfield  {author} {\bibinfo {author} {\bibfnamefont {G.~A.}\ \bibnamefont
  {Bird}},\ }\href@noop {} {\emph {\bibinfo {title} {Molecular {{Gas Dynamics}}
  and the {{Direct Simulation}} of {{Gas Flows}}}}}\ (\bibinfo  {publisher}
  {{Oxford University Press}},\ \bibinfo {address} {{New York, USA}},\ \bibinfo
  {year} {1994})\BibitemShut {NoStop}%
\bibitem [{\citenamefont {Skullerud}(1968)}]{skullerud_stochastic_1968}%
  \BibitemOpen
  \bibfield  {author} {\bibinfo {author} {\bibfnamefont {H.~R.}\ \bibnamefont
  {Skullerud}},\ }\href {\doibase 10.1088/0022-3727/1/11/423} {\bibfield
  {journal} {\bibinfo  {journal} {Journal of Physics D: Applied Physics}\
  }\textbf {\bibinfo {volume} {1}},\ \bibinfo {pages} {1567} (\bibinfo {year}
  {1968})}\BibitemShut {NoStop}%
\bibitem [{\citenamefont {LeVeque}(2002)}]{leveque_finite_2002}%
  \BibitemOpen
  \bibfield  {author} {\bibinfo {author} {\bibfnamefont {R.~J.}\ \bibnamefont
  {LeVeque}},\ }\href {\doibase 10.1017/CBO9780511791253} {\emph {\bibinfo
  {title} {Finite {{Volume Methods}} for {{Hyperbolic Problems}}}}}\ (\bibinfo
  {publisher} {{Cambridge University Press}},\ \bibinfo {address} {{Cambridge,
  UK}},\ \bibinfo {year} {2002})\BibitemShut {NoStop}%
\bibitem [{\citenamefont
  {Kollath}(1956)}]{kollath_sekundarelektronen-emission_1956}%
  \BibitemOpen
  \bibfield  {author} {\bibinfo {author} {\bibfnamefont {R.}~\bibnamefont
  {Kollath}},\ }in\ \href {\doibase 10.1007/978-3-642-45844-6_3} {\emph
  {\bibinfo {booktitle} {Electron-{{Emission Gas Discharges I}} /
  {{Elektronen}}-{{Emission Gasentladungen I}}}}},\ \bibinfo {series}
  {Encyclopedia of {{Physics}} / {{Handbuch}} Der {{Physik}}}, Vol.~\bibinfo
  {volume} {21},\ \bibinfo {editor} {edited by\ \bibinfo {editor}
  {\bibfnamefont {W.~B.}\ \bibnamefont {Nottingham}}, \bibinfo {editor}
  {\bibfnamefont {R.~H.}\ \bibnamefont {Good}}, \bibinfo {editor}
  {\bibfnamefont {E.~W.}\ \bibnamefont {M{\"u}ller}}, \bibinfo {editor}
  {\bibfnamefont {R.}~\bibnamefont {Kollath}}, \bibinfo {editor} {\bibfnamefont
  {G.~L.}\ \bibnamefont {Weissler}}, \bibinfo {editor} {\bibfnamefont {W.~P.}\
  \bibnamefont {Allis}}, \bibinfo {editor} {\bibfnamefont {L.~B.}\ \bibnamefont
  {Loeb}}, \bibinfo {editor} {\bibfnamefont {A.}~\bibnamefont {{von Engel}}}, \
  and\ \bibinfo {editor} {\bibfnamefont {P.~F.}\ \bibnamefont {Little}}}\
  (\bibinfo  {publisher} {{Springer}},\ \bibinfo {address} {{Berlin,
  Germany}},\ \bibinfo {year} {1956})\ pp.\ \bibinfo {pages}
  {232--303}\BibitemShut {NoStop}%
\bibitem [{\citenamefont {Ruzic}\ \emph {et~al.}(1982)\citenamefont {Ruzic},
  \citenamefont {Moore}, \citenamefont {Manos},\ and\ \citenamefont
  {Cohen}}]{ruzic_secondary_1982}%
  \BibitemOpen
  \bibfield  {author} {\bibinfo {author} {\bibfnamefont {D.}~\bibnamefont
  {Ruzic}}, \bibinfo {author} {\bibfnamefont {R.}~\bibnamefont {Moore}},
  \bibinfo {author} {\bibfnamefont {D.}~\bibnamefont {Manos}}, \ and\ \bibinfo
  {author} {\bibfnamefont {S.}~\bibnamefont {Cohen}},\ }\href {\doibase
  10.1116/1.571569} {\bibfield  {journal} {\bibinfo  {journal} {Journal of
  Vacuum Science and Technology}\ }\textbf {\bibinfo {volume} {20}},\ \bibinfo
  {pages} {1313} (\bibinfo {year} {1982})}\BibitemShut {NoStop}%
\bibitem [{\citenamefont {Janev}(1991)}]{janev_atomic_1991}%
  \BibitemOpen
  \bibinfo {editor} {\bibfnamefont {R.~K.}\ \bibnamefont {Janev}},\ ed.,\
  \href@noop {} {\emph {\bibinfo {title} {Atomic and {{Plasma}}-{{Material
  Interaction Data}} for {{Fusion}}}}},\ Vol.~\bibinfo {volume} {1}\ (\bibinfo
  {publisher} {{International Atomic Energy Agency}},\ \bibinfo {address}
  {{Vienna, Austria}},\ \bibinfo {year} {1991})\BibitemShut {NoStop}%
\bibitem [{\citenamefont {Walker}\ \emph {et~al.}(2008)\citenamefont {Walker},
  \citenamefont {El-Gomati}, \citenamefont {Assa'd},\ and\ \citenamefont
  {Zadra{\v z}il}}]{walker_secondary_2008}%
  \BibitemOpen
  \bibfield  {author} {\bibinfo {author} {\bibfnamefont {C.~G.~H.}\
  \bibnamefont {Walker}}, \bibinfo {author} {\bibfnamefont {M.~M.}\
  \bibnamefont {El-Gomati}}, \bibinfo {author} {\bibfnamefont {A.~M.~D.}\
  \bibnamefont {Assa'd}}, \ and\ \bibinfo {author} {\bibfnamefont
  {M.}~\bibnamefont {Zadra{\v z}il}},\ }\href {\doibase 10.1002/sca.20124}
  {\bibfield  {journal} {\bibinfo  {journal} {Scanning}\ }\textbf {\bibinfo
  {volume} {30}},\ \bibinfo {pages} {365} (\bibinfo {year} {2008})}\BibitemShut
  {NoStop}%
\bibitem [{\citenamefont {Tolias}(2014)}]{tolias_secondary_2014}%
  \BibitemOpen
  \bibfield  {author} {\bibinfo {author} {\bibfnamefont {P.}~\bibnamefont
  {Tolias}},\ }\href {\doibase 10.1088/0741-3335/56/12/123002} {\bibfield
  {journal} {\bibinfo  {journal} {Plasma Physics and Controlled Fusion}\
  }\textbf {\bibinfo {volume} {56}},\ \bibinfo {pages} {123002} (\bibinfo
  {year} {2014})}\BibitemShut {NoStop}%
\bibitem [{\citenamefont {Young}(1957)}]{young_dissipation_1957}%
  \BibitemOpen
  \bibfield  {author} {\bibinfo {author} {\bibfnamefont {J.~R.}\ \bibnamefont
  {Young}},\ }\href {\doibase 10.1063/1.1722794} {\bibfield  {journal}
  {\bibinfo  {journal} {Journal of Applied Physics}\ }\textbf {\bibinfo
  {volume} {28}},\ \bibinfo {pages} {524} (\bibinfo {year} {1957})}\BibitemShut
  {NoStop}%
\bibitem [{\citenamefont {Lye}\ and\ \citenamefont
  {Dekker}(1957)}]{lye_theory_1957}%
  \BibitemOpen
  \bibfield  {author} {\bibinfo {author} {\bibfnamefont {R.~G.}\ \bibnamefont
  {Lye}}\ and\ \bibinfo {author} {\bibfnamefont {A.~J.}\ \bibnamefont
  {Dekker}},\ }\href {\doibase 10.1103/PhysRev.107.977} {\bibfield  {journal}
  {\bibinfo  {journal} {Physical Review}\ }\textbf {\bibinfo {volume} {107}},\
  \bibinfo {pages} {977} (\bibinfo {year} {1957})}\BibitemShut {NoStop}%
\bibitem [{\citenamefont {Hunger}\ and\ \citenamefont
  {K{\"u}chler}(1979)}]{hunger_measurements_1979}%
  \BibitemOpen
  \bibfield  {author} {\bibinfo {author} {\bibfnamefont {H.-J.}\ \bibnamefont
  {Hunger}}\ and\ \bibinfo {author} {\bibfnamefont {L.}~\bibnamefont
  {K{\"u}chler}},\ }\href {\doibase 10.1002/pssa.2210560157} {\bibfield
  {journal} {\bibinfo  {journal} {Physica Status Solidi (a)}\ }\textbf
  {\bibinfo {volume} {56}},\ \bibinfo {pages} {K45} (\bibinfo {year}
  {1979})}\BibitemShut {NoStop}%
\bibitem [{\citenamefont {Brown}(1967)}]{brown_basic_1967}%
  \BibitemOpen
  \bibfield  {author} {\bibinfo {author} {\bibfnamefont {S.~C.}\ \bibnamefont
  {Brown}},\ }\href@noop {} {\emph {\bibinfo {title} {Basic {{Data}} of
  {{Plasma Physics}}: {{The Fundamental Data}} on {{Electrical Discharges}} in
  {{Gases}}}}}\ (\bibinfo  {publisher} {{MIT Press}},\ \bibinfo {address}
  {{Massachusetts, USA}},\ \bibinfo {year} {1967})\BibitemShut {NoStop}%
\bibitem [{\citenamefont {Large}\ and\ \citenamefont
  {Whitlock}(1962)}]{large_secondary_1962}%
  \BibitemOpen
  \bibfield  {author} {\bibinfo {author} {\bibfnamefont {L.~N.}\ \bibnamefont
  {Large}}\ and\ \bibinfo {author} {\bibfnamefont {W.~S.}\ \bibnamefont
  {Whitlock}},\ }\href {\doibase 10.1088/0370-1328/79/1/319} {\bibfield
  {journal} {\bibinfo  {journal} {Proceedings of the Physical Society}\
  }\textbf {\bibinfo {volume} {79}},\ \bibinfo {pages} {148} (\bibinfo {year}
  {1962})}\BibitemShut {NoStop}%
\bibitem [{\citenamefont {Svensson}\ and\ \citenamefont
  {Holm{\'e}n}(1982)}]{svensson_electron_1982}%
  \BibitemOpen
  \bibfield  {author} {\bibinfo {author} {\bibfnamefont {B.}~\bibnamefont
  {Svensson}}\ and\ \bibinfo {author} {\bibfnamefont {G.}~\bibnamefont
  {Holm{\'e}n}},\ }\href {\doibase 10.1103/PhysRevB.25.3056} {\bibfield
  {journal} {\bibinfo  {journal} {Physical Review B}\ }\textbf {\bibinfo
  {volume} {25}},\ \bibinfo {pages} {3056} (\bibinfo {year}
  {1982})}\BibitemShut {NoStop}%
\bibitem [{\citenamefont {Zalm}\ and\ \citenamefont
  {Beckers}(1985)}]{zalm_ion-induced_1985}%
  \BibitemOpen
  \bibfield  {author} {\bibinfo {author} {\bibfnamefont {P.~C.}\ \bibnamefont
  {Zalm}}\ and\ \bibinfo {author} {\bibfnamefont {L.~J.}\ \bibnamefont
  {Beckers}},\ }\href {\doibase 10.1016/0039-6028(85)90136-0} {\bibfield
  {journal} {\bibinfo  {journal} {Surface Science}\ }\textbf {\bibinfo {volume}
  {152-153}},\ \bibinfo {pages} {135} (\bibinfo {year} {1985})}\BibitemShut
  {NoStop}%
\bibitem [{\citenamefont {Szapiro}, \citenamefont {Rocca},\ and\ \citenamefont
  {Prabhuram}(1988)}]{szapiro_electron_1988}%
  \BibitemOpen
  \bibfield  {author} {\bibinfo {author} {\bibfnamefont {B.}~\bibnamefont
  {Szapiro}}, \bibinfo {author} {\bibfnamefont {J.~J.}\ \bibnamefont {Rocca}},
  \ and\ \bibinfo {author} {\bibfnamefont {T.}~\bibnamefont {Prabhuram}},\
  }\href {\doibase 10.1063/1.100401} {\bibfield  {journal} {\bibinfo  {journal}
  {Applied Physics Letters}\ }\textbf {\bibinfo {volume} {53}},\ \bibinfo
  {pages} {358} (\bibinfo {year} {1988})}\BibitemShut {NoStop}%
\bibitem [{\citenamefont {Pancheshnyi}\ \emph {et~al.}(2012)\citenamefont
  {Pancheshnyi}, \citenamefont {Biagi}, \citenamefont {Bordage}, \citenamefont
  {Hagelaar}, \citenamefont {Morgan}, \citenamefont {Phelps},\ and\
  \citenamefont {Pitchford}}]{pancheshnyi_lxcat_2012}%
  \BibitemOpen
  \bibfield  {author} {\bibinfo {author} {\bibfnamefont {S.}~\bibnamefont
  {Pancheshnyi}}, \bibinfo {author} {\bibfnamefont {S.}~\bibnamefont {Biagi}},
  \bibinfo {author} {\bibfnamefont {M.~C.}\ \bibnamefont {Bordage}}, \bibinfo
  {author} {\bibfnamefont {G.~J.~M.}\ \bibnamefont {Hagelaar}}, \bibinfo
  {author} {\bibfnamefont {W.~L.}\ \bibnamefont {Morgan}}, \bibinfo {author}
  {\bibfnamefont {A.~V.}\ \bibnamefont {Phelps}}, \ and\ \bibinfo {author}
  {\bibfnamefont {L.~C.}\ \bibnamefont {Pitchford}},\ }\href {\doibase
  10.1016/j.chemphys.2011.04.020} {\bibfield  {journal} {\bibinfo  {journal}
  {Chemical Physics}\ }\textbf {\bibinfo {volume} {398}},\ \bibinfo {pages}
  {148} (\bibinfo {year} {2012})}\BibitemShut {NoStop}%
\bibitem [{\citenamefont {Biagi}()}]{biagi_lxcat_nodate}%
  \BibitemOpen
  \bibfield  {author} {\bibinfo {author} {\bibfnamefont {S.~F.}\ \bibnamefont
  {Biagi}},\ }\href@noop {} {\enquote {\bibinfo {title} {{{LXCat Biagi}}
  database},}\ }\bibinfo {type} {{{www.lxcat.net}}, Retrieved on {{March}} 15,
  2019, {{Fortran}} Program {{Magboltz}}, Version 8.9 and after,
  {{https://magboltz.web.cern.ch/magboltz/}}}\BibitemShut {NoStop}%
\bibitem [{\citenamefont {Alves}(2014)}]{alves_ist-lisbon_2014}%
  \BibitemOpen
  \bibfield  {author} {\bibinfo {author} {\bibfnamefont {L.~L.}\ \bibnamefont
  {Alves}},\ }\href {\doibase 10.1088/1742-6596/565/1/012007} {\bibfield
  {journal} {\bibinfo  {journal} {Journal of Physics: Conference Series}\
  }\textbf {\bibinfo {volume} {565}},\ \bibinfo {pages} {012007} (\bibinfo
  {year} {2014})}\BibitemShut {NoStop}%
\bibitem [{\citenamefont {Alves}\ and\ \citenamefont
  {Guerra}()}]{alves_lxcat_nodate}%
  \BibitemOpen
  \bibfield  {author} {\bibinfo {author} {\bibfnamefont {L.~L.}\ \bibnamefont
  {Alves}}\ and\ \bibinfo {author} {\bibfnamefont {V.}~\bibnamefont {Guerra}},\
  }\href@noop {} {\enquote {\bibinfo {title} {{{LXCat IST}}-{{Lisbon}}
  database},}\ }\bibinfo {type} {{{www.lxcat.net}}, Retrieved on {{November}}
  29, 2018}\BibitemShut {NoStop}%
\bibitem [{\citenamefont {Phelps}(2008)}]{phelps_span_2008}%
  \BibitemOpen
  \bibfield  {author} {\bibinfo {author} {\bibfnamefont {A.~V.}\ \bibnamefont
  {Phelps}},\ }\href@noop {} {\enquote {\bibinfo {title}
  {{{ftp://jila.colorado.edu/collision\_data/electronneutral/ELECTRON.TXT}}},}\
  }\bibinfo {type} {Private Communication}\ (\bibinfo {year}
  {2008})\BibitemShut {NoStop}%
\bibitem [{\citenamefont {van Wingerden}\ \emph {et~al.}(1977)\citenamefont
  {van Wingerden}, \citenamefont {de~Heer}, \citenamefont {Weigold},\ and\
  \citenamefont {Nygaard}}]{wingerden_elastic_1977}%
  \BibitemOpen
  \bibfield  {author} {\bibinfo {author} {\bibfnamefont {B.}~\bibnamefont {van
  Wingerden}}, \bibinfo {author} {\bibfnamefont {F.~J.}\ \bibnamefont
  {de~Heer}}, \bibinfo {author} {\bibfnamefont {E.}~\bibnamefont {Weigold}}, \
  and\ \bibinfo {author} {\bibfnamefont {K.~J.}\ \bibnamefont {Nygaard}},\
  }\href {\doibase 10.1088/0022-3700/10/7/023} {\bibfield  {journal} {\bibinfo
  {journal} {Journal of Physics B: Atomic and Molecular Physics}\ }\textbf
  {\bibinfo {volume} {10}},\ \bibinfo {pages} {1345} (\bibinfo {year}
  {1977})}\BibitemShut {NoStop}%
\bibitem [{\citenamefont {Rapp}, \citenamefont {Englander-Golden},\ and\
  \citenamefont {Briglia}(1965)}]{rapp_cross_1965}%
  \BibitemOpen
  \bibfield  {author} {\bibinfo {author} {\bibfnamefont {D.}~\bibnamefont
  {Rapp}}, \bibinfo {author} {\bibfnamefont {P.}~\bibnamefont
  {Englander-Golden}}, \ and\ \bibinfo {author} {\bibfnamefont {D.~D.}\
  \bibnamefont {Briglia}},\ }\href {\doibase 10.1063/1.1695897} {\bibfield
  {journal} {\bibinfo  {journal} {The Journal of Chemical Physics}\ }\textbf
  {\bibinfo {volume} {42}},\ \bibinfo {pages} {4081} (\bibinfo {year}
  {1965})}\BibitemShut {NoStop}%
\bibitem [{\citenamefont {Rapp}\ and\ \citenamefont
  {Englander-Golden}(1965)}]{rapp_total_1965}%
  \BibitemOpen
  \bibfield  {author} {\bibinfo {author} {\bibfnamefont {D.}~\bibnamefont
  {Rapp}}\ and\ \bibinfo {author} {\bibfnamefont {P.}~\bibnamefont
  {Englander-Golden}},\ }\href {\doibase 10.1063/1.1696957} {\bibfield
  {journal} {\bibinfo  {journal} {The Journal of Chemical Physics}\ }\textbf
  {\bibinfo {volume} {43}},\ \bibinfo {pages} {1464} (\bibinfo {year}
  {1965})}\BibitemShut {NoStop}%
\bibitem [{\citenamefont {Gorse}\ \emph {et~al.}(1987)\citenamefont {Gorse},
  \citenamefont {Capitelli}, \citenamefont {Bacal}, \citenamefont {Bretagne},\
  and\ \citenamefont {Lagan{\`a}}}]{gorse_progress_1987}%
  \BibitemOpen
  \bibfield  {author} {\bibinfo {author} {\bibfnamefont {C.}~\bibnamefont
  {Gorse}}, \bibinfo {author} {\bibfnamefont {M.}~\bibnamefont {Capitelli}},
  \bibinfo {author} {\bibfnamefont {M.}~\bibnamefont {Bacal}}, \bibinfo
  {author} {\bibfnamefont {J.}~\bibnamefont {Bretagne}}, \ and\ \bibinfo
  {author} {\bibfnamefont {A.}~\bibnamefont {Lagan{\`a}}},\ }\href {\doibase
  10.1016/0301-0104(87)80120-9} {\bibfield  {journal} {\bibinfo  {journal}
  {Chemical Physics}\ }\textbf {\bibinfo {volume} {117}},\ \bibinfo {pages}
  {177} (\bibinfo {year} {1987})}\BibitemShut {NoStop}%
\bibitem [{\citenamefont {Tawara}\ \emph {et~al.}(1990)\citenamefont {Tawara},
  \citenamefont {Itikawa}, \citenamefont {Nishimura},\ and\ \citenamefont
  {Yoshino}}]{tawara_cross_1990}%
  \BibitemOpen
  \bibfield  {author} {\bibinfo {author} {\bibfnamefont {H.}~\bibnamefont
  {Tawara}}, \bibinfo {author} {\bibfnamefont {Y.}~\bibnamefont {Itikawa}},
  \bibinfo {author} {\bibfnamefont {H.}~\bibnamefont {Nishimura}}, \ and\
  \bibinfo {author} {\bibfnamefont {M.}~\bibnamefont {Yoshino}},\ }\href
  {\doibase 10.1063/1.555856} {\bibfield  {journal} {\bibinfo  {journal}
  {Journal of Physical and Chemical Reference Data}\ }\textbf {\bibinfo
  {volume} {19}},\ \bibinfo {pages} {617} (\bibinfo {year} {1990})}\BibitemShut
  {NoStop}%
\bibitem [{\citenamefont {{\v S}imko}\ \emph {et~al.}(1997)\citenamefont {{\v
  S}imko}, \citenamefont {Marti{\v s}ovit{\v s}}, \citenamefont {Bretagne},\
  and\ \citenamefont {Gousset}}]{simko_computer_1997}%
  \BibitemOpen
  \bibfield  {author} {\bibinfo {author} {\bibfnamefont {T.}~\bibnamefont {{\v
  S}imko}}, \bibinfo {author} {\bibfnamefont {V.}~\bibnamefont {Marti{\v
  s}ovit{\v s}}}, \bibinfo {author} {\bibfnamefont {J.}~\bibnamefont
  {Bretagne}}, \ and\ \bibinfo {author} {\bibfnamefont {G.}~\bibnamefont
  {Gousset}},\ }\href {\doibase 10.1103/PhysRevE.56.5908} {\bibfield  {journal}
  {\bibinfo  {journal} {Physical Review E}\ }\textbf {\bibinfo {volume} {56}},\
  \bibinfo {pages} {5908} (\bibinfo {year} {1997})}\BibitemShut {NoStop}%
\bibitem [{\citenamefont {Korolov}\ and\ \citenamefont
  {Donk{\'o}}(2015)}]{korolov_breakdown_2015}%
  \BibitemOpen
  \bibfield  {author} {\bibinfo {author} {\bibfnamefont {I.}~\bibnamefont
  {Korolov}}\ and\ \bibinfo {author} {\bibfnamefont {Z.}~\bibnamefont
  {Donk{\'o}}},\ }\href {\doibase 10.1063/1.4929858} {\bibfield  {journal}
  {\bibinfo  {journal} {Physics of Plasmas}\ }\textbf {\bibinfo {volume}
  {22}},\ \bibinfo {pages} {093501} (\bibinfo {year} {2015})}\BibitemShut
  {NoStop}%
\bibitem [{\citenamefont {Aanesland}\ and\ \citenamefont
  {Fredriksen}(2003)}]{aanesland_electron_2003}%
  \BibitemOpen
  \bibfield  {author} {\bibinfo {author} {\bibfnamefont {A.}~\bibnamefont
  {Aanesland}}\ and\ \bibinfo {author} {\bibfnamefont {{\AA}.}~\bibnamefont
  {Fredriksen}},\ }\href {\doibase 10.1063/1.1611613} {\bibfield  {journal}
  {\bibinfo  {journal} {Review of Scientific Instruments}\ }\textbf {\bibinfo
  {volume} {74}},\ \bibinfo {pages} {4336} (\bibinfo {year}
  {2003})}\BibitemShut {NoStop}%
\end{thebibliography}%

\end{document}